\documentclass[11pt]{amsart}
\usepackage{amssymb,amsthm}
 \newtheorem{thm}{Theorem}[subsection]
 \newtheorem{cor}[thm]{Corollary}
 \newtheorem{lem}[thm]{Lemma}
 \newtheorem{prop}[thm]{Proposition}
 \theoremstyle{definition}
 \newtheorem{defn}[thm]{Definition}
 \newtheorem{rem}[thm]{Remark}
 \numberwithin{equation}{subsection}

\theoremstyle{definition}

 \newcommand{\s}{\mathbb{S}}

 \newcommand{\X}{\frak{X}}
\newcommand{\T}{\triangle}

 \newcommand{\PF}{\mbox{Pf}}

 \newcommand{\K}{\mbox{K}}

  \newcommand{\TX}{\tilde{X}}
\newcommand{\TZ}{\tilde{Z}}
\DeclareMathOperator{\Prob}{Prob} \DeclareMathOperator{\sgn}{sgn}
\DeclareMathOperator{\Img}{Im} \DeclareMathOperator{\const}{const}
\DeclareMathOperator{\Airy}{Ai} \DeclareMathOperator{\Max}{Max}
 \DeclareMathOperator{\PFAFF}{Pf}
\DeclareMathOperator{\RES}{Res}
\begin{document}
\title[Characteristic Polynomials]
 {\bf{Averages of characteristic polynomials in Random Matrix Theory}}

\author{ A. Borodin and E. Strahov }

\address{Department of Mathematics,  253-37,  Caltech, Pasadena, CA 91125 }

\email{borodin@caltech.edu, strahov@caltech.edu}







\begin{abstract}
We compute averages of products and ratios of characteristic
polynomials associated with Orthogonal, Unitary, and Symplectic
Ensembles of Random Matrix Theory. The pfaffian/determinantal
formulas for these averages are obtained, and the bulk scaling
asymptotic limits are found for ensembles with Gaussian weights.
Classical results for the correlation functions of the random
matrix ensembles and their bulk scaling limits are deduced from
these formulas by a simple computation.

We employ a discrete approximation method: the problem is solved
for discrete analogues of random matrix ensembles originating from
representation theory, and then a limit transition is performed.
Exact pfaffian/determinantal formulas for the discrete averages
are proved using standard tools of linear algebra; no application
of orthogonal or skew-orthogonal polynomials is needed.
\end{abstract}

\maketitle

\section{Introduction}
\subsection{The problem} Consider the linear space of real
symmetric, Hermitian, or quaternion real Hermitian square matrices
$H$ with the Gaussian measure
\begin{equation}
P(dH)=\operatorname{const}\cdot \exp(-Tr H^2)\,dH
\end{equation}
These probability spaces are the basic objects of interest of
Random Matrix Theory (RMT, for short); they are known as {\it
Gaussian Orthogonal, Unitary} and {\it Symplectic ensembles},
respectively (GOE, GUE, and GSE). The goal of this paper is to
study the averages of products and ratios of characteristic
polynomials
\begin{equation}\label{INtroductionAverages}
\left\langle \frac{\det(\alpha_1-H)\cdots\det(\alpha_k-H)}
{\det(\beta_1-H)\cdots\det(\beta_m-H)} \right\rangle
\end{equation}
with respect to these ensembles and their generalizations.

Despite the fact that the Gaussian ensembles have been extensively
studied, the progress on evaluating  averages
(\ref{INtroductionAverages}) remained rather limited until very
recently. Starting from 1995 there appeared a number of papers by
different authors where the case of unitary ensembles (exact
definitions are below) was essentially settled both for the random
matrices of finite size and for their scaling limits in the bulk
of spectrum as the size of matrices tends to infinity, see Refs.
\cite{andreev,brezin0,brezin1,mehta1,fyodorov2,fyodorov4,fyodorov5,fyodorov6,strahov,fyodorov1,baik,akemann1,akemann2,vanlessen}.
The results turned out to be closely related to some problems of
the classical number theory, see Refs. \cite{conrey},
\cite{hughes1}-\cite{keating3} for details.

Some progress have also been achieved in the orthogonal and
symplectic cases: Brezin-Hikami \cite{brezin2,brezin3} computed
the bulk scaling limit asymptotics of the moments
$\langle\det^k(\alpha-H)\rangle$ and also provided some asymptotic
expressions for averages (\ref{INtroductionAverages}) with small
number of factors, and the asymptotics of the negative moments
$\langle 1/\det^{m}(\beta-H)\rangle$ has been obtained by
Fyodorov-Keating \cite{fyodorov3} and Forrester-Keating
\cite{forrester1}.

However, the problem of computing the bulk scaling limit
asymptotics of  general averages (\ref{INtroductionAverages}),
despite considerable interest of physicists, see e.g.
Andreev-Simon \cite{andreev}, Gronqvist, Guhr and Kohler
\cite{guhr}, Fyodorov \cite{fyodorov0}, Szabo \cite{szabo},
Splttorff-Verbaarschot \cite{verbaarschot}, Zirnbauer
\cite{zirnbauer1,zirnbauer2}, remained open.

The main goal of this paper is to provide explicit (determinantal
or pfaffian) expressions of averages (\ref{INtroductionAverages})
in terms of those that involve only one or two determinants, and
to evaluate the asymptotics of (\ref{INtroductionAverages}) in the
bulk scaling limit regime of the Gaussian ensembles in the middle
of the spectrum. (In the case of real symmetric matrices we
consider only the case when the matrices are of even size.) It is
worth noting that the standard correlation functions of the matrix
ensembles can be easily extracted from  averages
(\ref{INtroductionAverages}). Thus, we obtain the classical
determinantal and pfaffian formulas for the correlation functions
(see e.g. Refs. \cite{mehta}, \cite{forrester0}) as a corollary.
\subsection{The results: algebraic part} In order to state the
results we need to introduce some notation. Take a positive
measure $\mu$ on $\mathbb{R}$ with finite moments and infinite
support. Then
\begin{equation}\label{IntroductionNormalizationConstant}
C^{(\beta)}_{N}:=\frac{1}{N!}\int\limits_{\mathbb{R}^N}\prod_{1\le
i<j\le N}|x_i-x_j|^\beta \mu(dx_1)\otimes\cdots\otimes\mu(dx_N)\ne
0
\end{equation}
for any $N\ge 1$ and $\beta>0$. Take the probability measure on
$\mathbb{R}^N$ given by
\begin{equation}\label{IntroductionProbabilityMeasure}
p_{N}^{(\beta)}(dx_1,\dots ,dx_N)=\frac 1{N!C^{(\beta)}_{N}}
\prod_{1\le i<j\le N}|x_i-x_j|^\beta\,
\mu(dx_1)\otimes\cdots\otimes\mu(dx_N).
\end{equation}
and for any symmetric function $g:\mathbb{R}^N\to \mathbb{C} $ set
$$
\langle g
\rangle_{\triangle_N^{(\beta)}}:=\int\limits_{\mathbb{R}^N}
g(x_1,\dots,x_N)\,p^{(\beta)}_{N}(dx_1,\dots, dx_N)
$$
provided that the integral converges. Also, for
$\zeta\in\mathbb{C}$ set $ D(\zeta):=\prod_{i\ge 1}(\zeta-x_i). $
The number of factors in such products will always be clear from
the context.

It is well known (see e.g. \cite{mehta}) that the  radial parts
(that is, projections onto different eigenvalues $\{x_i\}$) of the
probability measures
$$
P(dH)=\operatorname{const}\cdot \exp(-Q(H))\,dH,
$$
where $Q(x)$ is an even degree polynomial with positive highest
coefficient, and $H$ belongs to the linear space of $2N\times 2N$
real symmetric, $N\times N$ Hermitian,  or $N\times N$ quaternion
real Hermitian matrices\footnote{Matrix elements of a quaternion
real matrix are $2\times 2$ matrices of the form $\bmatrix
z&w\\-\bar{w}&\bar{z}\endbmatrix$ where $z$ and $w$ are complex
numbers. An $N\times N$ quaternion real Hermitian matrix has $2N$
eigenvalues which come in pairs of coinciding real numbers.}, are
exactly the measures $p_{2N}^{(1)}$, $p_{N}^{(2)}$, and
$p_{N}^{(4)}$ with our measure $\mu$ being equal
$$
\mu(dx)=\begin{cases} \exp(-Q(x))dx, &\beta=1,2,\\\exp(-2Q(x))dx,
&\beta=4.\end{cases}
$$
In the cases $\beta=1$ and $2$, the characteristic polynomial
$\det(\zeta-H)$ of the random matrix $H$ is exactly our
product--function $D(\zeta)$, while in the case $\beta=4$ we have
$\det(\zeta-H)=D^2(\zeta)$. For any finite sets
$A=\{a_1,\dots,a_p\}$, $B=\{b_1,\dots,b_q\}$ denote
$$
\prod(A;B)=\prod_{i=1,j=1}^{p,q}(a_i-b_j), \quad V(A)=\prod_{1\le
i<j\le p}(a_i-a_j),\quad V(B)=\prod_{1\le i<j\le q}(b_i-b_j).
$$
\begin{thm}\label{IntroductionTheoremB14}(orthogonal and symplectic cases) (i) For any
integers $N\ge 1$ and $S>1-N$, and finite sets of mutually
distinct complex numbers
\begin{equation}\label{ALPHABETAA}
\alpha=\{\alpha_1,\dots,\alpha_{k}\},\qquad
\beta=\{\beta_1,\dots,\beta_{m}\},\qquad k-m=2S,
\end{equation}
such that $\beta\cap\mathbb{R}=\varnothing$, one has
\begin{equation}\label{MaincontB1}
\biggl\langle\dfrac{\prod_{i=1}^k D(\alpha_i)}{\prod_{i=1}^m
D(\beta_i)}\biggr\rangle_{\triangle_{2N}^{(1)}}
=\dfrac{C_{2N+2S}^{(1)}}{C_{2N}^{(1)}}\;
\dfrac{\prod(\alpha;\beta)}{V(\alpha)V(\beta)}\PFAFF\;
\biggl[W_N^{(1)}(\alpha,\beta\arrowvert\alpha,\beta)\biggr]
\end{equation}
where $W_N^{(1)}$ is a skew-symmetric $(k+m)\times(k+m)$ matrix
with rows and columns parameterized by elements of $\alpha$ and
$\beta$, and with matrix elements given by
\begin{equation}
\begin{split}
&W^{(1)}_N(\alpha_i,\alpha_j)
=\frac{C_{2N+2S-2}^{(1)}}{C_{2N+2S}^{(1)}}\;(\alpha_i-\alpha_j)\left\langle
D(\alpha_i)D(\alpha_j)\right\rangle_{\T_{2N+2S-2}^{(1)}}
\\
&W^{(1)}_N(\alpha_i,\beta_j)=\dfrac{1}{\alpha_i-\beta_j}\left\langle
\dfrac{D(\alpha_i)}{D(\beta_j)}\right\rangle_{\T_{2N+2S}^{(1)}}
\\
&W^{(1)}_N(\beta_i,\beta_j)=\frac{C_{2N+2S+2}^{(1)}}{C_{2N+2S}^{(1)}}\;
(\beta_i-\beta_j)\left\langle
\dfrac{1}{D(\beta_i)D(\beta_j)}\right\rangle_{\T_{2N+2S+2}^{(1)}}
\end{split}
\end{equation}
(ii) For any integers $N\ge 1$ and $S>1-N$, and finite sets of
mutually distinct complex numbers \ref{ALPHABETAA} such that
$\beta\cap \mathbb{R}=\varnothing$, one has
\begin{equation}\label{MainContB4}
\biggl\langle\dfrac{\prod_{i=1}^k D^2(\alpha_i)}{\prod_{i=1}^m
D^2(\beta_i)}\biggr\rangle_{\T_{N}^{(4)}}
=\dfrac{C_{N+S}^{(4)}}{C_{N}^{(4)}}\;
\dfrac{\prod(\alpha;\beta)}{V(\alpha)V(\beta)}\PFAFF\;
\biggl[W_N^{(4)}(\alpha,\beta\arrowvert\alpha,\beta)\biggr]
\end{equation}
where $W_N^{(4)}$ is a skew-symmetric matrix with rows and columns
parameterized by elements of $\alpha$ and $\beta$, and with matrix
elements given by
\begin{equation}
\begin{split}
&W^{(4)}_N(\alpha_i,\alpha_j)=\frac{C^{(4)}_{N+S-1}}{C^{(4)}_{N+S}}\;
(\alpha_i-\alpha_j)\left\langle
D^2(\alpha_i)D^2(\alpha_j)\right\rangle_{\T_{N+S-1}^{(4)}}
\\
&W^{(4)}_N(\alpha_i,\beta_j)=\dfrac{1}{\alpha_i-\beta_j}\left\langle
\dfrac{D^2(\alpha_i)}{D^2(\beta_j)}\right\rangle_{\T_{N+S}^{(4)}}
\\
&W^{(4)}_N(\beta_i,\beta_j)=\frac{C_{N+S+1}^{(4)}}{C_{N+S}^{(4)}}\;
(\beta_i-\beta_j)\left\langle \dfrac{1}{D^2(\beta_i)D^2(\beta_j)}
\right\rangle_{\T_{N+S+1}^{(4)}}
\end{split}
\end{equation}
\end{thm}
Note that the formulas for the orthogonal and symplectic ensembles
above are essentially identical, except for the number of distinct
eigenvalues, which in orthogonal averages is equal to twice that
number in the symplectic averages, and for the form of the
characteristic polynomial, which is equal to $D(\zeta)$ in the
orthogonal case and to $D^2(\zeta)$ in the symplectic case.

The formulas in the theorem above require that the total number
$k+m$ of factors in  average (\ref{INtroductionAverages})  is
even. These formulas can be easily extended to the situation when
the total number of factors is odd. In order to do that, one needs
to take the even case and send one of the parameters
$\alpha_i,\beta_j$ to $\infty$. After the limit transition, some
of the two-point averages in the matrix elements of
$W_N^{(\beta)}$ will turn into one-point averages of the form
$\langle \det(\zeta-H)\rangle, \langle 1/\det(\zeta-H)\rangle$.

Interestingly enough, the analogous result in the unitary case
looks more complicated! (Although, its proof is, actually,
simpler.) A large part of the next result is contained in the
previous work of Brezin-Hikami \cite{brezin0,brezin1},
Strahov-Fyodorov \cite{strahov}, Baik, Deift and Strahov
\cite{baik}.
\begin{thm}\label{IntroductionTheoremUnitaryCase}(unitary case) For any integers $N\ge 1$
and $S>1-N$, and finite sets of mutually distinct complex numbers
\begin{equation}
\begin{split}
\alpha^-=\{\alpha^-_1,\dots,\alpha^-_{m_1}\},\qquad
\alpha^+=\{\alpha^+_1,\dots,\alpha^-_{k_1}\},\\
\beta^-=\{\beta_1^-,\dots,\beta^-_{m_2}\},\qquad
\beta^+=\{\beta_1^+,\dots,\beta^+_{k_2}\},\nonumber
\end{split}
\end{equation}
with $|\alpha^-|-|\alpha^+|=|\beta^-|-|\beta^+|=S$, such that
$\alpha^+\cap\mathbb{R}=\varnothing$,
$\beta^+\cap\mathbb{R}=\varnothing$, one has
\begin{equation}\label{MainContB2}
\begin{split}
\left\langle\dfrac{\prod_{i=1}^{m_1}D(\alpha_i^-)\prod_{i=1}^{m_2}
D(\beta_i^-)}{\prod_{j=1}^{k_1}D(\alpha_j^+)\prod_{j=1}^{k_2}D(\beta_j^+)}
\right\rangle_{\triangle_N^{(2)}}=
(-1)^{\tfrac{(|\alpha^-|+|\beta^-|)^2+(|\beta^-|-|\alpha^-|)}{2}}\;\\
\times\dfrac{C_{N+S}^{(2)}}{C_N^{(2)}}\;\dfrac{\prod(\alpha^-;\alpha^+)
\prod(\beta^-;\beta^+)}{V(\alpha^-)V(\alpha^+)V(\beta^-)V(\beta^+)}
\; \det\left[W_N^{(2)}(\alpha^-,\beta^+\vert
\beta^-,\alpha^+)\right].
\end{split}
\end{equation}
Here $W_N^{(2)}(\alpha^-,\beta^+\vert \beta^-,\alpha^+)$ is a
matrix with rows parameterized by elements of $\alpha^-$ and
$\beta^+$, columns parameterized by elements of $\beta^-$ and
$\alpha^+$, and with matrix elements
\begin{equation}
\begin{split}
 &W_N^{(2)}(\alpha_i^-,\beta_j^-)=\dfrac{C^{(2)}_{N+S-1}}{C^{(2)}_{N+S}}\;
  \left\langle
    D(\alpha_i^-)D(\beta_j^-)\right\rangle_{\triangle_{N+S-1}^{(2)}}\\
&W_N^{(2)}(\alpha_i^-,\alpha_j^+)=\dfrac{1}{\alpha_i^--\alpha_j^+}\;
\left\langle\dfrac{D(\alpha_i^-)}{D(\alpha_j^+)}\right\rangle_{\triangle_{N+S}^{(2)}}
\\
&W_N^{(2)}(\beta_i^+,\beta_j^-)=\dfrac{1}{\beta_i^+-\beta_j^-}\;
\left\langle\dfrac{D(\beta_j^-)}{D(\beta_i^+)}\right\rangle_{\triangle_{N+S}^{(2)}}
\\
&W_N^{(2)}(\beta_i^+,\beta_j^+)=\dfrac{C^{(2)}_{N+S+1}}{C^{(2)}_{N+S}}\;
\left\langle\dfrac{1}{D(\beta_i^+)D(\beta_j^+)}\right\rangle_{\triangle_{N+S+1}^{(2)}}
\end{split}
\end{equation}
\end{thm}
Once again, the formula above holds for the even total number of
determinants, but the odd case is easily obtained by sending one
of the parameters $\alpha_i^\pm,\beta_j^\pm$ to infinity.

It is worth noting that Theorem
\ref{IntroductionTheoremUnitaryCase} provides many different
expressions for the same averages, depending on how we split the
factors in the numerator and denominator into groups. The
resulting identities are often not easy to prove independently.

Define the {\it $n$th correlation measure} of $p_N^{(\beta)}$ by
$$
\rho_{n,N}^{(\beta)}(dx_1,\dots,dx_n)=\frac{N!}{(N-n)!}
\int\limits_{x_{n+1},x_{n+2},\dots,x_N}p_N^{(\beta)}
(dx_1,\dots,dx_N).
$$
For a function $f(\zeta)$ of a complex variable $\zeta$, which is
continuous in both half-planes $\Img \zeta>0$ and $\Img \zeta<0$
up to the real axis, we will denote by $[f(\zeta)]_{\zeta=x}$,
$x\in \mathbb{R}$, the difference of the limit values of
$f(\zeta)$ as $\zeta\to x$ from bottom and from top divided by
$2\pi i$:
$$
[f(\zeta)]_x=\frac 1{2\pi i}\,(f(x-i0)-f(x+i0)).
$$
The next statement is an easy corollary of Theorems
\ref{IntroductionTheoremUnitaryCase} and
\ref{IntroductionTheoremB14}.
\begin{cor}\label{IntroductionCorrolary}
 Take any $n\ge 1$ and assume that near $n$
points $x_1,\dots,x_n\in\mathbb{R}$ the measure $\mu$ is
absolutely continuous with respect to the Lebesgue measure, and
its density their is uniformly H\"older continuous. Then the $n$th
correlation measure of $\T_N^{(\beta)}$ (or $\T_{2N}^{(\beta)}$
for $\beta=1$) has a continuous density near $(x_1,\dots,x_n)$
which is given by \\
$\bullet$ For $\beta=2$
$$
\rho_{n,N}^{(2)}(x_1,\dots,x_n)=\det[K^{(2)}_N(x_i,x_j)]_{i,j=1}^n
$$
where for $x\ne y$ the kernel is given by
$$
K^{(2)}(x,y)=
\frac{1}{x-y}\left[\left\langle\frac{D(x)}{D(\zeta)}\right\rangle_{\T_N^{(2)}}
\right]_{\zeta=y},
$$
and for $x=y$ the kernel is defined by continuity. \\
$\bullet$ For $\beta=1$
$$
\rho_{n,2N}^{(1)}(x_1,\dots,x_n)=\PFAFF\;[K^{(1)}_{2N}(x_i,x_j)]_{i,j=1}^n
$$
where the skew-symmetric $2\times 2$ matrix kernel for $x\ne y$ is
given by
$$
 K_{2N}^{(1)}(x,y)=\\
 \bmatrix
\frac{C_{2N-2}^{(1)}}{C_{2N}^{(1)}}\,(x-y)\left\langle
D(x)D(y)\right\rangle_{\T_{2N-2}^{(1)}}&
\frac{1}{x-y}\left[\left\langle
\frac{D(x)}{D(\zeta)}\right\rangle_{\T_{2N}^{(1)}}\right]_{\zeta=y}
\\
-\frac{1}{x-y}\left[\left\langle
\frac{D(x)}{D(\zeta)}\right\rangle_{\T_{2N}^{(1)}}\right]_{\zeta=y}&
\frac{C_{2N+2}^{(1)}}{C_{2N}^{(1)}}\,(x-y) \left[\left\langle
\frac{1}{D(\zeta)D(\eta)}\right\rangle_{\T_{2N+2}^{(1)}}\right]_{\zeta=x,\,\eta=y}
\endbmatrix
$$
and for $x=y$ the kernel is defined by continuity.\\
 $\bullet$ For
$\beta=4$
$$
\rho_{n,N}^{(4)}(x_1,\dots,x_n)=\PFAFF\;[K^{(4)}_{N}(x_i,x_j)]_{i,j=1}^n
$$
where the skew-symmetric $2\times 2$ matrix kernel for $x\ne y$ is
given by
$$
 K_{N}^{(4)}(x,y)=\\
 \bmatrix
 \frac{C^{(4)}_{N-1}}{2C^{(4)}_{N}}\; (x-y)\left\langle
D^2(x)D^2(y)\right\rangle_{\T_{N+1}^{(4)}}&
\frac{1}{2(x-y)}\left[\left\langle
\frac{D^2(x)}{D^2(\zeta)}\right\rangle_{\T_{N}^{(4)}}\right]_{\zeta=y}\\
-\frac{1}{2(x-y)}\left[\left\langle
\frac{D^2(x)}{D^2(\zeta)}\right\rangle_{\T_{N}^{(4)}}\right]_{\zeta=y}&
\frac{C_{N+1}^{(4)}}{2C_{N}^{(4)}}\; (x-y)\left[\left\langle
\frac{1}{D^2(\zeta)D^2(\eta)}
\right\rangle_{\T_{N-1}^{(4)}}\right]_{\zeta=x,\,\eta=y}
\endbmatrix
$$
and for $x=y$ the kernel is defined by continuity.
\end{cor}

If instead of asking for the measure $\mu$ to have nice density
near $x_1,\dots,x_n$, we require that $\mu$ is purely atomic near
these points, then the formulas for the correlation functions
above will continue to hold if we understand the symbol
$[f(\zeta)]_x$ as the residue of the meromorphic function
$f(\zeta)$ at the point $x$.

Using standard techniques of RMT, one can evaluate the two-point
($k+m=2$) averages (\ref{INtroductionAverages}) via the
(skew)-orthogonal polynomials associated with the problem, see
Sections \ref{SectionAveragesCharacteristicPolynomials},
\ref{PSectionDSPE}, and \ref{PSectionDOE} below. Then the formulas
above yield a new proof of the well-known evaluation of the
correlation functions of $p_N^{(\beta)}$ with $\beta=1,2,4$, in
terms of (skew)-orthogonal polynomials (see e.g. \cite{tracy}).

Let us emphasize that our approach provides a proof of the
determinantal and pfaffian formulas for the correlation functions
which does not use the (skew)-orthogonal polynomials.
\subsection{The results: analytic part}
Theorems \ref{IntroductionTheoremB14} and
\ref{IntroductionTheoremUnitaryCase} are very convenient for
computing the limits of averages (\ref{INtroductionAverages}) as
the size of matrices goes to infinity. In what follows we will use
the notation $GOE_N$, $GUE_N$, $GSE_N$ to denote the measures
$p_N^{(\beta)}$ with
$$
\beta=1,\ \mu(dx)=e^{-\frac {x^2}2}dx,\qquad \beta=2,\
\mu(dx)=e^{-x^2}dx,\qquad \beta=4,\ \mu(dx)=e^{-x^2}dx,
$$
respectively. The normalization is chosen in such a way that the
needed scaling will be the same in all three cases. Note that in
the case of Gaussian weights, the constants $C_N^{(\beta)}$ are
explicitly computed as certain products of $\Gamma$-functions, see
e.g. Ref. \cite{mehta}. The values relevant for our three cases
will be given in Section \ref{SectionAsymptotics}.
\begin{thm}\label{IntroductionAnalyticTheorem1}(i) With the notation and assumptions of
Theorem \ref{IntroductionTheoremB14}(i), we have
\begin{equation}
\begin{split}
\lim_{N\to\infty}\frac{C_{2N}^{(1)}}{C_{2N+2S}^{(1)}}\,(2N)^{\frac{km}2-
\frac{k(k+1)}4-\frac{m(m-1)}4}\biggl\langle\dfrac{\prod_{i=1}^k
D(\alpha_i/\sqrt{2N})}{\prod_{i=1}^m
D(\beta_i/\sqrt{2N})}\biggr\rangle_{GOE(2N)}\\ =
\dfrac{\prod(\alpha;\beta)}{V(\alpha)V(\beta)}\PFAFF\;
\Bigl[\s_{GOE}(\alpha,\beta\arrowvert\alpha,\beta)\Bigr]
\end{split}\nonumber
\end{equation}
where $\s_{GOE}(\alpha,\beta\arrowvert\alpha,\beta)$ is a
skew-symmetric $(k+m)\times(k+m)$ matrix with rows and columns
parameterized by elements of $\alpha$ and $\beta$, and with matrix
elements given by
$$
\aligned &\s_{GOE}(\alpha_p,\alpha_q) =-\frac{1}{\pi
}\frac{\partial}{\partial
\alpha_i}\frac{\sin(\alpha_p-\alpha_q)}{\alpha_p-\alpha_q}\,,
\\
&\s_{GOE}(\alpha_p,\beta_q)= \begin{cases}
-\frac{\exp{i(\beta_q-\alpha_p)}}{\beta_q-\alpha_p},&\Img \beta_q>0,\\
\frac{\exp{i(\alpha_p-\beta_q)}}{\alpha_p-\beta_q},
&\Img\beta_q<0,
\end{cases}
\\
&\s_{GOE}(\beta_p,\beta_q)=2\pi i\begin{cases}
\int_{1}^{+\infty}\frac{\exp(i(\beta_p-\beta_q)t)}t\,dt,&\Img
\beta_p>0,\,\Img\beta_q<0,\\
-\int_{1}^{+\infty}\frac{\exp(i(\beta_q-\beta_p)t)}t\,dt,&\Img
\beta_p<0,\,\Img\beta_q>0,\\
0,&\text{  in all other cases}.
\end{cases}
\endaligned
$$
(ii) With the notation and assumptions of Theorem
\ref{IntroductionTheoremB14}(ii), we have
\begin{equation}
\begin{split}
\lim_{N\to\infty}\frac{C_{N}^{(4)}}{C_{N+S}^{(4)}}\,(2N)^{\frac{km}2-
\frac{k(k-1)}4-\frac{m(m+1)}4}\biggl\langle\dfrac{\prod_{i=1}^k
D^2(\alpha_i/\sqrt{2N})}{\prod_{i=1}^m
D^2(\beta_i/\sqrt{2N})}\biggr\rangle_{GSE(N)}\\ =
\dfrac{\prod(\alpha;\beta)}{V(\alpha)V(\beta)}\PFAFF\;
\Bigl[\s_{GSE}(\alpha,\beta\arrowvert\alpha,\beta)\Bigr]
\end{split}\nonumber
\end{equation}
where $\s_{GSE}(\alpha,\beta\arrowvert\alpha,\beta)$ is a
skew-symmetric $(k+m)\times(k+m)$ matrix with rows and columns
parameterized by elements of $\alpha$ and $\beta$, and with matrix
elements given by
$$
\aligned &\s_{GSE}(\alpha_p,\alpha_q) =\frac{1}{\pi
}\int_{0}^{1}\frac{\sin (\alpha_p-\alpha_q)t}t\,dt,
\\
&\s_{GSE}(\alpha_p,\beta_q)=  \begin{cases}
-\frac{\exp{i(\beta_q-\alpha_p)}}{\beta_q-\alpha_p},&\Img\beta_q>0,\\
\frac{\exp{i(\alpha_p-\beta_q)}}{\alpha_p-\beta_q},
&\Img\beta_q<0,
\end{cases}
\\
&\s_{GSE}(\beta_p,\beta_q)=2\pi i\begin{cases}
\frac{\partial}{\partial\beta_p}\frac {\exp
i(\beta_p-\beta_q)}{\beta_p-\beta_q},&\Img
\beta_p>0,\,\Img\beta_q<0,\\
-\frac{\partial}{\partial \beta_q}\frac {\exp
i(\beta_q-\beta_p)}{\beta_q-\beta_p},&\Img
\beta_p<0,\,\Img\beta_q>0,\\
0,&\text{  in all other cases}.
\end{cases}
\endaligned
$$

\end{thm}
Similarly, Theorem \ref{IntroductionTheoremUnitaryCase} yields
\begin{thm}\label{IntroductionAnalyticTheorem2} With the notation and
assumptions of Theorem \ref{IntroductionTheoremUnitaryCase} we
have
\begin{equation}
\begin{split}
\lim_{N\to\infty}\frac{C_{N}^{(2)}}{C_{N+S}^{(2)}}\,
(2N)^{-\tfrac{S^2}{2}}
\left\langle\dfrac{\prod_{i=1}^{k_1}D(\alpha_i^-/\sqrt{2N})\prod_{i=1}^{k_2}
D(\beta_i^-/\sqrt{2N})}{\prod_{j=1}^{m_1}D(\alpha_j^+/\sqrt{2N})\prod_{j=1}^{m_2}
D(\beta_j^+/\sqrt{2N})}
\right\rangle_{GUE(N)}=\\
(-1)^{\tfrac{(|\beta^-|+|\alpha^-|)^2+(|\beta^-|-|\alpha^-|)}{2}}
\dfrac{\prod(\alpha^-;\alpha^+)\prod(\beta^-;\beta^+)}
{V(\alpha^-)V(\alpha^+)V(\beta^-)V(\beta^+)}
\det\left[\s_{GUE}(\alpha^-,\beta^+\vert \beta^-,\alpha^+)\right]
\end{split}\nonumber
\end{equation}
Here $\s_{GUE}(\alpha^-,\beta^+\vert \beta^-,\alpha^+)$ is a
matrix with rows parameterized by elements of $\alpha^-$ and
$\beta^+$, columns parameterized by elements of $\beta^-$ and
$\alpha^+$, and with matrix elements

$$
\aligned
     &\s_{GUE}(\alpha_p^-,\beta_q^-)=\frac 1\pi\,
     \frac{\sin(\alpha_p^--\beta_q^-)}{\alpha_p^--\beta_q^-}\,,\\
     &\s_{GUE}(\alpha_p^-,\alpha_q^+)=\begin{cases}
-\frac{\exp{i(\alpha^+_q-\alpha_p^-)}}{\alpha_q^+-\alpha_p^-},&\Img \alpha_q^+>0,\\
\frac{\exp{i(\alpha_p^--\alpha_q^+)}}{\alpha_p^--\alpha_q^+},
&\Img\alpha_q^+<0,
\end{cases}
\\
&\s_{GUE}(\beta_p^+,\beta_q^-)=\begin{cases}
\frac{\exp{i(\beta^+_p-\beta_q^-)}}{\beta_p^+-\beta_q^-},&\Img \beta_p^+>0,\\
-\frac{\exp{i(\beta_q^--\beta_p^+)}}{\beta_q^--\beta_p^+},
&\Img\beta_p^+<0,
\end{cases}
\\
&\s_{GUE}(\beta_p^+,\alpha_q^+)=2\pi i\begin{cases}\frac{\exp
i(\beta_p^+-\alpha_q^+)}
 {\beta_p^+-\alpha_q^+},&\Img\beta_p^+>0,\,\Img\alpha_q^+<0,\\
-\frac{\exp i(\alpha_q^+-\beta_p^+)}{\alpha_q^+-\beta_p^+},&\Img\beta_p^+<0,\,\Img\alpha_q^+>0,\\
 0,&\text{  in all other cases}.
 \end{cases}
\endaligned
$$
\end{thm}
Observe that if in any of the three cases (GOE, GUE, GSE) the
number of determinants in the numerator is the same as the number
of factors in the denominator, then no prefactors of the form
$C_{N_1}^{(\beta)}/C_{N_2}^{(\beta)}$ or of the form $2N$ raised
to a power are needed, the limit exists without any normalization.
We believe that this is a deep fact, however, at the moment we do
not have a conceptual explanation for it.

Another intriguing observation is that if we take the same number
of determinants in the numerator and denominator, and the signs of
the imaginary parts of all the arguments in the denominator are
the same, the limit turns out to be the same in all three cases!
More exactly,
\begin{equation}
\begin{split}
\lim_{N\to\infty} \left\langle
\frac{\det(\alpha_1/\sqrt{2N}-H)\cdots\det(\alpha_k/\sqrt{2N}-H)}
{\det(\beta_1/\sqrt{2N}-H)\cdots\det(\beta_k/\sqrt{2N}-H)}
\right\rangle_{GOE(2N),\,GUE(N),\,GSE(N)} \\
=\exp\left[\pm
i\sum\limits_{j=1}^k(\alpha_j-\beta_j)\right]
\end{split}\nonumber
\end{equation}
where the positive sign  inside the exponential corresponds to the
case $\Img\beta_1,\dots,\Img\beta_k<0$, and the negative sign
 corresponds to the case $\Img\beta_1,\dots,\Img\beta_k>0$.
Again, it would be nice to have a conceptual explanation for such
an unexpected coincidence. In particular, one may ask if such a
limit exists for arbitrary $\beta>0$, and if so then whether it
depends on $\beta$.

It is worth noting that if one formally uses the asymptotics of
Theorems \ref{IntroductionAnalyticTheorem1},
\ref{IntroductionAnalyticTheorem2} in Corollary
\ref{IntroductionCorrolary} to compute the asymptotics of the
correlation functions then one easily recovers the well-known
results for the bulk scaling limit of the correlation kernels, see
e.g. Refs. \cite{forrester0}, \cite{mehta}. For details of this
computation see  Remark
\ref{RemarkAsymptoticsCorrelationFunctions}, Section
\ref{SectionStatementofAsymptoticResults} below.

An important feature of the bulk scaling limit in random matrix
models is that the limiting values of the correlation kernels turn
out to be {\it universal}: they depend on $\beta$ but do not
depend on the potential $Q(x)$. This universality property has
been verified in full generality for $\beta=2$ case in Refs.
\cite{deift3}, \cite{bleher}, and in the middle of the spectrum
for $\beta=1,4$ in Ref. \cite{deiftgioev}. We expect that the bulk
scaling limits of  averages (\ref{INtroductionAverages} which we
computed in Theorems \ref{IntroductionAnalyticTheorem1},
\ref{IntroductionAnalyticTheorem2} in the middle of the spectrum
for Gaussian potentials, are also universal in a similar sense.
For $\beta=2$ this was proved in  Ref. \cite{strahov}, and for
$\beta=1,4$ the question remains open.

\subsection{The method: discrete approximation} In
recent years we have seen a lot of progress in understanding {\it
discrete} probabilistic models of random matrix type which come
from various domains of mathematics. One important observation
that becomes clear from the point of view of those discrete models
is that the measures $p_N^{(\beta)}$ with $\beta=1,2,4$, may be
viewed as degenerate cases of more general objects, the so-called
{\it determinantal point processes\/} in the unitary case and {\it
pfaffian point processes\/} in the orthogonal and symplectic case.

Inside these more general classes of point processes one also
finds discrete analogs of the measures $p_N^{(\beta)}$. In the
unitary case, the discrete ensemble is different from the
continuous one only by the fact that the support of the measure
$\mu$ is discrete. In the orthogonal and symplectic cases the
difference is more substantial.

Let $\X$ be a finite subset of $\mathbb{R}$ with an even number of
elements. The discrete analog of the measure $p_{2N}^{(1)}$ is a
measure on $2N$-point subsets $X$ of $\X$ which have the following
property: for any $x\in X$ the number of points in $X$ that are
smaller than $x$ has the same parity as the number of point in
$\X$ that are smaller than $x$. The weight of one such subset is
given by the familiar formula, cf.
(\ref{IntroductionProbabilityMeasure})
\begin{equation}\label{IntroductionDiscretOrthEnsemble}
\operatorname{Prob}(x_1,\dots,x_{2N})=\const\prod_{1\le i<j\le
2N}|x_i-x_j|\cdot\prod_{i=1}^{2N}f(x_i),
\end{equation}
where $f$ is some positive weight function. We call this measure a
{\it discrete orthogonal ensemble}.

A {\it discrete symplectic ensemble} is defined by the same
formula (\ref{IntroductionDiscretOrthEnsemble}), but the
admissible subsets $X$ are different: with any point $x$ such
subset must contain the immediate predecessor of $x$ in $\X$.

It is not hard to see that if we take $\X$ to be a lattice then in
the limit when the step of this lattice goes to zero, the discrete
orthogonal/symplectic ensembles turn into measures similar to
$p_{2N}^{(1)}$ and $p_{N}^{(4)}$. (Indeed, if the points of $X$
are split into pairs, and in each pair the points are
infinitesimally close, then the Vandermonde determinant of size
$|X|$ in (\ref{IntroductionDiscretOrthEnsemble}) is asymptotically
given by the fourth power of the Vandermonde determinant of size
$|X|/2$ in scaled locations of the pairs.)

The discrete orthogonal and symplectic ensembles are related by
the {\it particle-hole involution}: if $X$ is from a discrete
orthogonal ensemble than $\widehat X=\X\setminus X$ is from a
discrete symplectic ensemble (with a different weight function and
different number of particles though) and {\it vice versa}. The
particle-hole involution of a discrete unitary ensemble (which is
a measure of the form $p_N^{(2)}$ with discretely supported $\mu$)
is again a discrete unitary ensemble. In the $\beta=2$ case the
particle-hole involution has been used before, see Refs.
\cite{borodin}, \cite{BorodinOlshanski2} (Section 5),
\cite{BorodinOlshanski4}, \cite{johansson}, \cite{baik0}.

The motivation for introducing these discrete objects comes from
representation theory; in certain models these are just different
ways of parameterizing partitions. If
$\lambda=(\lambda_1,\lambda_2,\dots)$ is a partition then
$\{2\lambda_i-i\}$ is an admissible point configuration for a
discrete orthogonal ensemble, while
$\{\lambda_i-2i,\lambda_i-2i+1\}$ is an admissible point
configuration for a discrete symplectic ensemble. The convenience
of coordinates $\{\lambda_i-\beta i/2\}$ in the models involving
Vandermonde determinant raised to the power $\beta$ can be
observed, e.g., in Refs. \cite{kerov}, \cite{BorodinOlshanski3}.
The fact that discrete orthogonal and symplectic ensembles are
dual to each other comes from two different ways of parameterizing
partitions: using rows or columns of the corresponding Young
diagram. We hope to give more details on this connection in a
future publication.

In the paper we actually prove analogs of Theorems
\ref{IntroductionTheoremB14} and
\ref{IntroductionTheoremUnitaryCase} for discrete orthogonal,
unitary, and symplectic ensembles and then obtain Theorems
\ref{IntroductionTheoremB14} and
\ref{IntroductionTheoremUnitaryCase} by limit transition. Note
that two parts of Theorem \ref{IntroductionTheoremB14} in the
discrete situation are simply equivalent
--- they are obtained one from the other by the particle-hole
involution. This explains the similarity of the formulas.

The proofs in the discrete case are obtained as follows. It turns
out that if we apply the particle-hole involution on a suitable
part of the phase space $\X$ then the orthogonal/symplectic and
unitary ensembles turn into what we call {\it $L$--ensembles}.
This is another subclass of pfaffian/determinantal point
processes; the processes in this subclass are defined using a
matrix of size $|\X|\times |\X|$ which is often denoted by $L$.
The discrete analogs of Theorems \ref{IntroductionTheoremB14} and
\ref{IntroductionTheoremUnitaryCase} interpreted in the language
of this matrix $L$ provide some relations between matrix elements
of $L$ and its resolvent. These relations are then proved using
standard linear algebraic facts.

We would like to emphasize that after the objects of interest are
represented in terms of $L$-ensembles, the proofs become very
simple. The nontrivial part of our approach is in {\it
constructing the discrete models which make the statements easy to
prove.}
\subsection{ The
asymptotics} Despite the fact that Theorems
\ref{IntroductionAnalyticTheorem1} and
\ref{IntroductionAnalyticTheorem2} look like rather simple--minded
corollaries of Theorems \ref{IntroductionTheoremB14} and
\ref{IntroductionTheoremUnitaryCase}, rigorous proofs of the
asymptotic formulas do require some efforts.

The first step of our computation consists of expressing the
two-point averages - matrix elements of $W_N^{(\beta)}$ in
Theorems \ref{IntroductionTheoremB14} and
\ref{IntroductionTheoremUnitaryCase} --- in terms of associated
(skew)-orthogonal polynomials. As is well known, in all three
Gaussian ensembles the corresponding polynomials are expressible
through classical Hermite polynomials, see e.g.
\cite{mehta,forrester0,adler,nagao1}. Also, certain summation
formulas of Christoffel--Darboux type for all three case
$\beta=1,2,4$ are known, see \cite{nagao1}, \cite{widom}. Using
these formulas and applying some algebraic transformations (which
are least simple in the symplectic case), we write the two-point
averages as finite expressions involving Hermite polynomials,
their derivatives, and integrals.

Further computations are easy to explain --- we replace the
Hermite polynomials of large degree by their asymptotic
oscillatory behavior near the origin. However, sometimes we need
to do that in double or triple integrals, and the resulting limit
integrals are often only conditionally convergent. Such actions
require accurate estimates of the error terms, and they constitute
a large part of Section \ref{SectionAsymptotics}. In these
estimates it is essential for us to have uniform asymptotics of
the Hermite polynomials on the real line. The needed results
follow from much more general results of \cite{deift2,deift3}
\footnote{The classical estimates proved in \cite{Szego} turn to
be to be insufficient for our purposes.}; we adopt their formulas
to the Hermite case.

\subsection{ Final remarks}

To conclude the introduction we would like to suggest the
following debatable point of view which we developed while working
on the subject:

{\it The averages of products and ratios of characteristic
polynomials are more fundamental characteristics of random matrix
models than the correlation functions.}

This is especially visible in the $\beta=1,4$ cases. Here are some
arguments to defend this thesis:

\noindent $\bullet$\  The pfaffian/determinantal formulas for
these averages are simpler than those for the correlation
functions. The formulas express many-point averages in terms of
one- and two-point ones, and they do not require the introduction
of (skew)-orthogonal polynomials or Christoffel-Darboux type
kernels.

\noindent $\bullet$\  The formulas for the correlation functions
are easily recovered from those for averages \thetag{1}.

\noindent $\bullet$\  Even though there is no clear probabilistic
sense in taking the bulk scaling limit of such averages, it does
exist after proper normalization, and it appears to be fairly
universal. The limiting correlation kernels are easily obtained
from these limits.

We hope to provide further arguments related to discrete
probabilistic models arising in representation theory in future
publications.

\subsection {Acknowledgements}
We are very grateful to Grigori Olshanski for numerous discussions
of the $\beta=2$ case. This research was partially conducted
during the period one of the authors (A.B.) served as a Clay
Mathematics Institute Research Fellow.

\section{Determinantal point ensembles}
\subsection{Point configurations}\label{PointConfigurations}
Let $\frak{X}$ be a finite  set. We will denote by $X$, $Y$,
$\ldots $ subsets of the set $\frak{X}$ and we will call them
"point configurations". Let $\frak{X}$ have a fixed splitting into
the union of two disjoint subsets ("positive" and "negative"),
\begin{equation}
\frak{X}=\frak{X}_-\sqcup\frak{X}_+ \nonumber
\end{equation}
Then any point configuration $X$ is  a unit of two disjoint sets
as well,
\begin{equation}
X=X_-\sqcup X_+ \nonumber
\end{equation}
\begin{equation}
X_-=X\cap\frak{X}_-,\;\;X_+=X\cap\frak{X}_+ \nonumber
\end{equation}
and we will say that the point configuration $X$ consists of
\textit{positive particles} (elements of $X_+$) and
\textit{negative particles} (elements of $X_-$). If a given
configuration includes an equal number of positive and negative
particles we will say that such a configuration is a
\textit{balanced} point configuration. Thus
\begin{equation}
\vert X_+\vert=\vert X_-\vert \nonumber
\end{equation}
if $X=(X_-\vert X_+)$ is a balanced configuration. Here $\vert
A\vert$ denotes the number of elements in the set $A$.

If $X$ is a point configuration of particles in  $\frak{X}$ its
complement $\frak{X}\setminus X$ in the set $\frak{X}$ will be a
point configuration as well. It is natural to refer to the point
configuration $\frak{X}\setminus X$ as the point configuration of
\textit{holes}. The set $(\frak{X}\setminus X)\cap \frak{X}_-$
will be called  the set of \textit{positive} holes, and the set
$(\frak{X}\setminus X)\cap \frak{X}_+$ will be called the set of
\textit{negative} holes.

To any configuration $X$ of particles there will correspond (in
many ways) a balanced point configuration $Z$, which consists of
both particles and holes. Namely let $X=(X_-\vert X_+)$ be the
splitting of the point configuration $X$ into positive and
negative particles, and assume that $|X_-|-|X_+|=S\geq 0$. Then
the corresponding balanced configuration can be constructed in
accordance with the formulas:
\begin{equation}
\begin{split}
Z=Z_-\sqcup
Z_+,\;&Z_+=Z\cap(\X_+\sqcup\X_0),\;Z_-=Z\cap(\X_-\setminus\X_0)\\
& Z_-=X_-\setminus\X_0\\
& Z_+=X_+\sqcup(\X_0\setminus X)
\end{split}\nonumber
\end{equation}
where $\frak{X}_0$ is an arbitrary subset of the set $\frak{X}_-$
consisting of $S$ points. Fig. 1 shows the decomposition of the
set $\X$, where the set $\X_0$ is chosen to be the right-hand
subset of the set $\X_-$. Fig. 2 represents the unbalanced
particle-particle configuration X, and Fig. 3 explains the
construction of the balanced particle-hole configuration $Z$.

It is not hard to see that the point configuration $Z$ constructed
in such a way is a balanced point configuration, i.e. $\vert
Z_+\vert=\vert Z_-\vert$. Indeed if $\vert X_+\vert=d$ then $\vert
X_-\vert=S+d$. Assume that the chosen subset $\frak{X}_0$ includes
$d_1$ negative particles of the configuration $X$. It means that
$\frak{X}_0$ includes $S-d_1$ positive holes. The set $Z_+$
consists of these positive holes and positive particles of the
point configuration $X$, thus $\vert Z_+\vert=S-d_1+d$. Since
$Z_-$ consists of the negative particles lying outside the set
$\frak{X}_0$, $\vert Z_-\vert=S+d-d_1$. Therefore $\vert
Z_+\vert=\vert Z_-\vert$, and the configuration $Z$ defined above
is a balanced configuration.

We will say that the balanced configuration $Z$ is obtained from
the configuration $X$ by the \textit{particle-hole involution} on
the set $\frak{X}_0$.
 \subsection{Definition of $L$-ensembles}\label{SectionLEnsembles}
 First we recall the notion
of determinantal $L$-ensembles  following Daley and Vere-Jones
\cite{Daley}, Borodin and Olshanski
\cite{BorodinOlshanski1,BorodinOlshanski2}. A random point process
on the space $\frak{X}$ is defined by an introduction of a
probability $\mbox{Prob}_L (X)$ for each  subset $X$ of
$\frak{X}$, so that
\begin{equation}
\sum\limits_{X\subset \;\frak{X}} \mbox{Prob}_L(X)=1,\;\;\;
\mbox{Prob}_L(X)\geq 0\nonumber
\end{equation}
Let $L$ be a $\frak{X}\times\frak{X}$ matrix whose rows and
columns are parameterized by the points of $\frak{X}$. We use the
notation $A(\alpha\vert\beta)$ for a matrix $A$ and subsets
$\alpha=(\alpha_1,\ldots ,\alpha_a)$, $\beta=(\beta_1,\ldots
,\beta_b)$ of its rows and columns, to denote the submatrix
$\vert\vert A_{\alpha_i\beta_j}\vert\vert$ of A. Then to any
subset $X$ (or for any point configuration) there will correspond
the diagonal minor $\mbox{det}\;L(X\vert X)$ of the matrix $L$.
Assume further that any such diagonal minor of $L$ is nonnegative.
\begin{defn} A random point process living on the space of point
configurations in $\frak{X}$ is called an $L$-ensemble, if
\begin{equation}
\mbox{Prob}_L(X)=\dfrac{\mbox{det}\; L(X\vert
X)}{\mbox{det}\;(1+L)}.\nonumber
\end{equation}
\end{defn}
It is clear that $\sum\limits_{X\subset \;\frak{X}}
\mbox{Prob}_L(X)=1$ as the normalization constant
$\mbox{det}(1+L)$ has the following well-known decomposition:
\begin{equation}
\mbox{det}\;(1+L)=\sum\limits_{X\subset \;\frak{X}}
\mbox{det}\;L(X\vert X).\nonumber
\end{equation}
According to the decomposition of the set $\frak{X}$ into the
union of positive and negative subspaces, $\X=\X_-\sqcup \X_+$ we
write the matrix $L$ in the block form:
\begin{equation}\label{matrixLstructure}
L=\left[\begin{array}{cc}
  L_{--} & L_{-+} \\
  L_{+-} & L_{++} \\
\end{array}\right].
\end{equation}
Let $h$ be a real function defined on $\frak{X}$. We are
interested in $L$-ensembles for which the matrix $L$ is given by
the formula
\begin{equation}\label{matrixL}
L=\left[\begin{array}{cc}
  0 & A \\
 -A^T & 0 \\
\end{array}\right],\;\;\; A(x,y)=\dfrac{h(x)h(y)}{x-y}.
\end{equation}
In that case we use the formula for Cauchy determinants to rewrite
the expression for the probability of a random point configuration
as
\begin{equation}\label{PROBX}
\mbox{Prob}_L(X)=
\left\{%
\begin{array}{ll}
    \dfrac{1}{\mbox{det}(1+L)}
\dfrac{V^2(X_+)V^2(X_-)}{\prod^2(X_-;X_+)}\;h^2(X), & \vert X_+\vert=\vert X_-\vert , \\
    \qquad 0, & \hbox{otherwise} , \\
\end{array}%
\right.
\end{equation}
where
$\prod(A;B)\equiv\prod\limits_{i=1}^k\prod\limits_{j=1}^l(a_i-b_j)$
for any two sets $A=(a_1,\ldots ,a_k)$, $B=(b_1,\ldots ,b_l)$,
$V(X)$ is the Vandermonde determinant associated with the set $X$,
\begin{equation}
V(X)=\prod\limits_{1\leq i<j\leq N}(x_i-x_j),\;\;\; X=(x_1,\ldots
,x_N)\nonumber
\end{equation}
 and $h(X)=\prod\limits_{j=1}^Nh(x_j)$.

\subsection{ $ \hat{L}$-ensemble}\label{subsectionLHat}

We have seen in the  section above that an $L$-ensemble is
completely determined by a splitting of the discrete set
$\frak{X}$ into positive and negative parts, and by a weight $h$
on $\frak{X}$. With a different splitting of the set $\frak{X}$,
$\frak{X}=\left(\hat{\frak{X}}_-\vert \hat{\frak{X}}_+\right)$ and
different weight $\hat{h}$ we will construct another ensemble,
which we will call $\hat{L}$-ensemble.

Let $\hat{\frak{X}}_- =\frak{X}_-\setminus \frak{X}_0$ and
$\hat{\frak{X}}_+ =\frak{X}_0\sqcup \frak{X}_+$, $\vert
\frak{X}_0\vert=S$.  As we have already seen in Section
\ref{PointConfigurations}, to the unbalanced configurations $X$ of
particles
\begin{equation}
X=(X_-\vert X_+),\; \vert X_-\vert-\vert X_+\vert=S \nonumber
\end{equation}
with respect to the splitting $\frak{X}=\left(\frak{X}_-\vert
\frak{X}_+\right)$ there  corresponds the balanced hole-particle
configuration $Z$,
\begin{equation}
Z=(Z_-\vert Z_+),\; \vert Z_-\vert=\vert Z_+\vert\nonumber
\end{equation}
 with respect to the new splitting
$\frak{X}=\left(\hat{\frak{X}}_-\vert \hat{\frak{X}}_+\right)$.
The point configurations $X$ and $Z$ are related by the
hole-particle involution on the set $\frak{X}_0$ as it is
described in Section \ref{PointConfigurations}.

Consider an $\hat{L}$-ensemble on $\X$ with the matrix $\hat{L}$
having the following block structure:
\begin{equation}\label{matrixhatL}
L=\left[\begin{array}{cc}
  0 & \hat{A} \\
 -\hat{A}^T & 0 \\
\end{array}\right],\;\;\; \hat{A}(x,y)=\dfrac{\hat{h}(x)\hat{h}(y)}{x-y}
\end{equation}
Here the decomposition of the matrix $\hat{L}$ into the blocks
corresponds to the new splitting
$\X=\left(\hat{\X}_-\vert\hat{\X}_+\right)$ of the discrete set
$\X$. As soon as the particle-hole configurations $Z$ are balanced
with respect to this splitting we find
\begin{equation}\label{ProbLHat}
\mbox{Prob}_{\hat{L}}(Z)=\dfrac{1}{\mbox{det}\;(1+\hat{L})}\;
\hat{h}^2(Z)\;\dfrac{V^2(Z_-)V^2(Z_+)}{\prod^2(Z_-;Z_+)}
\end{equation}
The balanced configurations $Z$ of particles and holes were
constructed from the unbalanced configurations $X$ of particles,
in which the number of negative particles is larger by $S$ then
the number of positive particles. Clearly, it is possible to
rewrite $\mbox{Prob}_{\hat{L}}(Z)$ in terms of the configurations
$X$. With a suitable choice of the weight $\hat{h}$ the expression
for $\mbox{Prob}_{\hat{L}}(Z)$ in terms of configurations $X$
takes the same form as the right-hand side of equation
(\ref{PROBX}) (up to a normalization constant). Namely, introduce
the new weight $\hat{h}$ in terms of the old weight $h$ according
to the formula:
\begin{equation}\label{NewWeighth}
\hat{h}(z)=
\left\{%
\begin{array}{ll}
    h(z)\prod\limits_{y\in\;\frak{X}_0}(z-y), & z\in \frak{X}_-\setminus\frak{X}_0, \\
     & \\
    \dfrac{1}{h(z)\prod_{y\in\;\frak{X}_0,y\neq z}(z-y)}, &  z\in \frak{X}_0, \\
     & \\
    \dfrac{h(z)}{\prod_{y\in\;\frak{X}_0}(z-y)}, & z\in \frak{X}_+. \\
\end{array}%
\right.
\end{equation}
\begin{prop}\label{PropositionProbabilityHatL}
 This choice of the weight
$\hat{h}$ (equation (\ref{NewWeighth})) gives
\begin{equation}\label{PROBZ}
\Prob_{\hat{L}}(Z)=\dfrac{1}{\det(1+\hat{L})}\;
\dfrac{1}{V^2(\frak{X}_0)  h^2(\frak{X}_0)}\; \;
h^2(X)\;\dfrac{V^2(X_-)V^2(X_+)}{\prod^2(X_-;X_+)}.
\end{equation}
\end{prop}
\begin{proof}
In order to see that equation (\ref{PROBZ}) is valid we rewrite
the expression
\begin{equation}
h^2(X)\;\dfrac{V^2(X_-)V^2(X_+)}{\prod^2(X_-;X_+)} \nonumber
\end{equation}
in terms of the balanced configuration $Z$ (the relation between
the unbalanced configuration $X$ and the balanced configuration is
shown in Tab. 1. We will also use the notation $Z_+^{I,II}$
introduced there). In particular we find
\\
\begin{itemize}
    \item $h^2(X)=h^2(Z_-)\cdot\dfrac{h^2(\frak{X}_0)}{h^2(Z_+^{I})}\;
    \cdot h^2(Z_+^{II})$
    \\
    \item $V^2(X_-)=V^2(\frak{X}_0\setminus Z_+^I)\cdot V^2(Z_-)\cdot\prod^2(Z_-;
    \frak{X}_0\setminus Z_+^I)$
    \\
    \item $V^2(X_+)=
    \dfrac{V^2(Z_+)}{V^2(Z_+^{I})\cdot\prod^2(Z_+^{I}; Z_+^{II})}$
    \\
    \item $\prod^2(X_-; X_+)=\prod^2(Z_-; Z_+^{II})\cdot\prod^2(\frak{X}_0\setminus Z_+^I;
    Z_+^{II})$
\end{itemize}
Thus
\begin{equation}
\dfrac{V^2(X_-)V^2(X_+)}{\prod^2(X_-;X_+)}=\dfrac{V^2(Z_-)V^2(Z_+)}{\prod^2(Z_-;Z_+)}
\;\dfrac{V^2(\frak{X}_0\setminus Z_+^I)\cdot\prod^2(Z_-
;{\frak{X}}_0)}{V^2(Z_+^I)\cdot\prod^2(Z_+^{I};
Z_+^{II})\cdot\prod^2(\frak{X}_0\setminus Z_+^I;
Z_+^{II})}\nonumber
\end{equation}
We rewrite $V^2(\frak{X}_0\setminus Z_+^I)$
\begin{eqnarray}
V^2(\frak{X}_0\setminus
Z_+^I)=\dfrac{V^2(\frak{X}_0)}{V^2(Z_+^I)\cdot\prod^2(Z_+^{I};
\frak{X}_0\setminus Z_+^I)} \nonumber
\end{eqnarray}
and note that the resulting denominator equals
\begin{equation}
\begin{split}
 &V^2(Z_+^I)\cdot
{\prod}^2(Z_+^{I};Z_+^{II})\cdot{\prod}^2(\frak{X}_0\setminus
Z_+^I; Z_+^{I})\cdot{\prod}^2(\frak{X}_0\setminus Z_+^I;
Z_+^{II})\cdot V^2(Z_+^I)\\
&=V^2(Z_+^I)\cdot{\prod}^2(Z_+^{II};
\frak{X}_0)\cdot{\prod}^2(\frak{X}_0\setminus Z_+^I;
Z_+^{I})\cdot V^2(Z_+^I)\\
&=\underset{y<x}{\prod\limits_{x,\;y\in\; Z_+^I}}(x-y)^2\cdot
\underset{y\in\;\frak{X}_0\setminus Z_+^I} {\prod\limits_{x\in\;
Z_+^I}}(x-y)^2\cdot
 \underset{y\in\;
\frak{X}_0}{\prod\limits_{x\in\; Z_+^{II}}}(x-y)^2\cdot
\underset{y>x}{\prod\limits_{x,\;y\in\; Z_+^I}}(x-y)^2
=\underset{y\in\; \frak{X}_0,\;y\neq x}{\prod\limits_{x\in\;
Z_+}}(x-y)^2. \nonumber
\end{split}
\end{equation}
Introducing $\hat h(z)$ by equation (\ref{NewWeighth}) we see that
equation (\ref{PROBZ}) is equivalent to equation (\ref{ProbLHat}).
\end{proof}
\subsection{Correlation functions for $L$-ensembles and averages of
characteristic polynomials}\label{CorrelationsofE}
By correlation functions for $L$-ensembles $\varrho(X)$ we mean
the probabilities that the random configurations include fixed
sets $X$. Thus
\begin{equation}
\varrho(X)=\sum\limits_{Y\supseteq X}\mbox{Prob}_{L}(Y).\nonumber
\end{equation}
It is known that the correlation functions $\varrho(X)$ are
expressed as symmetric minors of the correlation kernel
$K=L(1+L)^{-1}$, i.e.
\begin{equation}\label{correlationfunctionasdeterminantK}
\varrho(X)=\mbox{det}\; K(X\vert X).
\end{equation}
The proof of this fact can be found in the book by  Daley and
Vere-Jones   \cite{Daley} (Exercise 5.4.7), and in Borodin and
Olshanski  \cite{BorodinOlshanski1} (Proposition 2.1), Borodin,
Okounkov and Olshanski  \cite{BorodinOkounkovOlshanski}
(Appendix). Equation (\ref{correlationfunctionasdeterminantK})
tell us that random processes associated with $L$-ensembles are
discrete \textit{determinantal} point processes (see Borodin and
Olshanski \cite{BorodinOlshanski1, BorodinOlshanski2} for
definitions. A comprehensive survey on determinantal point
processes is given in Soshnikov \cite{soshnikov}).

In this Section we express the minors of the matrix $K$ in terms
of averages of  "characteristic polynomials" $E(\alpha, X)$
associated with $L$-ensembles. These objects were first introduced
in Borodin and Olshanski \cite{BorodinOlshanski3} and are
constructed as follows.
 With the above notations, for any fixed set $\alpha$ and point configurations $X$ and $Z$
 we set
\begin{equation}\label{LCharacteristicPolynomials}
\begin{split}
E(\alpha,X)=&\dfrac{\prod(\alpha;X_+)}{\prod(\alpha;X_-)},\;\;E(\alpha,Z)=\dfrac{\prod(\alpha;Z_+)}{\prod(\alpha;Z_-)},\\
&E(\alpha,X)=\frac{E(\alpha,Z)}{\prod(\alpha;\frak X_0)}.
\end{split}
\end{equation}
In order to compute the minors of the matrix $K$ we need the
following Lemma.
\begin{lem}\label{L(A,B)}
Let $A$, $B$ be two balanced configurations on the set $\frak{X}$,
\begin{equation}
\vert A_+\vert=\vert A_-\vert=a,\; \vert B_+\vert=\vert B_-\vert=b
\nonumber
\end{equation}
Then
\begin{equation}
 \det\;\left[L(A_-,B_+\vert B_-,A_+)\right] =
(-)^{w_{ab}}\; h(A)h(B)
\dfrac{V(A_-)V(A_+)}{\prod(A_-;A_+)}\dfrac{V(B_-)V(B_+)}{\prod(B_-;B_+)}
\nonumber
\end{equation}
where
\begin{equation}
w_{ab}=ab+b^2+\frac{a(a-1)}{2}+\frac{b(b-1)}{2}\nonumber
\end{equation}
\end{lem}
\begin{proof}
The formula above follows from the explicit structure of the
matrix $L$ (equation (\ref{matrixL})), from formulas for the
Cauchy determinants (equation (\ref{CauchyDeterminant})) and the
determinants of the block matrices (equation
(\ref{determinantoftheblockmatrix})).
\end{proof}
Introduce nonintersecting sets $\alpha^{\pm}$, $\beta^{\pm}$ of
complex numbers with nonequal elements,
\begin{alignat}{2}
\alpha^+= &\left(\alpha_1^+,\ldots ,\alpha_{k_1}^+\right),
&\qquad\beta^+=\left(\beta_1^+,\ldots ,\beta^+_{k_2}\right), \\
\alpha^-= &\left(\alpha_1^-,\ldots ,\alpha_{m_1}^-\right),
&\qquad\beta^-=\left(\beta_1^-,\ldots ,\beta^-_{m_2}\right).
\end{alignat}
 Assume that
$\alpha^{\pm}\cap\frak{X} =0$, $\beta^{\pm}\cap\frak{X}=0$, and
\begin{equation}
\vert\alpha^+\vert-\vert\alpha^-\vert=\vert\beta^+\vert-\vert\beta^-\vert=S.
\nonumber
\end{equation}
In what follows we extend the definitions of the matrices $K$, $L$
to the sets $\alpha^{\pm}$, $\beta^{\pm}$ in the following way. We
add to $L$ rows parameterized by $\alpha^-\sqcup\beta^+$ and
columns parameterized by $\alpha^+\sqcup\beta^-$, and then define
new matrix elements of $L$ according to
(\ref{matrixLstructure})-(\ref{matrixL}), where we assume that
$\alpha^-$ and $\beta^-$ are added to $\X_-$, and
$\alpha^+,\beta^+$ are added to $\X_+$. Then we set extended $K$
to be related to the extended $L$ by $K=L(1+L)^{-1}$.
\begin{prop}\label{LEMMAOKMINORAX}
The minor $\det\;\left[
K(\alpha^-,\beta^+\vert\beta^{-},\alpha^{+})\right]$ can be given
as a normalized average of a ratio of the functions $E(.,.)$
introduced above  with respect to the $\hat{L}$-ensemble. Namely,
\begin{equation}\label{EquationLemmOKMinorax}
\begin{split}
\det\;\left[ K(\alpha^-,\beta^+\vert\beta^{-},\alpha^{+})\right]&
=(-)^{w_{\alpha,\beta}}\left[\dfrac{\det\;(1+\hat{L})}{\det\;(1+L)}\;h^2(\frak{X}_0)V^2(\frak{X}_0)\right]\\
&\times\left[
h(\alpha)\dfrac{V(\alpha^{-})V(\alpha^+)}{\prod(\alpha^-;\alpha^+)}\right]
\left[
h(\beta)\dfrac{V(\beta^{-})V(\beta^+)}{\prod(\beta^-;\beta^+)}\right]
\\
&\times \left[\dfrac{\prod(\alpha^-;\frak{X}_0)\prod(\beta^-;
\frak{X}_0)}{\prod(\alpha^+;\frak{X}_0)\prod(\beta^+;\frak{X}_0)}\right]
\left\langle\dfrac{E(\alpha^+,Z)E(\beta^+,Z)}{E(\alpha^-,Z)E(\beta^-,Z)}\right\rangle_{\hat{L}}
\end{split}
\end{equation}
where
$w_{\alpha,\beta}=\dfrac{1}{2}\left[(|\alpha^-|+|\beta^-|)^2+|\beta^-|-|\alpha^-|\right]$.
\end{prop}
\begin{proof}
Equation (\ref{KMinor}) gives
\begin{equation}\label{DetKAsSum}
\mbox{det}\;\left[
K(\alpha^-,\beta^+\vert\beta^{-},\alpha^{+})\right]=
\sum\limits_{X\subset\;\X} \dfrac{\mbox{det}\;
L(\alpha^-,\beta^+,X\vert\beta^{-},\alpha^{+},X)}{\mbox{det}(1+L)}.
\end{equation}
We define the following sets
\begin{equation}\label{A}
A_+=\alpha^+\cup X_+ ,\;\;A_-=\alpha^-\cup X_-,
\end{equation}
\begin{equation}\label{B}
B_+=\beta^+\cup X_+ ,\;\;B_-=\beta^-\cup X_-.
\end{equation}
Only those configurations $X$ will contribute to the sum in
equation (\ref{DetKAsSum}) for which the sets $A_-$, $A_+$ have
equal number of elements and the sets $B_-$, $B_+$ have equal
number of elements. Otherwise
\begin{equation}
\mbox{det}\;\left[
L(\alpha^-,\beta^+,X\vert\beta^{-},\alpha^{+},X)\right]=0.\nonumber
\end{equation}
It means that the sum in equation (\ref{DetKAsSum}) runs
over unbalanced configurations $X$,
\begin{equation}
\vert X_-\vert-\vert X_+\vert=S.\nonumber
\end{equation}
Now we apply Lemma \ref{L(A,B)} to compute $\mbox{det}\;\left[
L(\alpha^-,\beta^+,X\vert\beta^{-},\alpha^{+},X)\right]$ with
$A_{\pm}$, $B_{\pm}$ given by equations (\ref{A}) and (\ref{B})
respectively. We find
\begin{equation}
\begin{split}
\mbox{det}\;\left[
K(\alpha^-,\beta^+\vert\beta^{-},\alpha^{+})\right]&=(-)^{w_{\alpha,\beta}}\left[\mbox{det}\;(1+L)\right]^{-1}\\
&\times\; \left[
h(\alpha)\dfrac{V(\alpha^{-})V(\alpha^+)}{\prod(\alpha^-;\alpha^+)}\right]
\left[
h(\beta)\dfrac{V(\beta^{-})V(\beta^+)}{\prod(\beta^-;\beta^+)}\right]\\
&\times\sum\limits_{X}
\left[\dfrac{E(\alpha^+,X)E(\beta^+,X)}{E(\alpha^-,X)E(\beta^-,X)}\right]
h^2(X)\dfrac{V(X^{-})V(X^+)}{{\prod^2}(X^-;X^+)}
\end{split}\nonumber
\end{equation}
Since the sum in this equation  runs over unbalanced
configurations, we cannot interpret the sum  as an average over
the $L$-ensemble. However, it is possible to rewrite this sum  in
terms of the balanced configurations $Z$ of particles and holes as
it was explained in Section \ref{subsectionLHat}. Then equation
(\ref{PROBZ}) tell us that the sum in equation above is just an
average over $\hat{L}$-ensemble. Replacing the sum over unbalanced
configurations $X$ by the sum over balanced configurations $Z$ and
using (\ref{LCharacteristicPolynomials}) we prove the Proposition.
\end{proof}
\begin{rem}
The Proposition above clearly gives the matrix elements of the
matrix $K$ in terms of averages of $E'^s $. For example,
\begin{equation}
K(\beta_i^+\vert\alpha_j^+)=
\dfrac{\mbox{det}\;(1+\hat{L})}{\mbox{det}\;(1+L)} \; h^2(x_0)\;
\dfrac{h(\beta_i^+)h(\alpha_j^+)}{(\beta_i^+-x_0)(\alpha_j^+-x_0)}
 \left\langle
E(\beta_i^+,Z)E(\alpha_j^+,Z)\right\rangle_{\hat{L}}\nonumber
\end{equation}
where the set $\frak{X}_0$ in the definition of the ensemble
$\hat{L}$ (see Tab.1) consists of only one point $x_0$,
$\frak{X}_0=\{x_0\}$. The point $x_0$ is an arbitrary point in
$\frak{X}_-$. The weight $\hat{h}$ of this $\hat{L}$-ensemble is
\begin{equation}\label{NewWeighth1}
\hat{h}(z)=
\left\{%
\begin{array}{ll}
    h(z)(z-x_0)\;, & z\in \frak{X}_-\setminus\{x\}_0 \\
     & \\
    \dfrac{1}{h(x_0)}\;, &  z=x_0 \\
     & \\
    \dfrac{h(z)}{(z-x_0)}\;, & z\in \frak{X}_+ \\
\end{array}%
\right.
\end{equation}
as it follows from expression (\ref{NewWeighth}).
\end{rem}
\subsection{Discrete polynomial ensembles}
Assume that a nonnegative function $f(x)$ is given on a finite set
$\frak{X}\subset\mathbb{R}$. We also require that $f$ does not
vanish at least at $N$ distinct points. Then the monic orthogonal
polynomials $\pi_0=1,\pi_1,\ldots ,\pi_{N-1}$ can be introduced by
orthogonalizing the system $(1,x,\ldots ,x^{N-1})$ in the Hilbert
space $L^2(\X, f\mu)$. Here $\mu $ denotes the counting measure on
the set $\X$. The orthogonality condition for the monic orthogonal
polynomials on $\X$ is
\begin{equation}\label{orthogonalityrelationforpolynomials}
\sum\limits_{x\in\;\frak{X}}\pi_j(x)\pi_k(x)f(x)=c_j^2\;\delta_{jk}.
\end{equation}
Let $Conf_N(\frak{X})$ denote the set of $N$-point configurations
(subsets) in $\frak{X}$. Such configurations will be denoted by
$X^{\triangle}$, $X^{\triangle}\in Conf_N(\frak{X})$. Consider a
point process on $\frak{X}$ which lives on $Conf_{N}(\frak{X})$
and for which the probability of a configuration $X$ is given by
\begin{equation}\label{DefinitionOfDiscretePolynomialEnsemble}
\mbox{Prob}(X^{\triangle})=\mbox{const}\;\prod\limits_{x\in\;
X^{\T}}f(x)\cdot V^2(X^{\T}),\;\;X^{\T}\in Conf_N(\frak{X}).
\end{equation}
We will denote this process by $\triangle_N(f)$.
\begin{defn}
The point process  $\triangle_N(f)$ will be called $N$-point
 ($\beta=2$)  discrete polynomial ensemble with the weight function
$f$. (The value of $\beta$ refers to the power of $V(X)$ in
(\ref{DefinitionOfDiscretePolynomialEnsemble}).)
\end{defn}
For symmetric functions
$g\left(X^{\triangle}\right)=g\left(x_1^{\triangle},\ldots
,x^{\triangle}_N\right)$ of points of the configuration
$X^{\triangle}$, the average $\left\langle\
g\right\rangle_{\triangle_N(f)}$ with respect to
$\mbox{Prob}(X^{\triangle})$ is
\begin{equation}
\left\langle\ g\right\rangle_{\triangle_N(f)}\equiv
\frac{\underset{X^{\triangle}\in\;Conf_N(\X)}{\sum}
g(X^{\T})f(X^{\T})\cdot
V^2(X^{\triangle})}{\sum\limits_{X^{\triangle}\in\;Conf_N(\X)}
f(X^{\T})\cdot V^2(X^{\triangle})}\nonumber
\end{equation}
Define the \textit{characteristic polynomial} $d(\xi)$ associated
with the point configuration $X^{\triangle}$ and the complex
parameter $\xi$:
\begin{equation}\label{DefinitionOfCharcateristicPolynomial}
d(\xi)=\prod(\xi;X^{\triangle}).
\end{equation}
Then the discrete variant of the classical Heine identity is given
by the following
\begin{prop}
\begin{equation}
\pi_N(\xi)=\left\langle
d(\xi)\right\rangle_{\triangle_N(f)}\nonumber
\end{equation}
where $\langle\,\cdot \,\rangle_{\T_N(f)}$ denotes the average
over the discrete polynomial ensemble $\T_N(f)$.
\end{prop}
\begin{proof} The proof of this relation is an application of
standard arguments of the theory of orthogonal polynomials (see,
for example, Deift \cite{deift}, Chapter 3).
\end{proof}
\begin{prop}\label{PROPOSITIONDISCRETEMPOINTCORR}
The $m$-point correlation function
\begin{equation}
\varrho_m(y_1,\ldots ,y_m)= \underset{X^{\T}\ni\;\{y_1,\ldots
,y_m\}}{\sum\limits_{X^{\T}\in\;Conf_N(\X)}} \Prob(X^{\T})
\nonumber
\end{equation}
for the discrete polynomial ensemble is given by the determinantal
formula
\begin{equation}\label{mpointcorrelationfunctionas
determinantofdckernel}
 \varrho_m(y_1,\ldots
,y_m)=\det\left[K_N^{CD}(y_i,y_j)\right]_{i,j=1}^m,\;\;
m=1,2,\ldots
\end{equation}
where $K_N^{CD}(x,y)$ stands for the normalized
Christoffel-Darboux kernel,
\begin{equation}
K_N^{CD}(x,y)=\sqrt{f(x)f(y)}\;\sum\limits_{j=0}^{N-1}\dfrac{\pi_j(x)\pi_j(y)}{c_j^2}.\nonumber
\end{equation}
\end{prop}
One possible proof of this proposition is to adopt standard
arguments of the random matrix theory to the case of the discrete
ensemble $\T_N(f)$ (see, for example, Mehta \cite{mehta}, Deift
\cite{deift}). In the following sections we present another proof.
First, we will show that the discrete polynomial ensembles are
equivalent to the $L$-ensembles (see Section \ref{equaivalence}).
Specifically, the discrete polynomial ensembles and the
$L$-ensembles can be transformed to each other by a suitable
particle-hole involution. Second, we compute the averages of
products and ratios of characteristic polynomials $E(\alpha, X)$
for the $L$-ensembles (see Section
\ref{SectionAveragesCharacteristicPolynomials}). From these
averages we deduce the averages of products and ratios of
characteristic polynomials $d(\xi)$ for the discrete orthogonal
ensemble $\triangle_N(f)$. Third, we  show in Section
\ref{SectionCorrelationFunctions} how formula
(\ref{mpointcorrelationfunctionas determinantofdckernel}) can be
obtained from the resulting expressions.
\subsection{Equivalence of $L$-ensembles and discrete polynomial
ensembles}\label{equaivalence} 
The connection between the discrete polynomial ensembles
$\triangle_N(f)$ and $L$-ensembles was demonstrated in Borodin and
Olshanski \cite{BorodinOlshanski2}, \S 5 . In the terminology of
Section \ref{PointConfigurations}, this relation is a consequence
of the particle-hole involution on an $N$-point subset of $\X$. To
be more specific, assume that the subset $\X_+$ in the definition
of the $L$-ensemble (see Section \ref{SectionLEnsembles}) is
finite, and consists of $N$ points. Thus $\vert
\frak{X}_+\vert=N$. We assume that the weight $h$ of the
$L$-ensemble is nonnegative on $\X$ and strictly positive on
$\X_+$. If the $L$-ensemble is defined by equation (\ref{matrixL})
then only balanced configurations $X$ with $\vert X_+\vert=\vert
X_-\vert$ have non-zero probabilities. Consider the particle-hole
configuration $X^{\T}$ constructed by the particle-hole involution
on the set $\X_+$. Thus
\begin{equation}
X^{\T}=X_-\sqcup\left(\X_+\setminus X_+\right). \nonumber
\end{equation}
Clearly, the configuration $X^{\T}$ consists of precisely $N-\vert
X_+\vert+\vert X_-\vert=N$ points where $N-\vert X_+\vert$ is the
number of negative holes, and $\vert X_-\vert$ is the number of
negative particles. Moreover, it is always possible to introduce a
weight $f$ on $\X$ in such a way that $\mbox{Prob}(X^{\T})$ takes
the same form as in the definition of the discrete polynomial
ensemble (equation
(\ref{DefinitionOfDiscretePolynomialEnsemble})). Namely, introduce
the weight $f$ in terms of the weight $h$ of the $L$-ensemble by
the formula
\begin{equation}\label{relation f and h}
f(x)=
\left\{%
\begin{array}{ll}
    \dfrac{h^2(x)}{\prod_{y\in\; \frak{X}_+}(x-y)^2}, & x\in\;\frak{X}_- ,\\
     \dfrac{1}{h^2(x)\;\prod_{y\in\; \frak{X}_+, y\neq x}(x-y)^2}, &
x\in\;\frak{X}_+.
\end{array}%
\right.
\end{equation}
\begin{prop}\label{PropositionRelationFandH}
The discrete polynomial ensemble with the weight $f$ and the $L$
-ensemble defined by equation (\ref{matrixL})-(\ref{PROBX}) are
connected by the particle-hole involution on the set $\X_+$.
\end{prop}
\begin{proof}
See Borodin and Olshanski \cite{BorodinOlshanski2}, Proposition
5.2.
\end{proof}
Once the $L$-ensemble is given, we can construct an
$\hat{L}$-ensemble with respect to the new splitting of the set
$\frak{X}$,
$\frak{X}=\left(\hat{\frak{X}}_-\vert\hat{\frak{X}}_+\right)$ as
in Section \ref{subsectionLHat} (see also Tab. 1). Clearly
$\vert\hat{\frak{X}}_+\vert=N+S$. This $\hat{L}$-ensemble induces
a discrete $(N+S)$-point polynomial ensemble under the
particle-hole involution on the set $\hat{\frak{X}}_+$. It follows
from Proposition \ref{PropositionRelationFandH} that the induced
polynomial ensemble with the weight $\hat{f}(x)$ will be
equivalent to the $\hat{L}$-ensemble with the weight $\hat{h}(x)$
if
\begin{equation}\label{relationhatfandhath}
\hat{f}(x)=
\left\{%
\begin{array}{ll}
    \dfrac{\hat{h}^2(x)}{\prod_{y\in\; \hat{\frak{X}}_+}(x-y)^2}, & x\in\;\hat{\frak{X}}_- \\
     \dfrac{1}{\hat{h}^2(x)\;\prod_{y\in\; \hat{\frak{X}}_+, y\neq x}(x-y)^2}, & x\in\;\hat{\frak{X}}_+
\end{array}%
\right.
\end{equation}
Denote by $\triangle_N(f)$ the discrete $N$-point polynomial
ensemble which is equivalent to the $L$-ensemble, and by
$\hat{\triangle}_{N+S}(\hat{f})$ the discrete $(N+S)$-point
polynomial ensemble which is equivalent to the $\hat{L}$-ensemble.
\begin{prop}\label{PROPOSITION O RAVENSTVE VESOV}
The ensembles $\triangle_N(f)$ and
$\hat{\triangle}_{N+S}(\hat{f})$ have the same weight,
\begin{eqnarray}
\hat{f}(x)=f(x),\;\; x\in \frak{X}. \nonumber
\end{eqnarray}
\end{prop}
\begin{proof}
We  use formula (\ref{NewWeighth}) to express the weight
$\hat{h}(x)$ in terms of the weight $h(x)$ in expression
(\ref{relationhatfandhath}), and after that formula (\ref{relation
f and h}) which expresses the weight $f(x)$ in terms of the weight
$h(x)$:
\begin{itemize}
    \item $x\in\;\frak{X}_-\setminus \frak{X}_0$
\end{itemize}
\begin{equation}
\begin{split}
\hat{f}(x)=\;\frac{\hat{h}^2(x)}{\prod_{y\in\;
\hat{\frak{X}}_+}(x-y)^2}&=\frac{h^2(x)\prod_{y\in\;\frak{X}_0}(x-y)^2}{\prod_{y\in\;
\hat{\frak{X}}_+}(x-y)^2}\\
&=\;\frac{h^2(x)}{\prod_{y\in\; \frak{X}_+}(x-y)^2}=f(x)
\end{split}\nonumber
\end{equation}
\begin{itemize}
    \item $x\in\;\frak{X}_0$
\end{itemize}
\begin{equation}
\begin{split}
\hat{f}(x)=\;\frac{1}{\hat{h}^2(x)\prod_{y\in\;
\hat{\frak{X}}_+,y\neq
x}(x-y)^2}&=\frac{h^2(x)\prod_{y\in\;\frak{X}_0,y\neq
x}(x-y)^2}{\prod_{y\in\;
\hat{\frak{X}}_+,y\neq x}(x-y)^2}\\
&=\;\frac{h^2(x)}{\prod_{y\in\; \frak{X}_+}(x-y)^2}=f(x)\nonumber
\end{split}
\end{equation}
\begin{itemize}
    \item $x\in\;\frak{X}_+$
\end{itemize}
\begin{equation}
\begin{split}
\hat{f}(x)=\;\frac{1}{\hat{h}^2(x)\prod_{y\in\;
\hat{\frak{X}}_+,y\neq
x}(x-y)^2}=\frac{\prod_{y\in\;\frak{X}_0}(x-y)^2}{h^2(x)\prod_{y\in\;
\hat{\frak{X}}_+,y\neq x}(x-y)^2}\\
=\frac{1}{h^2(x)\prod_{y\in\; \frak{X}_+,y\neq
x}(x-y)^2}=f(x)\nonumber
\end{split}
\end{equation}
\end{proof}
The correspondence between the $L$-ensembles and the discrete
orthogonal polynomial ensembles is summarized in Tab. 2.

\subsection{Averages of characteristic
polynomials}\label{SectionAveragesCharacteristicPolynomials}
 The
goal of this Section is to compute the averages of characteristic
polynomials for the discrete polynomial ensembles. We will use the
equivalence of the $L$-ensembles and the discrete polynomial
ensembles, and deduce from Proposition \ref{LEMMAOKMINORAX} the
averages of products and ratios of characteristic polynomials.
Specifically, we will express the right-hand side of equation
(\ref{EquationLemmOKMinorax}) in terms of averages and
normalization constants of the ensemble $\T_N(f)$. This ensemble
has the weight $f$ given by equation (\ref{relation f and h}).
Thus $\T_N(f)$ is connected with the $L$-ensemble by the
particle-hole involution on the set $\X_+$ in accordance with
Proposition \ref{PropositionRelationFandH}.

We begin with the constant in equation
(\ref{EquationLemmOKMinorax}). Denote by $C_N$ and $C_{N+S}$ the
normalization constants for $\T_N(f)$ and $\T_{N+S}(f)$:
\begin{align}
C_N &=\sum\limits_{X\in\; Conf_N(\X)} V^2(X)\cdot f(X),\nonumber\\
C_{N+S}&=\sum\limits_{X\in\; Conf_{N+S}(\X)} V^2(X)\cdot
f(X).\nonumber
\end{align}
\begin{prop}
The constant in equation (\ref{EquationLemmOKMinorax}) is equal to
the ratio of the normalization constants $C_N$ and $C_{N+S}$, i.e.
\begin{equation}
\frac{C_{N+S}}{C_N}=
V^2(\frak{X}_0)h^2(\frak{X}_0)\;\frac{\det\;(1+\hat{L})}{\det\;(1+L)}.\nonumber
\end{equation}
\end{prop}
\begin{proof}
The expression $\left[\mbox{det}\;(1+L)\right]^{-1}$ represents
the probability that a random configuration $X$ of the
$L$-ensemble is empty, i.e.
\begin{equation}
\mbox{Prob}\left(X=\emptyset\right)=\left[\mbox{det}\;(1+L)\right]^{-1}.
\nonumber
\end{equation}
The empty configuration (of particles) for the $L$-ensemble
corresponds to the configuration $X^{\triangle}=\frak{X}_+$ of the
$N$ positive holes for the $\triangle_N(f)$ ensemble. Since the
$L$-ensemble is equivalent to the $\triangle_N(f)$ ensemble we
have
\begin{equation}
\mbox{Prob}\left(X=\emptyset\right)=\mbox{Prob}\left(X^{\triangle}=\X_+\right)
 =\frac{1}{C_N}\; V^2(\frak{X}_+)f(\frak{X}_+).\nonumber
\end{equation}
Therefore
\begin{equation}
\left[\mbox{det}\;(1+L)\right]^{-1}=\frac{1}{C_N}\;
V^2(\frak{X}_+)f(\frak{X}_+).\nonumber
\end{equation}
We repeat the above considerations for the $\hat{L}$-ensemble and
conclude that
\begin{eqnarray}
\frac{\mbox{det}\;(1+\hat{L})}{\mbox{det}\;(1+L)}=
\frac{C_{N+S}}{C_N}\;
\frac{V^2(\frak{X}_+)f(\frak{X}_+)}{V^2\left(\X_0
\sqcup\X_+\right)f\left(\X_0\sqcup\X_+\right)}\nonumber\\
=\frac{C_{N+S}}{C_N}\; \frac{1}{V^2\left(\X_0
\right){\prod}^2(\X_0;\X_+)f\left(\X_0\right)}\nonumber\\
=\frac{C_{N+S}}{C_N}\; \frac{1}{V^2\left(\X_0
\right)h^2\left(\X_0\right)}.\nonumber
\end{eqnarray}
Here we have used formula (\ref{relation f and h}) to obtain the
last equation.
\end{proof}
Let the sets $\alpha^{\pm}$, $\beta^{\pm}$ be defined as in
Section \ref{CorrelationsofE}. Recall that $d(\xi)$ was defined in
equation (\ref{DefinitionOfCharcateristicPolynomial}).
\begin{thm}\label{THEOREMAvergesCharacteristicPolynomials}
For any integer $N\geq 1$, take an integer $S$ such that
$N-\vert\X\vert+1\leq S\leq N-1$, complex numbers
$\alpha^-=(\alpha_1^-,\ldots ,\alpha_{m_1}^-)$,
$\alpha^+=(\alpha_1^+,\ldots ,\alpha_{k_1}^+)$,
$\beta^-=(\beta_1^-,\ldots ,\beta_{m_2}^-)$,
$\beta^+=(\beta_1^+,\ldots ,\beta_{k_2}^+)$ such that $\vert
\alpha^-\vert-\vert\alpha^+\vert=\vert
\beta^-\vert-\vert\beta^+\vert=S$, in each set $\alpha^{\pm}$,
$\beta^{\pm}$ the numbers are pairwise distinct, and the sets
$\alpha^+,\beta^+$ do not intersect $\X$. Then the average of
products and ratios of characteristic polynomials with respect to
the discrete polynomial ensemble is given by the formula
\begin{multline}\label{FormulaAveragesCharacteristicPolynomials}
\left\langle\dfrac{\prod_{i=1}^{m_1}d(\alpha_i^-)\prod_{i=1}^{m_2}
d(\beta_i^-)}{\prod_{j=1}^{k_1}d(\alpha_j^+)\prod_{j=1}^{k_2}d(\beta_j^+)}\right\rangle_{\triangle_N(f)}=
\left[\dfrac{C_{N-S}}{C_N}\right]\\
\times\;(-)^{w_{\alpha\beta}}\;\dfrac{\prod(\alpha^-;\alpha^+)\prod(\beta^-;\beta^+)}{V(\alpha^-)V(\alpha^+)V(\beta^-)V(\beta^+)}
\; \;\det\;\left[W_N(\alpha^-,\beta^+\vert
\beta^-,\alpha^+)\right]
\end{multline}
where the kernel function $W_N$ is defined by
\begin{itemize}
    \item
    $W_N(\alpha_i^-,\beta_j^-)=\dfrac{C_{N-S-1}}{C_{N-S}}\;
    \left\langle d(\alpha_i^-)d(\beta_j^-)\right\rangle_{\triangle_{N-S-1}(f)}$
    \item
    $W_N(\alpha_i^-,\alpha_j^+)=\dfrac{1}{\alpha_i^--\alpha_j^+}\;
    \left\langle\dfrac{d(\alpha_i^-)}{d(\alpha_j^+)}\right\rangle_{\triangle_{N-S}(f)}$
    \item $W_N(\beta_i^+,\beta_j^-)=\dfrac{1}{\beta_i^+-\beta_j^-}\;
    \left\langle\dfrac{d(\beta_j^-)}{d(\beta_i^+)}\right\rangle_{\triangle_{N-S}(f)}$
    \item
    $W_N(\beta_i^+,\alpha_j^+)=\dfrac{C_{N-S+1}}{C_{N-S}}\;
   \left\langle\dfrac{1}{d(\beta_i^+)d(\alpha_j^+)}\right\rangle_{\triangle_{N-S+1}(f)}$
\end{itemize}
 and $S$, $w_{\alpha\beta}$ are expressed in terms of
$|\alpha^{\pm}|$, $|\beta^{\pm}|$ as in Proposition
\ref{LEMMAOKMINORAX}. Note that $S$ above can be \textbf{any}
integer, positive or negative.
\end{thm}
\begin{proof}
Assume first that $S\geq 0$. Then the proof is based on the
expression for the minors of the $K$ matrix in terms of averages
of $E'^{s}$ over the $\hat{L}$-ensemble (see Proposition
\ref{LEMMAOKMINORAX}). The equivalence of the $L$-ensembles and
orthogonal polynomial ensembles enables us to rewrite averages of
$E'^s$ as averages over characteristic polynomials with respect to
orthogonal polynomial ensembles.  We note that
\begin{eqnarray}
E(\alpha,Z)\equiv \dfrac{\prod(\alpha;Z^+)}{\prod(\alpha;Z^-)}
=\dfrac{\prod(\alpha;\frak{X}_0)\prod(\alpha;\frak{X}_+)}{\prod(\alpha;Z^{\triangle})}
\nonumber
\end{eqnarray}
where $Z^{\triangle}$ is a random point configuration for the
ensemble $\hat{\triangle}_{N+S}(f)$. We insert the expression
above for $E(\alpha,Z)$ to the formula for the minors of the
matrix $K$ in Proposition \ref{LEMMAOKMINORAX}. According to
Proposition \ref{PropositionRelationFandH} and Proposition
\ref{PROPOSITION O RAVENSTVE VESOV} the average in the formula for
$\mbox{det}\;\left[K(\alpha^-,\beta^+\vert\beta^-,\alpha^+)\right]$
can be understood as that over the discrete polynomial ensemble
$\triangle_{N+S}(f)$. Thus the formula for the minors of the
matrix $K$ takes the form
\begin{equation}\label{KKKK}
\begin{split}
\mbox{det}\;\left[
K(\alpha^-,\beta^+\vert\beta^{-},\alpha^{+})\right]
&=(-)^{w_{\alpha,\beta}}\left[\dfrac{C_{N+S}}{C_N}\right] \left[
h(\alpha)\dfrac{V(\alpha^{-})V(\alpha^+)}{\prod(\alpha^-;\alpha^+)}\right]\left[
h(\beta)\dfrac{V(\beta^{-})V(\beta^+)}{\prod(\beta^-;\beta^+)}\right]\\
&
\times\dfrac{\prod(\alpha^+;\frak{X}_+)\prod(\beta^+;\frak{X}_+)}{\prod(\alpha^-;\frak{X}_+)\prod(\beta^-;\frak{X}_+)}
\left\langle\dfrac{\prod_{i=1}^{m_1}d(\alpha_i^-)\prod_{i=1}^{m_2}
d(\beta_i^-)}{\prod_{j=1}^{k_1}d(\alpha_j^+)\prod_{j=1}^{k_2}d(\beta_j^+)}\right\rangle_{\triangle_{N+S}(f)}.
\end{split}
\end{equation}
It is clear that the formula above expresses the matrix elements
of the matrix $K$ as averages of corresponding products and ratios
of characteristic polynomials. In particular we find
\begin{equation}
\begin{split}
    K(\beta_i^+\vert\alpha_j^+)&=h(\alpha_j^+)h(\beta_i^+)\prod(\alpha^+_j;\frak{X}_+)\prod(\beta^+_i;\frak{X}_+)
   \\&\times
     \dfrac{C_{N+1}}{C_N}\;
    \left\langle\dfrac{1}{d(\beta_i^+)d(\alpha_j^+)}\right\rangle_{\triangle_{N+1}(f)},
\end{split}\nonumber
\end{equation}
\begin{equation}
\begin{split}
K(\alpha_i^-\vert\beta_j^-)&=\dfrac{C_{N-1}}{C_N}\;h(\alpha_i^-)h(\beta_j^-){\prod}^{-1}(\alpha^-_i;\frak{X}_+){\prod}^{-1}(\beta^-_j;\frak{X}_+)\\
   &\times \left\langle d(\alpha_i^-)
    d(\beta_j^-)\right\rangle_{\triangle_{N-1}(f)},
\end{split}\nonumber
\end{equation}
\begin{equation}
\begin{split}
K(\alpha_i^-\vert\alpha_j^+)&=h(\alpha_i^-)h(\alpha_j^+){\prod}^{-1}(\alpha^-_i;\frak{X}_+){\prod}(\alpha^+_j;\frak{X}_+)\\
  &\times \dfrac{1}{\alpha_i^--\alpha_j^+}
    \left\langle\dfrac{d(\alpha_i^-)}{
    d(\alpha_j^+)}\right\rangle_{\triangle_{N}(f)},
\end{split}\nonumber
\end{equation}
\begin{equation}
\begin{split}
K(\beta_i^+\vert\beta_j^-)&=
h(\beta_i^+)h(\beta_j^-)\prod(\beta_i^+;\frak{X}_+){\prod}^{-1}(\beta^-_j;\frak{X}_+)
\\
 &\times \dfrac{1}{\beta_i^+ -\beta_j^-}
    \left\langle\dfrac{d(\beta_j^-)}{
    d(\beta_i^+)}\right\rangle_{\triangle_{N}(f)}.
\end{split}\nonumber
    \end{equation}
Expressing the elements of the matrix $K$ in the left-hand side of
equation (\ref{KKKK}) in accordance with these formulas we prove
the Theorem after some simplifications.

If $S<0$ we can prove Proposition \ref{LEMMAOKMINORAX} and Theorem
\ref{THEOREMAvergesCharacteristicPolynomials} in a similar way
considering $\X_0$ in the construction of balanced particle-hole
configurations  as a subset of $\X_+$.  Another possibility to see
that formula (\ref{FormulaAveragesCharacteristicPolynomials}) for
$S<0$ holds is to perform the particle-hole involution on the
whole  $\X$. Under this particle-hole involution formula
(\ref{FormulaAveragesCharacteristicPolynomials}) does note change.
However, $S>0$ is transformed to $S<0$.
\end{proof}
\begin{rem} In formula
(\ref{FormulaAveragesCharacteristicPolynomials}) the number of
characteristic polynomials is greater or smaller than that in the
numerator by $2\vert S\vert$. The case when these numbers are
different by an odd integer is obtained from
(\ref{FormulaAveragesCharacteristicPolynomials}) by taking the
limit $\alpha_i\rightarrow \infty$ or $\beta_j\rightarrow\infty$
for some $i$ or $j$, and using relation
$d(\xi)=\xi^N(1+\mathcal{O}(1)), \xi\rightarrow\infty$.
\end{rem}
\begin{rem}
Theorem \ref{THEOREMAvergesCharacteristicPolynomials} combines the
different formulas of two-point function type obtained in Brezin
and Hikami \cite{brezin1}, Strahov and Fyodorov \cite{strahov}
into one formula (equation
(\ref{FormulaAveragesCharacteristicPolynomials})).
\end{rem}
\begin{prop}The kernels $W_N$ in the theorem above can
be expressed in terms of the monic discrete orthogonal polynomials
$\pi_k(x)$ and their Cauchy type transforms $h_k(\epsilon)$,
\begin{equation}\label{CauchyTransforms}
h_k(\epsilon)=\dfrac{1}{2\pi i}\;\sum\limits_{x\in
\X}\dfrac{\pi_k(x) f(x)}{x-\epsilon}
\end{equation}
In particular we find
\begin{equation}
W_N(\alpha_i^-,\beta_j^-)=\dfrac{C_{N-S-1}}{C_{N-S}}
\;\dfrac{\pi_{N-S}(\alpha_i^-)\pi_{N-S-1}(\beta_j^-)-
\pi_{N-S}(\beta_j^-)\pi_{N-S-1}(\alpha_i^-)}{\alpha^-_i-\beta_j^-}
\nonumber
\end{equation}
\begin{equation}
W_N(\alpha_i^-,\alpha_k^+)=\gamma_{N-S-1}\;\dfrac{\pi_{N-S}(\alpha_k^+)h_{N-S-1}(\alpha_i^-)-
\pi_{N-S-1}(\alpha_k^+)h_{N-S}(\alpha_i^-)}{\alpha^-_i-\alpha_k^+}\nonumber
\end{equation}
\begin{equation}
W_N(\beta_i^+,\beta_j^-)=\gamma_{N-S-1}\;
\dfrac{\pi_{N-S}(\beta_i^+)h_{N-S-1}(\beta_j^-)-
\pi_{N-S-1}(\beta_i^+)h_{N-S}(\beta_j^-)}{\beta^+_i-\beta^-_j}\nonumber
\end{equation}
\begin{equation}
W_N(\beta_i^+,\beta_j^-)=\dfrac{C_{N-S+1}}{C_N}\;\gamma_{N-S}\gamma_{N-S-1}\;\dfrac{h_{N-S}(\beta_i^+)h_{N-S-1}(\beta_j^+)-
h_{N-S}(\beta_j^+)h_{N-S-1}(\beta_i^+)}{\beta^+_i-\beta^+_j}\nonumber
\end{equation}
Here we have introduced the coefficients $\gamma_k$,
\begin{equation}
\gamma_k=-\dfrac{2\pi i}{c_k^2}\nonumber
\end{equation}
\end{prop}
\begin{proof} The computations for the case of the discrete orthogonal
ensemble $\triangle_N(f)$ are similar to that for the case of the
ensemble of Hermitian matrices (see Strahov and Fyodorov
\cite{strahov}, Baik, Deift and Strahov \cite{baik}).
\end{proof}

\subsection{Correlation
functions of discrete polynomial ensemble}\label{SectionCorrelationFunctions} 
This Section shows the relation between $m$-point correlation
function $\varrho_m(y_1,\ldots ,y_m)$ for the discrete polynomial
ensembles and averages of characteristic polynomials. The
Proposition below gives $\varrho_m(y_1,\ldots ,y_m)$ in terms of
these averages.
\begin{prop}\label{PropositionCorrelationFunctionAsaverage}
For any $y_1,\ldots ,y_m\in \X$ we have
\begin{equation}\label{CorrelationFunctionAsaverage}
\varrho_m (y_1,\ldots , y_m)=
\underset{v_1=y_1}{\RES}\ldots\underset{v_m=y_m}{\RES}
\left[\left[\dfrac{\partial^m}{\partial u_1\ldots \partial u_m}\;
\left\langle\dfrac{d(u_1)\ldots d(u_m)}{d(v_1)\ldots
d(v_m)}\right\rangle_{\T_N(f)}\right]_{u=v}\right]
\end{equation}
\end{prop}
\begin{proof}
In order to see that that the expression for $\varrho_m(y_1,\ldots
,y_m)$ is correct we differentiate the product of characteristic
polynomials in the numerator,
\begin{equation}
\dfrac{\partial^m}{\partial u_1\ldots \partial u_m}\;
\left[\prod\limits_{i=1}^m\prod\limits_{j=1}^N[u_i-
x_j^{\T}]\right]
 = \sum\limits_{j_1,\ldots
,j_m=1}^N\left[\dfrac{1}{\prod_{k=1}^m(u_k-x^{\T}_{j_k})}\right]
\prod\limits_{i=1}^m\prod\limits_{j=1}^N[u_i-x_j^{\T}]\nonumber .
\end{equation}
Thus the formula in the proposition is equivalent to the following
one:
\begin{equation}
\varrho_m(y_1,\ldots ,y_m)=\underset{v_1=y_1}{\RES}\ldots
\underset{v_m=y_m}{\mbox{Res}} \biggl[\sum\limits_{X\in\;
Conf_N(\X)} \mbox{Prob}(X) \biggl[\sum\limits_{j_1,\ldots,j_m=1}^N
\dfrac{1}{\prod_{k=1}^m(v_k-x^{\T}_{j_k})} \biggr]\biggr]\nonumber
,
\end{equation}
which is evidently equivalent to the definition of $\varrho_m$.
\end{proof}
\begin{rem}
The above argument applies to any point process on $\X$, not just
$\triangle_N(f)$.
\end{rem}
\begin{prop}\label{KernelAsAnaveragesOftwocharacteristicpolynomials}
The $m$-point correlation function of the discrete ensemble
$\triangle_N(f)$ is given by the formula
\begin{equation}
\varrho_m(y_1,\ldots ,y_m)=\det\;\left[K(y_i,y_j)\right]_{i,j=1}^m
\end{equation}
where
\begin{equation}\label{definitionofK}
K(x,y)=\left\{%
\begin{array}{ll}
    \underset{\xi=y}{\RES}\left[\dfrac{1}{x-\xi}\left\langle\dfrac{d(x)}{d(\xi)}\right\rangle_{\triangle_N(f)}\right],
& x\neq y ; \\
\underset{\xi=y}{\RES}\left[\dfrac{d}{d\eta}\left\langle\dfrac{d(\eta)}{d(\xi)}\right\rangle\biggr|_{\eta=\xi}\right], & x=y. \\\end{array}%
\right.
\end{equation}
\end{prop}
\begin{proof}
We note first that if we take
\begin{equation}
\vert\beta^+\vert=\vert\beta^-\vert=0,\;\;
\vert\alpha^+\vert=\vert\alpha^-\vert=k \nonumber
\end{equation}
in Theorem \ref{THEOREMAvergesCharacteristicPolynomials} the
formula for averages of characteristic polynomials takes the
following form:
\begin{equation}
\left\langle\mbox{det}\;\left[\dfrac{1}{u_i-v_j}\;\dfrac{d(u_i)}{d(v_j)}\right]_{i,j=1}^k\right\rangle_{\triangle_N(f)}
=
\mbox{det}\;\left[\dfrac{1}{u_i-v_j}\left\langle\dfrac{d(u_i)}{d(v_j)}\right\rangle_{\triangle_N(f)}\right]_{i,j=1}^k
\end{equation}
We multiply the right-hand and left-hand sides of the equation
above by the product $\prod\limits_{j=1}^k(u_j-v_j)$. After that
we differentiate $k$ times with respect to $u_1,\ldots u_k$, take
$u_1=v_1,\ldots ,u_k=v_k$ and find (after simple manipulations)
\begin{multline}\label{RavenstvoDeterminantov}
\left\langle\left\vert
\begin{array}{cccc}
  G(v_1) & \dfrac{1}{v_1-v_2} & \ldots & \dfrac{1}{v_1-v_k} \\
  \dfrac{1}{v_2-v_1} & G(v_2) & \ldots & \dfrac{1}{v_2-v_k} \\
  \vdots &  &  &  \\
  \dfrac{1}{v_k-v_1} & \dfrac{1}{v_k-v_2} & \ldots & G(v_k) \\
\end{array}%
\right\vert\right\rangle_{\triangle_N(f)}=
\\
\left\vert%
\begin{array}{cccc}
  \left\langle G(v_1)\right\rangle_{\triangle_N(f)} & \dfrac{1}{v_1-v_2}\;\left\langle \dfrac{d(v_1)}{d(v_2)}\right\rangle_{\triangle_N(f)}
   & \ldots & \dfrac{1}{v_1-v_k}\;\left\langle \dfrac{d(v_1)}{d(v_k)}\right\rangle_{\triangle_N(f)} \\
  \dfrac{1}{v_2-v_1}\;\left\langle \dfrac{d(v_2)}{d(v_1)}\right\rangle_{\triangle_N(f)}
   & \left\langle G(v_2)\right\rangle_{\triangle_N(f)} & \ldots & \dfrac{1}{v_2-v_k}\;\left\langle \dfrac{d(v_2)}{d(v_k)}\right\rangle_{\triangle_N(f)} \\
  \vdots &  &  &  \\
  \dfrac{1}{v_k-v_1}\;\left\langle \dfrac{d(v_k)}{d(v_1)}\right\rangle_{\triangle_N(f)}
  & \dfrac{1}{v_k-v_2}\;\left\langle \dfrac{d(v_k)}{d(v_2)}\right\rangle_{\triangle_N(f)} & \ldots & \left\langle G(v_k)\right\rangle_{\triangle_N(f)} \\
\end{array}%
\right\vert
\end{multline}
where we have introduced
\begin{equation}
G(v)=\sum\limits_{j=1}^N\dfrac{1}{v-x_j}.\nonumber
\end{equation}
We note
\begin{equation}
\begin{split}
\underset{v_1=y_1}{\mbox{Res}}&\ldots\underset{v_m=y_m}{\mbox{Res}}\left[\left
\langle G(v_1)\ldots
G(v_k)\right\rangle_{\triangle_N(f)}\right]=\\
&=\sum\limits_{X\in\;Conf_N(\X)}\left[\sum\limits_{j_1,\ldots
,j_m=1}^N\delta_{y_1,x_{j_1}}\ldots
\delta_{y_{m},x_{j_m}}\right]\mbox{Prob}(X)\\
&=\varrho_m(y_1,\ldots ,y_m)
\end{split}\nonumber
\end{equation}
Thus the residue of the left-hand side of equation
(\ref{RavenstvoDeterminantov}) is precisely the $m$-point
correlation function. Now we compute the residue of the right-hand
side of equation (\ref{RavenstvoDeterminantov}). It follows from
the definition of $K(x,y)$ (see equation (\ref{definitionofK}))
that
\begin{equation}
\underset{v=y}{\mbox{Res}}\left[\left\langle
G(v)\right\rangle_{\triangle_N(f)}\right]=K(y,y) \nonumber
\end{equation}
The determinant in the right-hand side of equation
(\ref{RavenstvoDeterminantov}) is a sum of products, and the
off-diagonal elements are included as the cyclic products. The
residues of the cyclic products are then the cyclic products of
the kernels $K(y_i,y_j)$, for example
\begin{equation}
\begin{split}
\underset{v_1=y_1}{\mbox{Res}}\ldots  &
\underset{v_j=y_j}{\mbox{Res}}
\biggl[\dfrac{1}{v_1-v_2}\dfrac{1}{v_2-v_3}\cdots
\dfrac{1}{v_j-v_1}
\left\langle\dfrac{d(v_1)}{d(v_2)}\right\rangle_{\triangle_N(f)}
\left\langle\dfrac{d(v_2)}{d(v_3)}\right\rangle_{\triangle_N(f)}
\cdots
\left\langle\dfrac{d(v_j)}{d(v_1)}\right\rangle_{\triangle_N(f)}
\biggr] \\
\\
& =K(y_1,y_2)K(y_2,y_3)\ldots K(y_j,y_1)
\end{split}\nonumber
\end{equation}
It means that the residues of the right-hand side of equation
(\ref{RavenstvoDeterminantov}) is
$\mbox{det}\;\left[K(y_i,y_j)\right]_{i,j=1}^m$ which proves the
Proposition.
\end{proof}
\subsection{Proof of Proposition
\ref{PROPOSITIONDISCRETEMPOINTCORR}} This standard result (see
Mehta \cite{mehta}, Deift \cite{deift}) follows from the formula
for the averages of two characteristic polynomials
\cite{fyodorov2, strahov, baik},
\begin{equation}
\dfrac{1}{x-\xi}\;\left\langle\dfrac{d(x)}{d(\xi)}\right\rangle_{\triangle_N(f)}=
\gamma_{N-1}\;\dfrac{\pi_N(x)h_{N-1}(\xi)-\pi_{N-1}(x)h_N(\xi)}{x-\xi}\nonumber
\end{equation}
Indeed,
\begin{equation}
\underset{\epsilon=y}{\mbox{Res}}\left[h_k(\epsilon)\right]=-\dfrac{1}{2\pi
i}\;f(y)\pi_k(y) \nonumber
\end{equation}
Therefore  Proposition
\ref{KernelAsAnaveragesOftwocharacteristicpolynomials} gives
\begin{equation}
K(x,y)=\dfrac{1}{c_{N-1}^2}\;f(y)\;\dfrac{\pi_{N}(x)\pi_{N-1}(y)-\pi_{N-1}(x)\pi_N(y)}{x-y}\nonumber
\end{equation}
when $x\neq y$, and
\begin{equation}
\begin{split}
K(x,x)&=\dfrac{1}{c_{N-1}^2}\;f(x)\left[\pi'_{N}(x)\pi_{N-1}(x)-\pi'_{N-1}(x)\pi_N(x)\right]\\
&=f(x)\sum\limits_{j=0}^{N-1}\dfrac{\pi_j(x)\pi_j(x)}{c_j^2}\nonumber
\end{split}
\end{equation}
when $x=y$.
 Taking into account the Christoffel-Darboux
summation formula (see \cite{Szego} ) we see that the determinant
with the kernel $K(x,y)$ is equal to the determinant with the
Christoffel-Darboux kernel $K^{CD}_N(x,y)$.
\section{Pfaffian point ensembles}
\subsection{Pfaffian
$L$-ensembles}\label{PSectionPfafianLEnsembles} Given a set $\X$
let us construct two copies of $\X$, and denote them by $\X'$ and
$\X''$. We then introduce a $2\times 2$ matrix valued function
$L(x,y)$, which depends on two variables $x,y\in \X$
\begin{equation}\label{PLl}
L(x,y)=\left[\begin{array}{cc}
  L(x',y') & L(x',y'')\\
  L(x'',y') & L(x'',y'') \\
\end{array}\right]
\end{equation}
Here we denote by the same letter $L$ $2\times 2$ matrix valued
function of two arguments $x,y\in\X$, and the scalar function
whose arguments are taken from $\X'\sqcup\X''$. Once $x,y$ take
values in $\X$, the variables $x'$, $y'$ ($x''$, $y''$) are the
elements of $\X'$ ($\X''$) corresponding to $x$, $y$. Assume that
this function $L$ is antisymmetric. Then $L(x,y)$ defines a
$2\times 2$ block antisymmetric matrix on $\X$.

To any $X\subset \X$ there will correspond a $2\times 2$ block
antisymmetric submatrix of $L$. We denote this submatrix by
$L(X\vert X)$. If $X$ consists of $m$ points,
\begin{equation}
X=\left(x_1,\ldots , x_m\right),\;\; X\subset \X\nonumber
\end{equation}
the submatrix $L(X\vert X)$ has the form
\begin{equation}\label{PLl1}
L(X\vert X)=
\left[%
\begin{array}{ccccc}
  0 & L(x_1',x_1'') & \ldots & L(x_1',x_m') & L(x_1',x_m'') \\
  -L(x_1',x_1'') & 0 &  &L(x_1'',x_m') & L(x_1'',x_m'') \\
  \vdots & & &  &  \\
  -L(x_1',x_m') & -L(x_1'',x_m')  &  & 0 & L(x_m',x_m'') \\
  -L(x_1',x_m'') & -L(x_1'',x_m'') &  & -L(x_m',x_m'') & 0 \\
\end{array}%
\right] \nonumber
\end{equation}
\\
Denote by $\mbox{Pf}\;A$ the Pfaffian of an even dimensional
antisymmetric matrix $A$. The definition of $\PF\;A$ is given in
Appendix, Section \ref{ASectionDefinitionOFPfaffian}. Assume that
the matrix $L$ has the property
\begin{equation}
\PF\;L(X\vert X)\geq 0,\;\;\;\forall X\subset\X . \nonumber
\end{equation}
Let $J$ be $2\times 2$ block matrix of format $\X\times \X$ with
matrix elements
\begin{equation}
J(x,y)=\left\{%
\begin{array}{ll}
    \left[\begin{array}{cc}
      0 & 1 \\
      -1 & 0 \\
    \end{array}\right], & x=y; \\
    0, & \hbox{otherwise}. \\
\end{array}%
\right.
\end{equation}
\begin{defn}
A point process on $\X$ defined by
\begin{equation}
\mbox{Prob}_{L}(X)=\dfrac{\PF\;L(X\vert
X)}{\PF\;(J+L)},\;\;\;\forall X\subset\X
\end{equation}
is called the Pfaffian $L$-ensemble.
\end{defn}
The fact that $\sum\limits_{X\subset\X}\mbox{Prob}_{L}(X)=1$
follows from the expansion of $\PF\;(J+L)$ into a sum of Pfaffians
of the symmetric $2\times 2$ block submatrices $L(X\vert X)$ of
$L$, i. e.
\begin{equation}
\PF(J+L)=\sum\limits_{X\subset\X}\PF\;L(X\vert X).
\end{equation}

The striking property of the Pfaffian $L$-ensembles is that the
$m$-point correlation function $\varrho_m(Y)$ is given by a
Pfaffian,
\begin{equation}
\varrho_m(Y)=\PF\left[\K(y_i,y_j)\right]_{i,j=1}^m,\;\;Y=(y_1,\ldots
,y_m)
\end{equation}
Here the matrix $\K$ is defined in terms of $L$ by the expression
\begin{equation}
\K=J+(J+L)^{-1}
\end{equation}
the Pfaffian expression for $m$-point correlation functions
reflects the fact that the Pfaffian $L$-ensembles is a special
class of Pfaffian point processes. Different Pfaffian processes
were considered previously by Tracy and Widom \cite{tracy}, Rains
\cite{rains}, Olshanski \cite{olshanski}, Soshnikov
\cite{soshnikov}.
\subsection{Special matrices $L$}\label{PSectionSpecialMatrices}
Here we assume that sets $\X$, $X$ are ordered. Thus if $X=(x_1,
x_2,\ldots )$ then $x_1<x_2<\ldots$. For any $x\in\X$ we denote by
$^lx$ the immediate left neighbor of $X$ and by $^rx$ the
immediate right neighbor of $x$. For example, if $\X=\mathbb{Z}$
then $^lx=x-1$ and $^rx=x+1$.

Given a fixed splitting of $\X$ into positive and negative parts,
$\X=\X_-\sqcup\X_+$, we denote by $\frak{x}$ the minimal (left
most) element  of $\X_+$. We introduce the \textit{parity} on the
sets $\X_{\pm}$, referring to the minimal elements of these sets
as \textit{odd} elements.

According to the decomposition of the set $\X$ , $\X=\X_-\sqcup
\frak{x}\sqcup\X_+\setminus \frak{x}$, we write the matrix $L$ in
the block form:
\begin{equation}\label{PstructureL}
L=\left[\begin{array}{ccc}
  L_{--} & L_{-0} & L_{-+} \\
  L_{0-} & L_{00} & L_{0+} \\
  L_{+-}& L_{+0} & L_{++} \\
\end{array}\right]
\end{equation}
We are interested in the matrices $L$ defined by
\begin{equation}\label{PmatrixL}
L=\left[\begin{array}{ccc}
  \mbox{E} & \mbox{A} & \mbox{B} \\
  -\mbox{A}^T & 0 & 0 \\
  -\mbox{B}^T & 0 & 0 \\
\end{array}\right]
\end{equation}
As usual, here E, A, B are the matrices with $2\times 2$ block
elements. Specifically,
\begin{equation}\label{PmatrixE}
\mbox{E}(x,y)=\left[\begin{array}{cc}
  \epsilon(x,y) & 0 \\
  0 & 0 \\
\end{array}\right],\;\;x,y\in\;\X_-
\end{equation}
\begin{equation}\label{PmatrixA}
\mbox{A}(x,y)=\left[\begin{array}{cc}
  \epsilon(x,y) & 0 \\
  0 & \dfrac{h(x)h(y)}{x-y} \\
\end{array}\right],\;\;x\in\;\X_-,\;\; y=\frak{x}.
\end{equation}
\begin{equation}\label{PmatrixB}
\mbox{B}(x,y)=\left[\begin{array}{cc}
  0 & 0 \\
 \dfrac{h(x)h(y)}{x-y}  & \dfrac{h(x)h({}^ly)}{x- {{}^ly}} \\
\end{array}\right],\;\;x\in\;\X_-, \;\;y\in\;\X_+\setminus\frak{x}.
\end{equation}
The two-point function $\epsilon(x,y)$ in equations above is
antisymmetric, $\epsilon(x,y)=-\epsilon(y,x)$. When $x<y$,
\begin{equation}\label{Pepsilon}
\epsilon(x,y)=\left\{%
\begin{array}{ll}
    1, & x-\hbox{odd}, y-\hbox{even}\\
    0, & \hbox{otherwise.} \\
\end{array}%
\right.
\end{equation}
The function $h$ is nonnegative on $\X$.

Configurations $X\in\X$ can be divided into two classes. The first
class consists of configuration which do not include the point
$\frak{x}$. Such configurations have the form $X=X_-\sqcup X_+$,
$X_+=(x_1^+,x_2^+,\ldots )$, $x_1^+>\frak{x}$. The second class
consists of configurations including the point $\frak{x}$. For
such configuration $X_+=(\frak{x},x_1^+,x_2^+,\ldots )$. For any
$X\subset\X$ denote by $\tilde{X}$ the configuration defined by
\begin{equation}
\TX=\TX_-\sqcup\TX_+ \nonumber
\end{equation}
\begin{equation}
\TX_-=X_-\nonumber
\end{equation}
\begin{equation}
\TX_+=\left\{%
\begin{array}{ll}
    ({^lx}_1^+,x_1^+,{^lx}_2^+,x_2^+,\ldots ), &\;\;\; X_+\cap\frak{x}=\emptyset \\
     (\frak{x},{^lx}_1^+,x_1^+,{^lx}_2^+,x_2^+,\ldots ), &\;\;\; X_+\cap\frak{x}\neq\emptyset\\
\end{array}%
\right. \nonumber
\end{equation}
\begin{defn}
We say that $X\in Conf^{L}(\X)$ if
\begin{itemize}
\item $X_+=(x_1^+<x_2^+<\ldots )$
    \item all points of $\TX_+$ are different
    \item $\vert \TX_-\vert=\vert \TX_+\vert$
    \item $\TX_-=(x_1^-<x_2^-<\ldots )$, where $x_i^-$ has the same
    parity as $i$.
\end{itemize}
\end{defn}
This definition is justified by the following statement.
\begin{thm}\label{PTheoremReductionToPfaffianProcesses} With $L$ given by equations
(\ref{PstructureL})-(\ref{Pepsilon}) we have
\begin{equation}\label{PReductionToPfaffianProcessesEquation}
\mbox{Prob}_{L}(X)=\dfrac{1}{\PFAFF\;(J+L)}\;\dfrac{V(\TX_-)V(\TX_+)}{\prod(\TX_+;\TX_-)}\;h(\TX)
\end{equation}
for $X\in Conf^{L}(\X)$ and $0$ for all other $X\in \X$.
\end{thm}
\begin{proof}
The positive integer $d=\vert\TX_-\vert=\vert\TX_+\vert$ can be
even or odd, depending on whether $X$ includes the point
$\frak{x}$ or not. According to that consider two cases.\\
\textbf{Case 1}. $X\cap \frak x=\emptyset$\\
Given copies $X'$, $X''$ of $X\in\X$ in $\X'$, $\X''$ we denote by
$X'\uplus X''$ the set $(x_1', x_1'', x_2', x_2'',\ldots )$. Then
we have
\begin{equation}
\begin{split}
\PF\;L(X\vert X)&=\PF\;L\biggl[X_-\sqcup X_+\arrowvert X_-\sqcup
X_+\biggr]\\
&=\PF\;L\biggl[\left(X_-'\uplus X_{-}''\right)\sqcup
\left(X_{+}'\uplus X_{+}''\right)\arrowvert \left(X_-'\uplus
X_{-}''\right)\sqcup \left(X_{+}'\uplus X_{+}''\right)\biggr]\\
&=(-)^{\frac{d(d-1)}{2}}\cdot \PF\;L\biggl[X_-'\arrowvert X_-'\biggr]\\
&\times \PF\;L\biggl[ X_{-}''\sqcup \left(X_{+}'\uplus
X_{+}''\right)\arrowvert  X_{-}''\sqcup \left(X_{+}'\uplus
X_{+}''\right)\biggr]
\end{split}
\nonumber
\end{equation}
as the function $L(x,y)=0$ for any $x\in X_-'$ and any $y$ which
does not belong to $X_-'$ (see equations (\ref{PLl})-(\ref{PLl1})
and (\ref{PmatrixL})-(\ref{Pepsilon})). We note that $L(x,y)=0$,
if $x,y\in\;\X_-''$, or if $x,y\in X_+'\uplus X_+''$ Therefore
$\vert X_-''\vert =\vert X_+'\arrowvert +\vert X_+''\vert $, or
$\vert X_-\vert=2\vert X_+\vert $, which means that
$\vert\TX_+\vert =\vert \TX_-\vert $.

Consider $\PF\;L\biggl[X_-'\vert X_-'\biggr]$. Note that the
matrix $L\biggl[X_-'\vert X_-'\biggr]$ is even dimensional, if
$\vert X_- \vert=2\vert X_+\vert$. Moreover the matrix
$L\biggl[X_-'\vert X_-'\biggr]$ is the matrix whose $(i,j)$ entry
is, by definition, given by $\epsilon (x_i^-,x_j^-)$. Clearly, if
$x_1^-$ is even, the first row of this matrix consists of only
zeros. Thus, if $\PF\;L\biggl[X_-'\vert X_-'\biggr]\neq 0$,
$x_1^-$ must be odd. Now assume that $x_{2i-1}^-$ and $x_{2i}^-$
have the same parity. In this case $(2i-1)^{\mbox{st}}$ and
$2i^{\mbox{th}}$ rows of the matrix $L\biggl[X_-'\vert
X_-'\biggr]$ are equal to each other. Therefore, if
$\PF\;L\biggl[X_-'\vert X_-'\biggr]\neq 0$ the elements of the set
$\TX_-=(x_1^-,x_2^-,\ldots )$ are such that $x_1^-$ is odd,
$x_2^-$ is even, $x_3^-$ is odd and so on. This proves the
condition on the parity for the configurations in $Conf^{L}(\X)$.
Moreover, using the definition of Pfaffian it is not hard to
conclude that $\PF\;L\biggl[X_-'\vert X_-'\biggr]=1$ for the
configurations with non-zero probabilities.

Since $\vert X_-''\vert =\vert X_+'\vert +\vert X_+''\vert $ the
matrix $L\biggl[ X_{-}''\sqcup \left(X_{+}'\uplus
X_{+}''\right)\arrowvert  X_{-}''\sqcup \left(X_{+}'\uplus
X_{+}''\right)\biggr]$ has the block structure:
\begin{equation}
\biggl[\begin{array}{cc}
  \mathbb{O}_{d\times d} & Q_{d\times d} \\
  -Q^T_{d\times d} & \mathbb{O}_{d\times d} \\
\end{array}\biggr]
\nonumber
\end{equation}
with
 \begin{equation}
 Q_{d\times d}=\left[\begin{array}{ccccc}
  \dfrac{h(x_1^-)h(x_1^+)}{x_1^--x_1^+} & \dfrac{h(x_1^-)h({^lx}_1^+)}{x_1^--{^lx}_1^+} &
   \ldots & \dfrac{h(x_1^-)h(x_{d/2}^+)}{x_1^--x_{d/2}^+} & \dfrac{h(x_1^-)h({^lx}_{d/2}^+)}{x_1^--{^lx}_{d/2}^+} \\
  \vdots &  & & &  \\
  \dfrac{h(x_d^-)h(x_1^+)}{x_d^--x_1^+} & \dfrac{h(d_1^-)h({^lx}_1^+)}{x_d^--{^lx}_1^+} &
   \ldots & \dfrac{h(x_d^-)h(x_{d/2}^+)}{x_d^--x_{d/2}^+} &  \dfrac{h(x_d^-)h({^lx}_{d/2}^+)}{x_d^--{^lx}_{d/2}^+}\\
\end{array}\right]\nonumber
\end{equation}
($d$ is even). Thus we have
\begin{equation}
\begin{split}
\PF\;L(X\vert X)&=(-)^{\frac{d(d-1)}{2}}
\PF\;\left[\begin{array}{cc}
  \mathbb{O}_{d\times d} & Q_{d\times d} \\
  -Q^T_{d\times d} & \mathbb{O}_{d\times d} \\
\end{array}\right]= \mbox{det}\;Q_{d\times
d}\\
&=(-)^{\frac{d}{2}}(-)^{\frac{d(d-1)}{2}}\;\dfrac{V(\TX_-)V(\TX_+)}{\prod(\TX_+;\TX_-)}\;h(\TX)
\end{split}\nonumber
\end{equation}
where we have used the formula for the Cauchy determinant, see
Appendix. Noting that
$(-)^{\frac{d(d-1)}{2}+\frac{d}{2}}=(-)^{\frac{d^2}{2}}=1$ (as
$d$ is even) we obtain the formula stated in the Theorem.\\
\textbf{Case 2.} $X\cap \frak{x}\neq 0$\\
The proof is very similar. We observe that any configuration $X$
has a form
\begin{equation}
X=X_-\sqcup\frak{x}\sqcup X_+ \nonumber
\end{equation}
Then
\begin{equation}
\begin{split}
&\PF\;L(X\vert X)=\PF\;L\biggl[X_-\sqcup\frak{x}\sqcup X_+\vert
X_-\sqcup\frak{x}\sqcup
X_+\biggr]=\\
&\PF\;L\biggl[\left(X_-'\uplus X_{-}''\right)\sqcup
(\frak{x}',\frak{x}'')\sqcup \left(X_{+}'\uplus
X_{+}''\right)\vert \left(X_-'\uplus
X_{-}''\right)\sqcup (\frak{x}',\frak{x}'')\sqcup\left(X_{+}'\uplus X_{+}''\right)\biggr]\\
&=(-)^{\frac{d(d-1)}{2}}\cdot \PF\;L\biggl[X_-',\frak{x}'\vert X_-',\frak{x}'\biggr]\\
&\times \PF\;L\biggl[ X_{-}''\sqcup \frak{x}''\sqcup
\left(X_{+}'\uplus X_{+}''\right)\vert  X_{-}''\sqcup
\frak{x}''\sqcup\left(X_{+}'\uplus X_{+}''\right)\biggr]
\end{split}
\nonumber
\end{equation}
Clearly, $\vert X_-''\vert=\vert\frak{x}'\sqcup\left(X_{+}'\uplus
X_{+}''\right)\vert$, otherwise $\PF\;L(X\vert X)=0$. Thus $\TX_-$
consists of odd number of elements, and
$\vert\TX_-\vert=\vert\TX_+\vert=d$, $d$ is odd, and we repeat the
same computations as in the previous case.
\end{proof}
\subsection{Pfaffian $ \hat{L} $ -ensembles}\label{PSectionHatL}
Given $\X=\X_-\sqcup\X_+$ consider a different splitting of $\X$,
$\X=\hat\X_-\sqcup\hat\X_+$. Here $\hat\X_+=\X_+\sqcup\X_0$,
$\hat\X_-=\X_-\setminus \X_0$, and the set $\X_0$ is a subset of
$\X_-$. Assume that $\X_0$ consists of $2S$ rightmost points of
$\X_-$.

Define a new matrix $\hat{L}$ on $\X$ in such a way that it has
the same structure with respect to the decomposition
$\X=\hat\X_-\sqcup\hat{\frak{x}}\sqcup\hat\X_+\setminus\hat{\frak{x}}$
as the matrix $L$ with respect to the decomposition
$\X=\X_-\sqcup\frak{x}\sqcup\X_+\setminus\frak{x}$. (Here
$\hat{\frak{x}}$ stands for the minimal (left most) element of
$\hat\X_+$). Then $\hat{L}$ is given by
\begin{equation}\label{PmatrixhatL}
\hat L=\left[\begin{array}{ccc}
  \hat{\mbox{E}} & \hat{\mbox{A}} & \hat{\mbox{B}} \\
  -\hat{\mbox{A}}^T & 0 & 0 \\
  -\hat{\mbox{B}}^T & 0 & 0 \\
\end{array}\right]
\end{equation}
Here $\hat{\mbox{E}}$, $\hat{\mbox{A}}$, $\hat{\mbox{B}}$ are the
matrices with $2\times 2$ block elements. Specifically,
\begin{equation}\label{PmatrixhatE}
\hat{\mbox{E}}(x,y)=\left[\begin{array}{cc}
  \epsilon(x,y) & 0 \\
  0 & 0 \\
\end{array}\right],\;\;x,y\in\;\hat{\X}_-
\end{equation}
\begin{equation}\label{PmatrixhatA}
\hat{\mbox{A}}(x,y)=\left[\begin{array}{cc}
  \epsilon(x,y) & 0 \\
  0 & \dfrac{\hat{h}(x)\hat{h}(y)}{x-y} \\
\end{array}\right],\;\;x\in\;\hat\X_-,\;\; y=\hat{\frak{x}}.
\end{equation}
\begin{equation}\label{PmatrixhatB}
\hat{\mbox{B}}(x,y)=\left[\begin{array}{cc}
  0 & 0 \\
 \dfrac{\hat{h}(x)\hat{h}(y)}{x-y}  & \dfrac{\hat{h}(x)\hat{h}(^ly)}{x- {^ly}} \\
\end{array}\right],\;\;x\in\;\hat\X_-, \;\;y\in\;\hat\X_+\setminus\frak{\hat x}.
\end{equation}
If $\hat{h}$ is nonnegative on $\X$ the matrix $\hat{L}$ defines a
Pfaffian ensemble on $\X$ which we call $\hat{L}$-ensemble.

We introduce the set of point configurations $Conf^{\hat{L}}(\X)$
in the same way as the set $Conf^{L}(\X)$ for the Pfaffian
$L$-ensemble was introduced. Namely, for any $Z\in \X$ denote by
$\tilde{Z}$ the configuration defined by
\begin{equation}
\TZ=\TZ_-\sqcup\TZ_+ \nonumber
\end{equation}
\begin{equation}
\TZ_-=Z_-\nonumber
\end{equation}
\begin{equation}
\TZ_+=\left\{%
\begin{array}{ll}
    ({^lz}_1^+,z_1^+,{^lz}_2^+,z_2^+,\ldots ), &\;\;\; Z_+\cap\hat{\frak{x}}=\emptyset \\
     (\hat{\frak{x}},{^lz}_1^+,z_1^+,{^lz}_2^+,z_2^+,\ldots ), &\;\;\; Z_+\cap\hat{\frak{x}}\neq\emptyset\\
\end{array}%
\right. \nonumber
\end{equation}
Here $Z_{\pm}=Z\cap\hat{\X}_{\pm}$.
\begin{defn}
We say that $Z\in Conf^{\hat{L}}(\X)$ if
\begin{itemize}
    \item all points of $\TZ_+$ are different
    \item $\vert \TZ_-\vert=\vert \TZ_+\vert$
    \item $\TZ_-=(z_1^-<z_2^-<\ldots )$, where $z_i^-$ has the same
    parity as $i$.
\end{itemize}
\end{defn}
Theorem \ref{PTheoremReductionToPfaffianProcesses} says that with
$\hat{L}$ given by equations
(\ref{PmatrixhatL})-(\ref{PmatrixhatB}) we obtain
\begin{equation}
\mbox{Prob}_{\hat{L}}(Z)=\dfrac{1}{\PF\;(J+\hat{L})}\;\dfrac{V(\TZ_-)V(\TZ_+)}{\prod(\TZ_+;\TZ_-)}\;\hat{h}(\TZ)
\end{equation}
for $Z\in Conf^{\hat{L}}(\X)$ and $0$ for all other $Z\subset\X$.
Given $Z\in Conf^{\hat{L}}(\X)$ we build from $\TZ$ a
configuration $\TX$ by the particle-hole involution on $\X_0$ (see
Section \ref{PointConfigurations} for the definition). We note
that the corresponding configuration $X$ does not belong to $
Conf^{L}(\X)$, since the configuration $\TX$ is unbalanced with
respect to the splitting $\X=\X_-\sqcup\X_+$. Moreover,
\begin{itemize}
    \item $\vert\TX_-\vert-\vert\TX_+\vert=2S$
\item $\TX_+=({^lx}_1^+,x_1^+,{^lx}_2^+,x_2^+,\ldots ), \;\;\;
X_+\cap\frak{x}=\emptyset$ \item $
\TX_+=(\frak{x},{^lx}_1^+,x_1^+,{^lx}_2^+,x_2^+,\ldots ), \;\;\;
X_+\cap\frak{x}\neq\emptyset$ \item $\TX_-=(x_1^-,x_2^-,\ldots )$,
$x_i^-$ has the same parity as $i$,
\end{itemize}
and all points of $\TX_{\pm}$ are different.

Define the weight $\hat h$ of the $\hat{L}$-ensemble in terms of
the weight $h$ of the $L$-ensemble by the formula:
\begin{equation}\label{PNewWeightHath}
\hat{h}(z)=
\left\{%
\begin{array}{ll}
    h(z)\prod\limits_{y\in\;\frak{X}_0}\vert z-y\vert, & z\in \frak{X}_-\setminus\frak{X}_0, \\
     & \\
    \dfrac{1}{h(z)\prod_{y\in\;\frak{X}_0,y\neq z}\vert z-y\vert}, &  z\in \frak{X}_0, \\
     & \\
    \dfrac{h(z)}{\prod_{y\in\;\frak{X}_0}\vert z-y\vert}, & z\in \frak{X}_+. \\
\end{array}%
\right.
\end{equation}
Note that this formula is identical to (\ref{NewWeighth}) except
for the absolute values. We did not need absolute values in
(\ref{NewWeighth}) because there the formulas only contain $h^2$
and not $h$.
\begin{prop}\label{PfaffianPropositiononProbZ}
With $\hat{h}(z)$ defined by equation (\ref{PNewWeightHath}) and
$\TX$ constructed from $\TZ$ by the particle-hole involution on
$\X_0$,
\begin{equation}
\begin{split}
\Prob_{\hat{L}}(Z)&=\dfrac{1}{\PFAFF\;(J+\hat{L})}\;
\dfrac{h(\TX)}{\vert V(\X_0)\vert h(\X_0)} \;\frac{\vert
V(\TX_-)\vert\vert V(\TX_+)\vert}{\prod(\TX_+,\TX_-)},\;\;Z\in
Conf^{\hat{L}}(\X)
\end{split}
\end{equation}
\end{prop}
\begin{proof}
The probability of the configuration $Z$ can be rewritten in terms
of $X$ by the same method as it was done for the case of
determinantal ensembles, see Proposition
\ref{PropositionProbabilityHatL}.
\end{proof}
\subsection{Correlation functions of Pfaffian $L$-ensembles and
averages of characteristic polynomials} The goal of this Section
is to express the kernel $\K$ of the $m$-point correlation
function $\varrho_m(Y)$ in terms of averages of characteristic
polynomials associated with Pfaffian $L$-ensembles. Namely we want
to prove the Pfaffian analog of Proposition \ref{LEMMAOKMINORAX}.

Introduce nonintersecting sets $\alpha^{\pm}$ of complex numbers
with nonequal elements,
\begin{equation}
\alpha_+=\left(\alpha_1^+,\ldots
,\alpha^+_k),\;\;\alpha_-=(\alpha_1^-,\ldots ,\alpha_m^-\right).
\end{equation}
Assume that $\alpha^{\pm}\cap\X=0$, and
\begin{equation}
k-m=\arrowvert\alpha_+\arrowvert-\arrowvert\alpha_-\arrowvert=2S,\;\;\;S\in
\mathbb{Z}.
\end{equation}
Similarly to the case of the determinantal $L$-ensembles we extend
the definition of the matrices $\K,L$ as follows. Let $\alpha'$,
$\alpha''$ denote two copies of the set
$\alpha=\alpha_-\sqcup\alpha_+$. We add to $L$ rows and columns
parameterized by $\alpha',\alpha''$, and then define new matrix
elements of $L$ in accordance with equations
(\ref{PmatrixL})-(\ref{PmatrixB}), where we assume that $\alpha_-$
is added to $\X_-$, and $\alpha_+$ is added to $\X_+$. Then we
consider the  matrix
\begin{equation}
\K(\alpha\sqcup\X\arrowvert\alpha\sqcup\X)=J+(J+L(\alpha\sqcup\X\arrowvert\alpha\sqcup\X))^{-1}.
\end{equation}
\begin{prop}\label{PPROPOSITION}
The Pfaffian of the symmetric $(k+m)\times (k+m)$  submatrix
$\K(\alpha'\arrowvert\alpha')$ of the matrix
$\K\biggl(\alpha'\uplus\alpha''\sqcup\X'\uplus\X''\arrowvert\alpha'\uplus\alpha''\sqcup\X'\uplus\X''\biggr)
$ can be given as a normalized average of the functions
$E(\cdot,\cdot)$ defined by equations
(\ref{LCharacteristicPolynomials}) with respect to the Pfaffian
$\hat{L}$-ensemble. Namely,
\begin{equation}\label{PEquationPfaffiansofKMinors}
\begin{split}
\PFAFF\;\left[ K(\alpha'\arrowvert\alpha')\right]& =
h(\alpha)\dfrac{V(\alpha_{-})V(\alpha_+)}{\prod(\alpha_-;\alpha_+)}
\left[\dfrac{\PFAFF\;(J+\hat{L})}{\PFAFF\;(J+L)}\;h(\frak{X}_0)\arrowvert V(\frak{X}_0)\arrowvert\right]\\
&\times\left[\dfrac{\prod(\alpha_+;\frak{X}_0)}{\prod(\alpha^-;\frak{X}_0)}\right]\left\langle
\dfrac{E(\alpha^+,Z)}{E(\alpha^-,Z)}\right\rangle_{\hat{L}}
\end{split}
\end{equation}
\end{prop}
\begin{proof}
Combining equations (\ref{AppendixPffafianJA}) and
(\ref{AppendixPfaffianOfSubmatrixOfTheInverse}) of Appendix we
obtain the following  formula:
\begin{equation}\label{PfaffianKExpansion}
\PF\;\K[\alpha'\arrowvert\alpha']=\dfrac{1}{\PF\;(J+L)}\sum\limits_{X\in\;\X}\PF\;L\biggl[\alpha''\sqcup
X'\uplus X''\arrowvert \alpha''\sqcup X'\uplus X''\biggr]
\end{equation}
Here $\alpha $ is a set with even $\vert\alpha\vert=k+m$, and
$\alpha', \alpha''$ are two copies of $\alpha$. Now we compute the
Pfaffian in the sum using similar arguments as in the proof of
Theorem \ref{PTheoremReductionToPfaffianProcesses}.

Assume first that $X\cap\frak{x}=0$. We denote $\L(\beta)\equiv
L(\beta\arrowvert\beta)$ for any set $\beta$. Then the Pfaffian in
the sum is
\begin{equation}
\begin{split}
\PF\;\L\biggl[&\alpha''\sqcup X'\uplus X''\biggr]\\
&=\PF\;\L\biggl[\alpha''_-\sqcup\alpha''_+\sqcup X'_-\uplus
X''_-\sqcup
X'_+\uplus X''_+\biggr]\\
&=\PF\;\L\biggl[\alpha''_-\sqcup X'_-\uplus X''_-\sqcup
X'_+\uplus X''_+\sqcup\alpha_+''\biggr]\\
&=(-)^{\frac{|X_-|(|X_-|-1)}{2}}(-)^{|\alpha_-||X_-|}
\PF\;\L\biggl[ X'_-\sqcup\alpha''_-\sqcup X''_-\sqcup
X'_+\uplus X''_+\sqcup\alpha_+''\biggr]\\
&=(-)^{\frac{|X_-|(|X_-|-1)}{2}}(-)^{|\alpha_-||X_-|}
\PF\;\L\biggl[X'_-\biggr] \PF\;\L\biggl[\alpha''_-\sqcup
X''_-\sqcup
X'_+\uplus X''_+\sqcup\alpha_+''\biggr]\\
\end{split}
\nonumber
\end{equation}
By equations (\ref{PmatrixL})-(\ref{PmatrixB}) $L(*_-,*_+)=0$.
Thus, the last expression does not equal to zero only if
$\arrowvert\alpha_-\arrowvert+ \arrowvert
X_-\arrowvert=\arrowvert\alpha_+\arrowvert+2\arrowvert
X_+\arrowvert$, and $X_-=(x_1^-,x_2^-,\ldots )$ is such that
$x_i^-$ has the same parity as $i$. Since
$|\alpha_+|-|\alpha_-|=2S$, $X_-$ consists of even number of
elements. We note that $\PF\;L(X_-')=1$. Then
\begin{equation}
\begin{split}
\PF\;\L\biggl[\alpha''\sqcup X'\uplus X''\biggr]&=(-)^{\frac{|X_-|(|X_-|-1)}{2}}(-)^{\frac{|\TX_+|}{2}}\\
&\times \dfrac{V(\alpha_-\sqcup \TX_-)V(\TX_+\sqcup
\alpha_+)}{\prod(\alpha_-\sqcup \TX_-;\TX_+\sqcup
\alpha_+)}\\
&\times h(\TX)h(\alpha)\\
&=(-)^{\frac{|X_-|(|X_-|-1)}{2}}(-)^{\frac{|\TX_+|}{2}}\\
&\times\biggl[h(\alpha)\dfrac{V(\alpha_{-})V(\alpha_+)}{\prod(\alpha_-;\alpha_+)}\biggr]\\
&\times\left[\dfrac{E(\alpha_+,\TX)}{E(\alpha_-,\TX)}\right]\dfrac{V(\TX_-)V(\TX_+)}{\prod(\TX_+;\TX_-)}\;h(\TX)
\end{split}
\nonumber
\end{equation}
It remains to see that the sign cancels out. We know that
$\arrowvert
\TX_-\arrowvert-\arrowvert\TX_+\arrowvert=|\alpha_+|-|\alpha_-|=2S$
and $|\TX_-|$ is even. Then
\begin{equation}
\begin{split}
&(-)^{\frac{|X_-|(|X_-|-1)}{2}}(-)^{\frac{|\TX_+|}{2}}=(-)^S\\
&\dfrac{ V(\TX_-)
V(\TX_+)}{\prod(\TX_+;\TX_-)}=(-)^S\dfrac{\arrowvert
V(\TX_-)\arrowvert\arrowvert
V(\TX_+)\arrowvert}{\prod(\TX_+;\TX_-)}
\end{split}
\nonumber
\end{equation}
and the result is
\begin{equation}
\begin{split}
\PF\;\L\biggl[\alpha''\sqcup X'\uplus & X''\biggr]= h(\alpha)\dfrac{V(\alpha_{-})V(\alpha_+)}{\prod(\alpha_-;\alpha_+)}\\
&\times\;\left[\dfrac{E(\alpha_+,\TX)}{E(\alpha_-,\TX)}\right]\;\dfrac
{\arrowvert V(\TX_-)\arrowvert\arrowvert
V(\TX_+)\arrowvert}{\prod(\TX_+;\TX_-)}\;h(\TX).
\end{split}\nonumber
\end{equation}
Now we apply Proposition \ref{PfaffianPropositiononProbZ} and
obtain formula (\ref{PEquationPfaffiansofKMinors}). The case
$X\cap\frak{x}\neq 0$ is considered in the same way.
\end{proof}
\subsection{Discrete symplectic ensemble}\label{PSectionDSPE}
Given $\X$ denote by $Conf_{2N}^{(4)}(\X)$ the following set of
point configurations:
\begin{equation}
Conf_{2N}^{(4)}(\X)=\biggl\{X\subset \X\arrowvert
X=(^lx_1<x_1<\ldots <{}^lx_N<x_N)\biggr\}\nonumber
\end{equation}
Assume that a nonnegative function $f$ is given on $\X$, which
does not vanish at least at $2N$ distinct points.
\begin{defn}
The point process which lives on $Conf_{2N}^{(4)}(\X)$ and for
which the probability of a configuration $X$ is given by
\begin{equation}\label{PDEFINITIONofSymplecticEnsembleEquation}
\mbox{Prob}(X)=\left[c_N^{(4)}\right]^{-1}\;\prod\limits_{i=1}^Nf(x_i)\;\arrowvert
V(X)\arrowvert,\;\;\; X\in\;Conf_{2N}^{(4)}(\X)
\end{equation}
will be called $2N$ point \textit{discrete symplectic ensemble}
and will be denoted by $\T_{2N}^{(4)}(f)$. Here $c_N^{(4)}$ is a
normalization constant.
\end{defn}
\begin{rem}
Rewrite $\arrowvert V(X)\arrowvert$ as follows
\begin{equation}
\arrowvert
V(X)\arrowvert=\biggl\arrowvert\prod\limits_{i<j}(x_i-x_j)(^lx_i-^lx_j)(^lx_i-x_j)(x_i-^lx_j)\prod\limits_{i=1}^N({}^lx_i-x_i)\biggr\arrowvert
.
\end{equation}
In the continuous limit the points $^lx$ and $x$ get closer and
closer to each other, and all differences ${}^lx_i-x_i$ turns into
the same small constant, say $\epsilon$. Thus the degree of the
Vandermonde determinant turns into four, and the probability
distribution (\ref{PDEFINITIONofSymplecticEnsembleEquation}) takes
the same form as the probability distribution of eigenvalues for
the symplectic ensemble of the Random Matrix Theory (see, for
example, Mehta \cite{mehta}, Chapter 3).
\end{rem}
For a symmetric function $g(X)=g(^lx_1,x_1,\ldots ,^lx_N,x_N)$ of
points of the configuration $X\in\;Conf_{2N}^{(4)}(\X)$, the
average with respect to the discrete symplectic ensemble is
defined by
\begin{equation}
\langle
g\rangle_{\T_{2N}^{(4)}(f)}=\left[c_N^{(4)}\right]^{-1}\sum\limits_{X\in\;Conf_{2N}^{(4)}(\X)}
g(X)\arrowvert V(X)\arrowvert\;f(X)
\end{equation}
If $\frak X$ is a lattice then in the continuous limit  we
introduce the skew symmetric inner product for arbitrary functions
$g_1$, $g_2$ on $\X$
\begin{equation}\label{PDefinitionSkewSymmetricInnerProductSymplectic}
\left\langle g_1,
g_2\right\rangle=\sum\limits_{x\in\;\dot{\X}}\biggl(g_1(^lx)g_2(x)-g_2(^lx)g_1(x)\biggr)f({}^lx)f(x)
\end{equation}
where $\dot{\X}$ denotes $\X$ without the leftmost point. If
\begin{equation}
\mbox{det}\;||\langle x^i, x^j\rangle||_{i,j=1}^{2n}\neq 0
\end{equation}
then a family of monic skew orthogonal polynomials associated with
the discrete symplectic ensemble can be constructed.
\begin{defn}\label{PDefinitionOfSkewsymmetricOrthogonalPolynomials}
For $i=0,1,2,\ldots$ let $p_{2i}$, $p_{2i+1}$ be monic polynomials
of the degrees $2i$ and $2i+1$, which satisfy the conditions
\begin{itemize}
    \item $\langle p_{2i}, p_{2j}\rangle =0$, $\langle p_{2i+1}, p_{2j+1}\rangle =0$
\item $\langle p_{2i}, p_{2j+1}\rangle
=h_i\delta_{ij}$.\end{itemize} The family $\{p_{2i}, p_{2i+1}\}$
will be called the family of the skew orthogonal polynomials.
\end{defn}
\begin{rem}
The skew orthogonal polynomials are defined up to the replacement
\begin{equation}
p_{2j+1}\rightarrow p_{2j+1}+\mbox{const}\;p_{2j}. \nonumber
\end{equation}
\end{rem}
\begin{lem}For arbitrary functions $\phi_i(x)$, $i=1,\ldots ,2N$ and an antisymmetric two
point function $\epsilon(x,y)$ the following identity is valid
\begin{equation}
\sum\limits_{x_1,\ldots
,x_{2N}\in\;\X}\PFAFF\biggl[\epsilon(x_i,x_j)
\biggr]_{i,j=1}^{2N}\det\biggl[\phi_i(x_j)\biggr]_{i,j=1}^{2N}=(2N)!\;\PFAFF\biggl[\left\langle\phi_i,\phi_j\right\rangle\biggr]_{i,j=1}^{2N}
\end{equation}
where
\begin{equation}
\left\langle\phi_i,\phi_j\right\rangle=\sum\limits_{x,y\in\;\X}\epsilon(x,y)\phi_i(x)\phi_j(y)
\end{equation}
\end{lem}
This is a well known de Bruijn identity, see e.g. de Bruijn
\cite{bruijn}, Tracy and Widom \cite{tracy}, Baik and Rains
\cite{baik1}, Rains \cite{rains}. This statement readily implies,
(see Rains \cite{rains}, Corollary 1.3):
\begin{cor}\label{PLEMMAAboutPFFafianA}
Let $\phi_1,\ldots ,\phi_{2N}$; $\psi_1,\ldots ,\psi_{2N}$ be
arbitrary functions on $\X$. We set
\begin{equation}
\underline{\phi}(.)=\left[\begin{array}{c}
  \phi_1(.) \\
  \vdots \\
  \phi_{2N}(.)\\
\end{array}\right],
\;\;\;\underline{\psi}(.)=\left[\begin{array}{c}
  \psi_1(.) \\
  \vdots \\
  \psi_{2N}(.)\\
\end{array}\right],
\nonumber
\end{equation}
and introduce the matrix $A=\biggl[ A_{ij}\biggr]_{i,j=1}^{2N}$:
\begin{equation}
A_{ij}=\sum\limits_{x\in\;\X}\phi_i(x)\psi_j(x)-\phi_j(x)\psi_i(x).
\end{equation}
Then
\begin{equation}
\sum_{\substack{x_1<\ldots <x_N\\x_i\in\;\X,\;
i=\overline{1,N}}}\det\;\biggl[\underline{\phi}(x_1),\underline{\psi}(x_1),\ldots
,\underline{\phi}(x_N),\underline{\psi}(x_N)\biggr]=\PFAFF\;A
\end{equation}
\end{cor}
The following statement is well known as well, and we give a proof
for the reader's convenience.
\begin{prop}
The normalization constant $c_N^{(4)}$ is equal to the product of
$\langle p_{2i},p_{2i+1}\rangle$, i.e.
\begin{equation}
c_{N}^{(4)}=\prod\limits_{i=0}^{N-1}h_i,\;\;\;h_i=\langle
p_{2i},p_{2i+1}\rangle
\end{equation}
\end{prop}
\begin{proof}
 Set
\begin{equation}
\pi_{i-1}(x)=x^{i-1}+\ldots ,\; i=1,\ldots ,2N.\nonumber
\end{equation}
This gives a system of polynomials of degrees $0,\ldots ,2N-1$
with the highest coefficients equal to one. The constant
$c_{N}^{(4)}$ is the average over point configurations which
includes the absolute value of the Vandermonde determinant and the
product of weights. Rewrite the absolute value of the Vandermonde
determinant in terms of the polynomials $\pi_j$. In particular
these polynomials can be chosen to be the skew symmetric
orthogonal polynomials with respect to the weight $f$ in the
average. Then we obtain
\begin{equation}
c_N^{(4)}=\sum_{\substack{x_1<\ldots <x_N\\x_i\in\;\dot{\X},\;
i=\overline{1,N}}}\mbox{det}\;\biggl[\underline{\phi}(x_1),\underline{\psi}(x_1),\ldots\;
,\underline{\phi}(x_N),\underline{\psi}(x_N)\biggr]
\end{equation}
where we have introduced
\begin{equation}\label{PPHIPSI}
\phi_i(x)=p_{i-1}(^lx)f({}^lx),\;\;\psi_i(x)=p_{i-1}(x)f(x),\;\;
i=\overline{1,2N}
\end{equation}
Now we apply Corollary \ref{PLEMMAAboutPFFafianA} and prove the
Proposition.
\end{proof}
Given $X\in Conf_{2N}^{(4)}(\X)$ and a complex parameter $\zeta$
we define the characteristic polynomial
\begin{equation}
d(\zeta)=\prod(\zeta;X).
\end{equation}
The following analog of Heine's identity is also well known, see
e. g. Eynard \cite{eynard}, Forrester \cite{forrester}.
\begin{prop}
\begin{equation}
p_{2N}(\zeta)=\langle d(\zeta)\rangle_{\T_{2N}^{(4)}(f)}
\end{equation}
\end{prop}
\begin{proof}
The average is  a sum over the variables $x_1,\ldots ,x_N$ which
take values in $\dot{\X}$. Set
\begin{equation}\label{PPHIPSI1}
\phi_i(x)=p_{i-1}(^lx)f({}^lx),\;\;\psi_i(x)=p_{i-1}(x)f(x),\;\;
i=\overline{1,2N+1}
\end{equation}
Then the expression under the sum is equal to the  determinant of
size $2N+1\times 2N+1$, whose odd columns are
$\underline{\phi}(x_i), i=\overline{1,N}$, the even columns are
$\underline{\psi}(x_i), i=\overline{1,N}$, and the last $2N+1
^{th}$ column is $p_{i-1}(\zeta)$. We represent this determinant
as a sum over permutations $\sigma\in\; S_{2N+1}$. Note that the
sum over all permutations which do not satisfy the following two
conditions
\begin{itemize}
    \item $\sigma(2N+1)=2N+1$
\item $\biggl\{\{\sigma (1),\sigma (2)\},\ldots
,\{\sigma(2N-1),\sigma(2N)\}\biggr\}=\biggl\{\{1,2\},\ldots
,\{2N-1, 2N\}\biggr\}$\end{itemize} vanishes after averaging. This
is so because of the orthogonality relations for the skew
symmetric orthogonal polynomials. On the other hand, there are
$N!$ permutations which do satisfy the conditions above. All of
them give the same contribution, which is proportional to
$p_{2N}(\zeta)$. Obviously, $\langle d(\zeta)\rangle$ is a monic
polynomial. This completes the proof.
\end{proof}
\begin{prop}\label{PPAVERAGEOFINVERSED}
For $\zeta\not\in\X$ the average of $1/d(\zeta)$ is given by the
skew symmetric inner product  of $R_{\zeta}(x)=(\zeta-x)^{-1}$ and
$p_{2N-2}(x)$,
\begin{equation}\label{PAVERAGEOFINVERSED}
\langle 1/d(\zeta)\rangle_{\T_{2N}^{(4)}(f)}=h^{-1}_{N-1}\langle
p_{2N-2},R_{\zeta}\rangle.
\end{equation}
\end{prop}
\begin{proof}
The average  over the discrete symplectic ensemble is the
normalized sum over configurations from the set
$Conf_{2N}^{(4)}(\X)$. The normalization constant is $c_N^{(4)}$.
This normalized sum is that over the ordered variables
$x_i\in\;\dot{\X}, i=\overline{1, N}$. We have a symmetric
function of variables $x_i$ under the sum, so  we can remove the
ordering of the variables changing the normalization constant from
$c_N^{(4)}$ to $N!\;c_N^{(4)}$.

The key observation which helps to compute the average is the
following one. The expression under the sum contains the absolute
value of the Vandermonde determinant, the product of weights and
$1/d(\zeta)$. This expression can be simplified if we expand
$1/d(\zeta)$ into partial fractions,
\begin{equation}
\begin{split}
d^{-1}(\zeta)&=\sum\limits_{\nu=1}^N\dfrac{1}{\zeta-{}^lx_{\nu}}\;
\dfrac{1}{\prod\limits_{j=1}^N({}^lx_{\nu}-x_j)\prod\limits^N_{\substack{j=1\\j\neq
\nu}}({}^lx_{\nu}-{}^lx_j)}\\
&+\sum\limits_{\nu=1}^N\dfrac{1}{\zeta-x_{\nu}}\;
\dfrac{1}{\prod\limits_{j=1}^N(x_{\nu}-x_j)\prod\limits^N_{\substack{j=1\\j\neq
\nu}}(x_{\nu}-{}^lx_j)}
\end{split}
\end{equation}
We note that each term of the first sum and each term of the
second sum gives the same contribution to the average. Indeed, we
always can make the change of variables under which $x_{\nu}$
becomes $x_1$, and $^lx_{\nu}$ becomes $^lx_1$, thanks to the
symmetry of the involved functions under permutations of
variables. Therefore  the average will consist of two terms. The
first term is
\begin{equation}
\left[c_N^{(4)}N!\right]^{-1}\sum\limits_{x_1,\ldots
,x_N}-N\;R_{\zeta}(^lx_1)\;\hat d(x_1)\;\biggl|V(^lx_2,x_2,\ldots
,^lx_N,x_N)\biggr|\cdot f(X)\nonumber
\end{equation}
where we have denoted $\hat d(x_1)=
\prod\limits_{j=2}^N(x_1-x_j)(x_1-^lx_j)$.  The second term is
\begin{equation}
\left[c_N^{(4)}N!\right]^{-1}\sum\limits_{x_1,\ldots
,x_N}N\;R_{\zeta}(x_1)\;\hat d(^lx_1)\;\biggl|V(^lx_2,x_2,\ldots
,^lx_N,x_N)\biggr|\cdot f(X)\nonumber
\end{equation}
The computation of the sums is reduced to the computation of
averages of characteristic polynomials over configurations
$X\in\;Conf_{2N-2}^{(4)}(\X)$. The previous Proposition says that
these averages are skew symplectic polynomials of the degree
$2N-2$. The remaining sum over $x_1$ gives us the skew symmetric
product in the righthand side of formula
(\ref{PAVERAGEOFINVERSED}).
\end{proof}
Given the family of the skew symmetric orthogonal polynomials with
respect to the weight $f$ introduce the Christoffel-Darboux type
kernel
\begin{equation}\label{PChristoffel4}
 K_N^{CD,4}(\zeta,\eta)=\sum\limits_{i=0}^{N-1}\frac{p_{2i+1}(\zeta)p_{2i}(\eta)-
p_{2i+1}(\eta)p_{2i}(\zeta)}{h_i}
\end{equation}
\begin{prop}For $\zeta,\eta\not\in\X$,
the averages of products and ratios of \textbf{two} characteristic
polynomials with respect to the discrete symplectic ensembles are
given in terms of $K_N^{CD,4}$ and its pairings  with
$R_{\zeta}=(\zeta-x)^{-1}$, $R_{\eta}=(\eta-x)^{-1}$:
\begin{equation}\label{PSymplecticAverageOftwo}
\langle
d(\zeta)d(\eta)\rangle_{\T_{2N}^{(4)}(f)}=\frac{h_N}{\zeta-\eta}\;K_{N+1}^{CD,4}(\zeta,\eta)
\end{equation}
\begin{equation}\label{PSymplecticAverageOftwoPR}
\begin{split}
\left\langle
\frac{d(\eta)}{d(\zeta)}\right\rangle_{\T_{2N}^{(4)}(f)}&=h_{N-1}^{-1}
\left\langle\frac{1}{R_{\eta}(.)}\left\langle
d(\eta)d(.)\right\rangle_{\T_{2N-2}^{(4)}(f)},\frac{R_{\zeta}(.)}{R_{\eta}(.)}\right\rangle
\end{split}
\end{equation}
\begin{equation}\label{PSymplecticAverageOftwoRR}
\begin{split}
\left\langle
\frac{1}{d(\eta)d(\zeta)}\right\rangle_{\T_{2N}^{(4)}(f)}&=h^{-1}_{N-1}\left\langle
R_{\eta}(.)\left\langle\frac{d(.)}{d(\eta)}\right\rangle_{\T_{2N-2}^{(4)}(f)},R_{\eta}(.)R_{\zeta}(.)\right\rangle
\end{split}
\end{equation}
\end{prop}
\begin{rem}
Clearly, the first equation remains valid if $\zeta, \eta\in\X$,
and the second equation remains valid for $\eta\in\X$.
\end{rem}
\begin{proof}
Basically we proceed as in the computations of averages of the
characteristic polynomial and its inverse. Let us first prove the
formula for the average of product of characteristic polynomials.
This average can be represented as a sum over the variables
$x_i\in \dot{\X}, i\in\;\overline{1,N}$, divided by the
normalization constant $c^{(4)}_N\;N!$. (The ordering of $x_1,
x_2, \ldots, x_N$ is removed by $N!$).  Set
\begin{equation}
\phi_i(x)=p_{i-1}({}^lx)f({}^lx),\;\;\psi_i(x)=p_{i-1}(x)f(x),\;\;
i=\overline{1,2N+2}\nonumber
\end{equation}
The expression under the sum can be rewritten as  the determinant
of the $2N+2\times 2N+2$ matrix whose first column is
$p_i(\zeta)$, the second column is $p_i(\eta)$, all other odd
columns are $\underline{\phi}(x_{2i-1})$, and all other even
columns are $\underline{\psi}(x_{2i})$. The average now is equal
to this sum multiplied by the factor
$\left[N!\;c_N^{(4)}\;(\zeta-\eta)\right]^{-1}$.

Now we rewrite the determinant as a sum over permutations
$\sigma\in\;S_{2N+2}$, and change the order of sums. The skew
symmetric product of $\phi_{2i-1}$ and $\psi_{2i}$ gives
$h_{i-1}$, and all other skew symmetric products constructed with
these functions are zero. Then the sum over all permutations
$\sigma$ which do not satisfy the relation
\begin{equation}
\begin{split}
\biggl\{\{\sigma(3),\sigma(4)\},\ldots
,&\{\sigma(2N+1),\sigma(2N+2)\}\biggr\}=\\
&\biggl\{\{1,2\},\ldots ,\widehat{\{2i-1,2i\}},\ldots ,\{2N+1,
2N\}\biggr\} \nonumber
\end{split}
\end{equation}
for some $1\leq i\leq N$ vanishes after the averaging. Consider
the permutations which do satisfy this condition. All these
permutations transfer the pair $(1,2)$ to the pair $(2i-1, 2i)$.
The sum over such permutations converts the sum over $x_1,\ldots
,x_N$ into the factor $N!\; c_{N+1}/h_i$, thanks to the
orthogonality relations between $\phi$ and $\psi$. The index $i$
takes values from $1$ to $N+1$, and we obtain equation
(\ref{PSymplecticAverageOftwo}).

The  second and the third equations are obtained from the first
one by the same procedure as that used in the computation of
$\langle 1/d(\zeta)\rangle_{\T_{2N}^{(4)}(f)}$, Proposition
\ref{PPAVERAGEOFINVERSED}.
\end{proof}

\begin{defn}\label{PDefinitionOfTheCuachyTransforms}
The \textbf{ Cauchy type transform} $h_k(\zeta)$ of the monic skew
orthogonal polynomial $p_k$ is the skew symmetric inner product of
$R_{\zeta}$ and $p_k$,
\begin{equation}
h_k(\zeta)=\left\langle p_k,R_{\zeta}\right\rangle
\end{equation}
Here $\zeta\not\in\X$, and  the skew symmetric product is defined
by equation
(\ref{PDefinitionSkewSymmetricInnerProductSymplectic}).
\end{defn}
\begin{prop}
The averages of \textbf{two} characteristic polynomials with
respect to the discrete symplectic ensemble are
Christoffel-Darboux type kernels constructed from the skew
orthogonal polynomials associated with this ensemble, and their
Cauchy type transforms:
\begin{itemize}
    \item $\left\langle d(\zeta)
d(\eta)\right\rangle_{\T_{2N}^{(4)}(f)}=
\dfrac{h_N}{\zeta-\eta}\;\sum\limits_{i=0}^N\dfrac{p_{2i+1}(\zeta)p_{2i}(\eta)-p_{2i}(\zeta)p_{2i+1}(\eta)}{h_i}$
\item
$\left\langle\dfrac{d(\eta)}{d(\zeta)}\right\rangle_{\T_{2N}^{(4)}(f)}=
(\eta-\zeta)\;\sum\limits_{i=0}^{N-1}\dfrac{p_{2i+1}(\eta)h_{2i}(\zeta)-p_{2i}(\eta)h_{2i+1}(\zeta)}{h_i}+1$
\item $\left\langle \dfrac{1}{d(\eta)
d(\zeta)}\right\rangle_{\T_{2N}^{(4)}(f)}=
\dfrac{1}{h_{N-1}}\dfrac{1}{\eta-\zeta}\;\biggl[
\sum\limits_{i=0}^{N-2}\dfrac{h_{2i+1}(\eta)h_{2i}(\zeta)-h_{2i}(\eta)h_{2i+1}(\zeta)}{h_i}+\left\langle
R_{\eta}, R_{\zeta}\right\rangle\biggr]$
\end{itemize}
\end{prop}
\begin{proof}
Straightforward computation of the skew symmetric products in the
righthand sides of equations
(\ref{PSymplecticAverageOftwo})-(\ref{PSymplecticAverageOftwoRR}).
\end{proof}
\subsection{Discrete orthogonal ensemble}\label{PSectionDOE}
Any point configuration from the set $Conf_{2N}^{(4)}(\X)$ defines
a configuration of holes on $\X$. All such configurations of holes
naturally form a new set of point configurations. We will denote
this set by $Conf_{2N}^{(1)}(\X)$. Thus,
\begin{equation}
Conf_{2N}^{(1)}(\X)=\biggl\{Y\subset\X\arrowvert Y=\X\setminus X,
X\in\;Conf_{2N}^{(4)}(\X)\biggr\}. \nonumber
\end{equation}
Point configurations from $Conf_{2N}^{(1)}$ admit an independent
description. Namely, recall that $\X$ is the ordered set, in which
the smallest point is the leftmost one. If we say that the
leftmost point of the set $\X$ is odd, any configuration from the
set $Conf_{2N}^{(1)}(\X)$ is such that its smallest  point is odd,
and any two neighboring points are always have different parity.
Assume once again that a nonnegative function $f$ is given on
$\X$, which does not vanish at least at $2N$ points.
\begin{defn}
The point process which lives on $Conf_{2N}^{(1)}(\X)$, and for
which the probability of a configuration $Y$ is given by
\begin{equation}\label{PDefinitonOfDiscreteOrthogonalEnsembleInitial}
\mbox{Prob}(Y)=\mbox{const}\;f(Y)\cdot\;\arrowvert
V(Y)\arrowvert,\;\;\;Y\in Conf_{2N}^{(1)}(\X)
\end{equation}
will be called $2N$ point \textit{discrete orthogonal ensemble}
and will be denoted by $\T_{2N}^{(1)}(f)$.
\end{defn}

Define a two point antisymmetric function $\epsilon(x,y)$ on $\X$
such that it can be equal only to one, zero, or minus one: for
$x<y$
\begin{equation}
\epsilon(x,y)=\left\{%
\begin{array}{ll}
    1, & \hbox{$x$ is odd;} \\
0, & \hbox{otherwise.} \\\end{array}%
\right.
\end{equation}
 Then the definition of $\T_{2N}^{(1)}(f)$ can be
rewritten as
\begin{equation}\label{PORTHOGONALENSEMBLEDEFINITION}
\mbox{Prob}(Y)=\left[c_N^{(1)}\right]^{-1}\;f(Y)\cdot\;\arrowvert
V(Y)\arrowvert\;\PF\left[\epsilon(y_i,y_j)\right]_{i,j=1}^N
\end{equation}
where $Y$ is \textit{any} configuration on $\X$ which consists of
$2N$ points. This is the consequence of the fact that the Pfaffian
in equation (\ref{PORTHOGONALENSEMBLEDEFINITION})  is zero for any
$Y\not\in\; Conf_{2N}^{(1)}(\X)$, and it is one for any $Y\in
\;Conf_{2N}^{(1)}(\X)$.
\begin{rem}
In the continuous limit the discrete orthogonal ensemble turns
into the orthogonal ensemble of the Random Matrix Theory (see
Mehta \cite{mehta}, Chapter 3, for the definition). This is so
because the parity condition  for the configurations with nonzero
probabilities becomes irrelevant in the continuous limit.
\end{rem}

For any functions $g_1, g_2$ introduce the skew symmetric inner
product:
\begin{equation}\label{POrthogonalSkewSymmetricProduct}
\left\langle g_1,
g_2\right\rangle=\sum\limits_{x_1,x_2\in\;\X}\epsilon(x_1,x_2)g_1(x_1)g_2(x_2)f(x_1)f(x_2)
\end{equation}
and the monic skew symmetric orthogonal polynomials $p_i$ with
respect to this inner product. (We assume that the weight $f$ is
such that $\mbox{det}||\langle x^i,x^j\rangle||_{i,j=1}^N\neq 0$).
These skew symmetric orthogonal polynomials are constructed so
that they satisfy the same orthogonality conditions as in
Definition \ref{PDefinitionOfSkewsymmetricOrthogonalPolynomials},
but with respect to the skew symmetric inner product defined by
equation (\ref{POrthogonalSkewSymmetricProduct}). Word-for-word
repetition of arguments of the previous Section gives
\begin{prop}
The normalization constant and the averages of characteristic
polynomials must be of the same form for the discrete orthogonal
and symplectic ensembles, up to the definition of the skew
symmetric inner product. In particular we have
\begin{equation}
c_{N}^{(1)}=\prod\limits_{i=0}^{N-1}h_i,\;\;\;h_i=\langle
p_{2i},p_{2i+1}\rangle
\end{equation}
\begin{equation}
p_{2N}(\zeta)=\langle d(\zeta)\rangle_{\T_{2N}^{(1)}(f)}
\end{equation}
\begin{equation}\label{PAVERAGEOFINVERSED1}
\langle
d^{-1}(\zeta)\rangle_{\T_{2N}^{(1)}(f)}=h^{-1}_{N-1}\langle
p_{2N-2},R_{\zeta}\rangle
\end{equation}
\begin{equation}\label{PSymplecticAverageOftwo1}
\langle
d(\zeta)d(\eta)\rangle_{\T_{2N}^{(1)}(f)}=\frac{h_N}{\zeta-\eta}\;K_{N+1}^{CD,1}(\zeta,\eta)
\end{equation}
\begin{equation}
\begin{split}
\left\langle
\frac{d(\eta)}{d(\zeta)}\right\rangle_{\T_{2N}^{(1)}(f)}=h_{N-1}^{-1}
\left\langle\frac{1}{R_{\eta}(.)}\left\langle
d(\eta)d(.)\right\rangle_{\T_{2N-2}^{(1)}(f)},\frac{R_{\zeta}(.)}{R_{\eta}(.)}\right\rangle
\end{split}
\end{equation}
\begin{equation}\label{PSymplecticAverageOftwoRR1}
\begin{split}
\left\langle
\frac{1}{d(\eta)d(\zeta)}\right\rangle_{\T_{2N}^{(1)}(f)}=h^{-1}_{N-1}\left\langle
R_{\eta}(.)\left\langle\frac{d(.)}{d(\eta)}\right\rangle_{\T_{2N-2}^{(1)}(f)},R_{\eta}(.)R_{\zeta}(.)\right\rangle
\end{split}
\end{equation}
Here the Christoffel-Darboux kernel $K_{N+1}^{CD,1}(\zeta,\eta)$
is given by formula (\ref{PChristoffel4}), where the skew
symmetric orthogonal polynomials $p_i$ are constructed with
respect to inner product (\ref{POrthogonalSkewSymmetricProduct}).
\end{prop}
Define the Cauchy type transforms of the skew orthogonal
polynomials by the same way as they were defined in the case of
the discrete symplectic ensembles, see Definition
\ref{PDefinitionOfTheCuachyTransforms}. Then we obtain the
following
\begin{prop}\label{PAveragesAsChristoffelDarby1}
The averages of \textbf{two} characteristic polynomials with
respect to the discrete orthogonal ensemble are
Christoffel-Darboux  type kernels constructed from the skew
orthogonal polynomials associated with this ensemble, and their
Cauchy type transforms:
\begin{itemize}
    \item $\left\langle d(\zeta)
d(\eta)\right\rangle_{\T_{2N}^{(1)}(f)}=
\dfrac{h_N}{\zeta-\eta}\;\sum\limits_{i=0}^N\dfrac{p_{2i+1}(\zeta)p_{2i}(\eta)-p_{2i}(\zeta)p_{2i+1}(\eta)}{h_i}$
\item
$\left\langle\dfrac{d(\eta)}{d(\zeta)}\right\rangle_{\T_{2N}^{(1)}(f)}=
(\eta-\zeta)\;\sum\limits_{i=0}^{N-1}\dfrac{p_{2i+1}(\eta)h_{2i}(\zeta)-p_{2i}(\eta)h_{2i+1}(\zeta)}{h_i}+1$
\item $\left\langle \dfrac{1}{d(\eta)
d(\zeta)}\right\rangle_{\T_{2N}^{(1)}(f)}=
\dfrac{1}{h_{N-1}}\dfrac{1}{\eta-\zeta}\;\biggl[
\sum\limits_{i=0}^{N-2}\dfrac{h_{2i+1}(\eta)h_{2i}(\zeta)-h_{2i}(\eta)h_{2i+1}(\zeta)}{h_i}+\left\langle
R_{\eta}, R_{\zeta}\right\rangle\biggr]$
\end{itemize}
\end{prop}
\begin{proof}
Straightforward computation of the skew symmetric products in the
righthand sides of equations
(\ref{PSymplecticAverageOftwo1})-(\ref{PSymplecticAverageOftwoRR1}).
\end{proof}
\subsection{Relation of ensembles}
This Section discusses the relation between Pfaffian
$L$-ensembles, discrete symplectic ensembles and the discrete
orthogonal ensembles, introduced in Section
\ref{PSectionSpecialMatrices}, Section \ref{PSectionDSPE}, and
Section \ref{PSectionDOE} respectively. Given $\X$ consider the
Pfaffian $L$-ensemble, defined by equations
(\ref{PstructureL})-(\ref{Pepsilon}). Denote
$\arrowvert\X_-\arrowvert=2M$, $\arrowvert\X_+\arrowvert=2N$,
$M,N\in\;\mathbb{Z}_{>0}$, and assume that the weight $h$ is
strictly positive on $\X$. Given $h$ introduce the weights
$f^{(4)}$, $f^{(1)}$ by
\begin{equation}\label{PrelationofWeightsSL}
f^{(4)}(x)=\left\{%
\begin{array}{ll}
    \dfrac{h(x)}{\prod_{y\in\;\X_-}|x-y|}, & x\in\;\X_+ \\
\dfrac{1}{h(x)\prod_{y\in\;\X_-,\;y\neq x}|x-y|}, & x\in \;\X_- \\\end{array}%
\right.
\end{equation}

\begin{equation}\label{PrelationofWeightOrth}
f^{(1)}(x)=\left\{%
\begin{array}{ll}
    \dfrac{h(x)}{\prod_{y\in\;\X_+}|x-y|}, & x\in\;\X_- \\
\dfrac{1}{h(x)\prod_{y\in\;\X_+,\;y\neq x}|x-y|}, & x\in \;\X_+ \\\end{array}%
\right.
\end{equation}
Note that the only difference in the definition of $f^{(1)}$ and
$f^{(4)}$ is the interchange of $\X_-$ and $\X_+$. It follows from
the two equations above that
\begin{equation}\label{PRelationofWeigthsSymplOrth}
f^{(1)}(x)\cdot f^{(4)}(x)=\prod_{y\in\;\X,\; y\neq
x}|x-y|^{-1},\;\;x\in\;\X
\end{equation}
\begin{thm}
The Pfaffian $L$-ensemble with the weight $h$ defined by equations
(\ref{PstructureL})-(\ref{Pepsilon}), the discrete symplectic
ensemble with the weight $f^{(4)}$ defined by equation
(\ref{PDEFINITIONofSymplecticEnsembleEquation}), and the discrete
orthogonal ensemble with the weight $f^{(1)}$, defined by equation
(\ref{PDefinitonOfDiscreteOrthogonalEnsembleInitial}) are
\textbf{equivalent} in the following sense:\\
a) There is a bijection between the sets $Conf^{L}(\X)$ and
$Conf_{2M}^{(4)}(\X)$ defined by the particle-hole involution on
$\X_+$. Under this bijection the probability distribution for the
Pfaffian  L-ensemble, equation
(\ref{PReductionToPfaffianProcessesEquation}), turns into the
probability distribution for the discrete symplectic ensemble,
equation (\ref{PDEFINITIONofSymplecticEnsembleEquation}).\\
b) There is a bijection between the sets $Conf^{L}(\X)$ and
$Conf_{2N}^{(1)}(\X)$ defined by the particle-hole involution on
$\X_-$. Under this bijection the probability distribution for the
Pfaffian  L-ensemble, equation
(\ref{PReductionToPfaffianProcessesEquation}), turns into the
probability distribution for the discrete orthogonal ensemble,
equation (\ref{PDefinitonOfDiscreteOrthogonalEnsembleInitial}).\\
c) There is a bijection between the sets $Conf_{2N}^{(1)}(\X)$ and
$Conf_{2M}^{(4)}(\X)$ defined by the particle-hole involution on
the whole set $\X$. Under this bijection the probability
distribution for the discrete orthogonal ensemble, equation
(\ref{PDefinitonOfDiscreteOrthogonalEnsembleInitial}) turns into
the probability distribution for the discrete symplectic ensemble,
(\ref{PDEFINITIONofSymplecticEnsembleEquation}).\\
In other words, the probability spaces defined by $Conf^{L}(\X)$,
$Conf_{2M}^{(4)}(\X)$, $Conf_{2N}^{(1)}(\X)$ with the associated
probability distributions are isomorphic.
\end{thm}
\begin{proof}
Given $Conf^{L}(\X)$, let $X\in\;Conf^{L}_{2N}(\X)$, and $\TX$ be
the point configuration associated with $X$, see  Section
\ref{PSectionSpecialMatrices}. Consider the particle-hole
involution on $\X_-$. Under this involution the configuration
$\TX$ turns into the configuration $Y$ related with the
configuration $\TX$ by the formula:
\begin{equation}\label{PYINTERMSOFTX}
Y=Y_-\sqcup Y_+,\;\;\;Y_-=\X_-\setminus \TX_-,\;\;\;Y_+=\TX_+.
\end{equation}
The key observation is that the configuration $Y$ is in
$Conf_{2M}^{(4)}(\X)$. It can be immediately seen from the
definition of $Conf_{2M}^{(4)}(\X)$. Clearly, the particle-hole
involution on $\X_-$ is a bijection between $Conf^{L}(\X)$ and
$Conf_{2M}^{(4)}(\X)$. In order to prove a), it remains to show
that the probability distribution on $Conf^{L}(\X)$,
 equation
(\ref{PReductionToPfaffianProcessesEquation}), turns into the
probability distribution on $Conf_{2M}^{(4)}(\X)$, equation
(\ref{PDEFINITIONofSymplecticEnsembleEquation}), provided the
weights are related by formula (\ref{PrelationofWeightsSL}). This
is  a consequence of two facts. The first one is  $|V(Y)|$ can be
rewritten as
\begin{equation}\label{somemodd}
\arrowvert
V(Y)\arrowvert=\mbox{const}\;\frac{V(\TX_+)V(\TX_-)}{\prod(\TX_+;
\TX_-)}\;\frac{\prod\limits_{x\in\;\TX_+}\prod\limits_{y\in
\X_-}|x-y|}{\prod\limits_{x\in\;\TX_-}\prod\limits_{y\in
\X_-,\;y\neq x}|x-y|},\;\;\;\mbox{const}=\arrowvert
V(\X_-)\arrowvert
\end{equation}
This follows from the relation $Y$ and $\TX$ (equation
(\ref{PYINTERMSOFTX})). The second  fact is the relation:
\begin{equation}
f(Y)=\mbox{const}\;\frac{f(\TX_+)}{f(\TX_-)},\;\;\;\mbox{const}=f(\X_-)
\nonumber
\end{equation}
We then see that the righthand side of equation
(\ref{PReductionToPfaffianProcessesEquation}) is identical to that
of equation (\ref{PDEFINITIONofSymplecticEnsembleEquation}), if
$h$ and $f$ are related by equation (\ref{PrelationofWeightsSL}).
Thus we have proved a).

The proof of b)  is constructed in the same way. The configuration
$Y$ obtained by the particle-hole involution on $\X_-$ from $\TX$
is
\begin{equation}\label{PYINTERMSOFTX1}
Y=Y_-\sqcup Y_+,\;\;\;Y_-=\TX_-,\;\;\;Y_+=\X_+\setminus \TX_+.
\end{equation}
Instead of formula (\ref{somemodd}) we obtain
\begin{equation}\label{somemodd1}
\arrowvert
V(Y)\arrowvert=\mbox{const}\;\frac{V(\TX_+)V(\TX_-)}{\prod(\TX_+;
\TX_-)}\;\frac{\prod\limits_{x\in\;\TX_-}\prod\limits_{y\in
\X_+}|x-y|}{\prod\limits_{x\in\;\TX_+}\prod\limits_{y\in
\X_+,\;y\neq x}|x-y|},\;\;\;\mbox{const}=\arrowvert
V(\X_+)\arrowvert
\end{equation}
which leads (together with the relation of the weights $h$ and
$f^{(1)}$, equation (\ref{PrelationofWeightOrth}) ) to the
equivalence of the probability distributions on $Conf^{L}(\X)$ and
$Conf_{2N}^{(1)}(\X)$.

The statement c) obviously follows from a) and b). However, it is
easy to give an independent argument. Let
$X\in\;Conf_{2M}^{(4)}(\X)$. Then the particle-hole involution on
$\X$ gives a configuration $Y$ which is in $Conf_{2N}^{(1)}(\X)$.
For $X$ and $Y$ related by the particle-hole involution we obtain
\begin{equation}
\arrowvert V(X)\arrowvert=\mbox{const}\;\arrowvert
V(Y)\arrowvert\;\prod\limits_{x\in\;X}\prod\limits_{y\in\;\X,\;y\neq
x}|x-y|,\;\;\;\mbox{const}=\arrowvert V(\X)\arrowvert^{-1}
\end{equation}
This expression (together with the relation of weights $f^{(1)}$
and $f^{(4)}$, equation (\ref{PRelationofWeigthsSymplOrth}))
leads to the equivalence of the discrete symplectic and the
discrete orthogonal ensembles.
\end{proof}
Assume that  the Pfaffian $L$-ensemble is given, and denote by
$\T_{2M}(f^{(4)})$ and $\T_{2N}(f^{(1)})$ the equivalent discrete
symplectic and orthogonal ensembles obtained from the Pfaffian
$L$-ensemble by the particle-hole involution. With the Pfaffian
$L$-ensemble we can construct a new ensemble, Pfaffian
$\hat{L}$-ensemble, by new splitting of $\X$,
$\X=\hat{X}_-\sqcup\hat{X}_+$,
$\arrowvert\hat{\X}_+\arrowvert=2N+2S$,
$\arrowvert\hat{\X}_-\arrowvert=2M-2S$, see Section
\ref{PSectionHatL}. This Pfaffian $\hat{L}$-ensemble induces a
discrete symplectic ensemble of $2(M-S)$ point configurations, and
a discrete orthogonal ensemble of $2(N+S)$ point configurations.
The first one is obtained by the particle-hole involution on
$\hat{\X}_-$, and the second one is obtained by the particle-hole
involution on $\hat{\X}_+$. Let $\hat{h}$ denotes the weight of
the Pfaffian $\hat{L}$-ensemble. Recall that it is given in terms
of the weight $h$ of the original Pfaffian $L$-ensemble by formula
(\ref{PNewWeightHath}). The Theorem above says that the Pfaffian
$\hat{L}$ ensemble with weight $\hat{h}$, the discrete symplectic
ensemble $\T_{2(M-S)}(\hat{f}^{(4)})$, and the discrete orthogonal
ensemble $\T_{2(N+S)}(\hat{f}^{(1)})$ are equivalent provided the
weights are related by formulae (\ref{PrelationofWeightsSL})-
(\ref{PrelationofWeightOrth}), with $\X_{\pm}\rightarrow
\hat{\X}_{\pm}$.
\begin{prop}
The discrete symplectic ensembles $\T_{2M}(f^{(4)})$ and
$\T_{2(M-S)}(\hat{f}^{(4)})$ have the same weight: $f^{(4)}\equiv
\hat{f}^{(4)}$. The discrete orthogonal ensembles
$\T_{2N}(f^{(1)})$ and $\T_{2N+2S}(\hat{f}^{(1)})$ have the same
weight: $f^{(1)}\equiv \hat{f}^{(1)}$.
\end{prop}
\begin{proof}
Application of the same arguments as in the case of the
determinantal point ensembles, see Proposition \ref{PROPOSITION O
RAVENSTVE VESOV}.
\end{proof}
\subsection{Averages of characteristic polynomials: symplectic and
orthogonal ensembles} Our goal now is to derive averages of
product and ratios for symplectic and orthogonal ensembles. We
will use the relation of these ensembles with the Pfaffian
$L$-ensembles discussed in the previous Section, and Proposition
\ref{PPROPOSITION}. We begin with symplectic ensembles.

Given equation (\ref{PEquationPfaffiansofKMinors}) let us rewrite
its righthand side in terms of the discrete symplectic ensembles.
We proceed as in the determinantal case, see Section
\ref{SectionAveragesCharacteristicPolynomials}.
\begin{prop}\label{PPropositionOnConstant}
The constant in equation (\ref{PEquationPfaffiansofKMinors}) is
the ratio of the normalization constants for the discrete
symplectic ensembles of $2N-2S$ particles and of $2N$ particles,
\begin{equation}\label{PConstat}
\frac{\PFAFF(J+\hat{L})}{\PFAFF(J+L)}\;h(\X_0)\arrowvert
V(\X_0)\arrowvert=\frac{c^{(4)}_{N-S}}{c_N^{(4)}}
\end{equation}
\end{prop}
\begin{proof}
There are two Pfaffian ensembles involved in Proposition
\ref{PPROPOSITION}. The first one, the Pfaffian $L$-ensemble, is
defined with respect to the splitting $\X=\X_-\sqcup\X_+$, and the
second one, the Pfaffian $\hat{L}$-ensemble, is defined with
respect to the splitting  $\hat{\X}=\hat{\X}_-\sqcup\hat{\X}_+$.
The expression $\left[\PF(J+L)\right]^{-1}$ is the probability of
the empty configuration for the first Pfaffian ensemble, and
$\left[\PF(J+\hat{L})\right]^{-1}$ is the probability of the empty
configuration for the second Pfaffian ensemble.

By the particle-hole involution on $\X_-$ and $\hat{\X_-}$
construct the equivalent discrete symplectic ensembles. Then the
empty configurations of the Pfaffian ensembles turn into certain
configurations of these symplectic ensembles. It is not hard to
see that these configurations are  $\X_-$ and $\hat{\X}_-$. Equate
the probabilities:
\begin{equation}
\frac{1}{\PF(J+L)}=\frac{1}{c_N^{(4)}}\;f(\X_-)\arrowvert
V(\X_-)\arrowvert,\;\;\;\frac{1}{\PF(J+\hat{L})}=\frac{1}{c_{N-S}^{(4)}}\;f(\hat{\X}_-)\arrowvert
V(\hat{\X}_-)\arrowvert\nonumber
\end{equation}
These equations, together  with the relation between weights,
equation (\ref{PrelationofWeightsSL}), give formula
(\ref{PConstat}).
\end{proof}
\begin{thm}\label{B4DISCRETEMAINTHEOREM}
For any integer $N\geq 1$ take an integer $S$ such that $2-2N\leq
2S\leq \arrowvert \X\arrowvert -2-2N$, complex numbers
$\alpha=(\alpha_-,\alpha_+)$, $\alpha_-=(\alpha_1^-,\ldots
\alpha_m^-)$, $\alpha_+=(\alpha_1^+,\ldots ,\alpha_k^+)$ such that
$|\alpha_+|-|\alpha_-|=k-m=2S$, in each set $\alpha_{\pm}$ the
numbers are pairwise distinct, and the set $\alpha$ does not
intersect $\X$. Then the average of products and ratios of
characteristic polynomials with respect to the discrete symplectic
ensemble $\T_{2N}^{(4)}(f)$ is given by the formula:
\begin{equation}
\biggl\langle\dfrac{\prod_{i=1}^k d(\alpha_i^+)}{\prod_{i=1}^m
d(\alpha_i^-)}\biggr\rangle_{\T_{2N}^{(4)}(f)}
=\dfrac{c_{N+S}^{(4)}}{c_{N}^{(4)}}\;\dfrac{\prod(\alpha_-;\alpha_+)}{V(\alpha_-)V(\alpha_+)}\PFAFF\;\biggl[W_N^{(4)}(\alpha\arrowvert\alpha)\biggr]
\end{equation}
where the kernel function $W^{(4)}_N$ is defined by
\begin{itemize}
    \item
$W^{(4)}_N(\alpha_i^+,\alpha_j^+)=h^{-1}_{N+S-1}(\alpha_i^+-\alpha_j^+)\left\langle
d(\alpha_i^+)d(\alpha_j^+)\right\rangle_{\T_{2N+2S-2}^{(4)}(f)}$
\item
$W_N^{(4)}(\alpha_i^-,\alpha_j^+)=-W_N^{(4)}(\alpha_j^+,\alpha_i^-)=\dfrac{1}{\alpha_i^--\alpha_j^+}\left\langle
\dfrac{d(\alpha_j^+)}{d(\alpha_i^-)}\right\rangle_{\T_{2N+2S}^{(4)}(f)}$
\item
$W^{(4)}_N(\alpha_i^-,\alpha_j^-)=h_{N+S}(\alpha_i^--\alpha_j^-)\left\langle
\dfrac{1}{d(\alpha_i^-)d(\alpha_j^-)}\right\rangle_{\T_{2N+2S+2}^{(4)}(f)}$\end{itemize}
\end{thm}
\begin{proof}
Assume first that $S\geq 0$. We use equation
(\ref{PEquationPfaffiansofKMinors}). Denote by
$\T^{(4)}_{2N-2S}(f^{(4)})$ the discrete symplectic ensemble
obtained by the particle-hole involution on $\hat{\X}$ from the
Pfaffian $\hat{L}$-ensemble involved in  equation
(\ref{PEquationPfaffiansofKMinors}). We want to rewrite the
righthand side of equation (\ref{PEquationPfaffiansofKMinors}) in
terms of $\T^{(4)}_{2N-2S}(f^{(4)})$. $\T^{(4)}_{2N-2S}(f^{(4)})$
and the Pfaffian $\hat{L}$-ensemble involved in  equation
(\ref{PEquationPfaffiansofKMinors}) are equivalent, the
corresponding probability measures are equal to each other. We
only need to express the functions $E(.,.)$ in terms of
characteristic polynomials associated with
$\T^{(4)}_{2N-2S}(f^{(4)})$. For any finite set $\alpha$,
$\alpha\cap \frak X=\varnothing$, we find
\begin{equation}
E(\alpha,X)\cdot\prod(\alpha;\hat{\X}_-)=\prod(\alpha;Y)\equiv\prod_i
d(\alpha_i)\nonumber
\end{equation}
where the configurations $X\in\; Conf^{\hat{L}}(\X)$ and
$Y\in\;Conf^{(4)}_{2N-2S}(\X)$ are related by the particle-hole
involution. This immediately follows from the definition of
$E(.,.)$, equation (\ref{LCharacteristicPolynomials}). This
observation and Proposition \ref{PPropositionOnConstant} give
\begin{equation}\label{PEquationPfaffiansofKMinorsI}
\mbox{Pf}\;\left[ K(\alpha'\arrowvert\alpha')\right]
=\frac{c^{(4)}_{N-S}}{c_N^{(4)}}\;
h(\alpha)\dfrac{V(\alpha_{-})V(\alpha_+)}{\prod(\alpha_-;\alpha_+)}
\left[\dfrac{\prod(\alpha_+;\X_-)}{\prod(\alpha^-;\X_-)}\right]
\biggl\langle\dfrac{\prod_{i=1}^k d(\alpha_i^+)}{\prod_{i=1}^m
d(\alpha_i^-)}\biggr\rangle_{\T_{2N-2S}^{(4)}(f)}
\end{equation}
Considering particular cases corresponding to $k=0, m=2$; $k=1,
m=1$, $k=2, m=0$ determine the kernel function $K$ in this
equation, and obtain the formula stated in the Theorem.

The case $S\leq 0$ corresponds to the splitting
\begin{equation}
\X=\hat{\X}_-\sqcup\hat{\X}_+,\;\;\hat{\X}_-=\X_-\sqcup\X_0,\;\;\hat{\X}_+=\X_+\setminus\X_0
\nonumber
\end{equation}
and is considered in the same way.
\end{proof}
 Applying the same arguments we prove the corresponding result
for the discrete orthogonal ensembles.
\begin{thm}\label{THEOREMAVERAGESCHARACTERISTICPOLYNOMIALSORTHOGON}
For any integer $N\geq 1$ take an integer $S$ such that
$2N+2-2\arrowvert\X\arrowvert\leq 2S\leq 2N-2$, complex numbers
$\alpha=(\alpha_-,\alpha_+)$, $\alpha_-=(\alpha_1^-,\ldots
\alpha_m^-)$, $\alpha_+=(\alpha_1^+,\ldots ,\alpha_k^+)$ such that
$|\alpha_+|-|\alpha_-|=k-m=2S$, in each set $\alpha_{\pm}$ the
numbers are pairwise distinct, and the set $\alpha$ does not
intersect $\X$. Then the average of products and ratios of
characteristic polynomials with respect to the discrete orthogonal
ensemble $\T_{2N}^{(1)}(f)$ is given by the formula:
\begin{equation}\label{EQUATIONAVERAGESCHARACTERISTICPOLYNOMIALSORTHOGON}
\biggl\langle\dfrac{\prod_{i=1}^m d(\alpha_i^-)}{\prod_{i=1}^k
d(\alpha_i^+)}\biggr\rangle_{\T_{2N}^{(1)}(f)}
=\dfrac{c_{N-S}^{(1)}}{c_{N}^{(1)}}\;\dfrac{\prod(\alpha_-;\alpha_+)}{V(\alpha_-)V(\alpha_+)}\PFAFF\;\biggl[W_N^{(1)}(\alpha\arrowvert\alpha)\biggr]
\end{equation}
where the kernel function $W^{(1)}_N$ is defined by
\begin{itemize}
    \item
$W^{(1)}_N(\alpha_i^+,\alpha_j^+)=h_{N-S}(\alpha_i^+-\alpha_j^+)\left\langle
\dfrac{1}{d(\alpha_i^+)d(\alpha_j^+)}\right\rangle_{\T_{2N-2S-2}^{(1)}(f)}$
\item
$W^{(1)}_N(\alpha_i^-,\alpha_j^+)=\dfrac{1}{\alpha_i^--\alpha_j^+}\left\langle
\dfrac{d(\alpha_i^-)}{d(\alpha_j^+)}\right\rangle_{\T_{2N-2S}^{(1)}(f)}$
\item
$W^{(1)}_N(\alpha_i^-,\alpha_j^-)=\dfrac{1}{h_{N-S-1}}\;(\alpha_i^--\alpha_j^-)\left\langle
d(\alpha_i^-)d(\alpha_j^-)\right\rangle_{\T_{2N-2S+2}^{(1)}(f)}$\end{itemize}
\end{thm}
\subsection{Computation of the $m$-point correlation function from
the averages of characteristic
polynomials}\label{SectionCorrFunctionsDiscrete} The aim of this
Section is to illustrate the convenience and the generality of the
averages of the characteristic polynomials. Similary to the case
of the polynomial ensembles we extract the standard $m$-point
correlation function from this averages.
\begin{prop}\label{PPropositionmpointfor1}
The $m$-point correlation function of the discrete orthogonal
ensemble $\T_{2N}^{(1)}(f)$ is given by the formula:
\begin{equation}
\varrho_m(z_1,\ldots ,z_m)=(-)^{\frac{m(m-1)}{2}}\;\PFAFF \left[
\begin{array}{cc}
  \left[D(z_i,z_j)\right]_{i,j=1}^m & \left[S(z_i,z_j)\right]_{i,j=1}^m  \\
 & \\
  \left[-S(z_j,z_i)\right]_{i,j=1}^m  & \left[I(z_i,z_j)\right]_{i,j=1}^m  \\
\end{array}\right]
\end{equation}
where
\begin{itemize}
    \item
$D(\zeta,\eta)=\dfrac{1}{h_{N-1}}\;(\zeta-\eta)\left\langle
d(\zeta)d(\eta)\right\rangle_{\T_{2N-2}^{(1)}(f)}$ \item
$I(\zeta,\eta)=\underset{y=\eta}{\underset{x=\zeta}{\RES}}\biggl[h_{N}\;(x-y)\left\langle
\dfrac{1}{d(x)d(y)}\right\rangle_{\T_{2N+2}^{(1)}(f)}\biggr]$
\item $S(\zeta,\eta)=\left\{%
\begin{array}{ll}
\underset{y=\eta}{\RES}\biggl[\dfrac{1}{\zeta-y}\;\left\langle
\dfrac{d(\zeta)}{d(y)}\right\rangle_{\T_{2N}^{(1)}(f)}\biggr], & \zeta\neq\eta \\
\underset{y=\eta}{\RES}\biggl[\dfrac{d}{dx}\biggl|_{x=y}\;\left\langle
\dfrac{d(x)}{d(y)}\right\rangle_{\T_{2N}^{(1)}(f)}\biggr], & \zeta=\eta\\\end{array}%
\right.$
\end{itemize}
\end{prop}
\begin{proof}
Set $\alpha_-=(x_1,\ldots ,x_m)$, $\alpha_+=(y_1,\ldots ,y_m)$ in
Theorem \ref{THEOREMAVERAGESCHARACTERISTICPOLYNOMIALSORTHOGON}.
Use the formula for the Cauchy determinant to rewrite equation
(\ref{EQUATIONAVERAGESCHARACTERISTICPOLYNOMIALSORTHOGON}) as
\begin{equation}
\left\langle\mbox{det}\left[\dfrac{1}{x_i-y_j}\;\dfrac{d(x_i)}{d(y_j)}\right]_{i,j=1}^m\right\rangle_{\T_{2N}^{(1)}(f)}=
(-)^{\frac{m(m-1)}{2}}\;\PF\left[W^{(1)}_N(x,y\arrowvert
x,y)\right] \nonumber
\end{equation}
Multiply the left-hand side of this equation by the product
$(x_1-y_1)\ldots (x_m-y_m)$. Differentiate the obtained expression
by $x_1,\ldots x_m$ and take parameters $x_i$ to be equal to the
parameters $y_i$. We then obtain some function of $y_1,\ldots
,y_m$. By taking residues at points $y_1=z_1,\ldots ,y_m=z_m$ from
this function we get $\varrho_m(z_1,\ldots ,z_m)$ as we have
learned previously in the determinantal case, Proposition
\ref{PropositionCorrelationFunctionAsaverage}.

We want to understand what happens with the righthand side under
the same transformations. Decompose the Pfaffian into the sum over
permutations. Now we multiply this sum by
$\prod\limits_{i=1}^m(x_i-y_i)$, differentiate and substitute
$x_i=y_i$, $i=\overline{1,m}$. We observe that all terms of this
sum which do not include $W^{(1)}_N(x_l,y_l)$ remain unchanged, we
only replace $x_i$ by $y_i$. As for the elements like
$W^{(1)}_N(x_l,y_l)$ they turn into the elements
\begin{equation}
W^{(1)}_N(x_l,y_l)\rightarrow\;\dfrac{d}{dx_l}\left\langle\dfrac{d(x_l)}{d(y_l)}\right\rangle_{\T_{2N}^{(1)}(f)}\biggl|_{x_l=y_l}
\nonumber
\end{equation}
Thus the obtained sum is equal to
\begin{equation}
\mbox{Pf} \left[
\begin{array}{cc}
  \left[\tilde{D}(y_i,y_j)\right]_{i,j=1}^m & \left[\tilde{S}(y_i,y_j)\right]_{i,j=1}^m  \\
 & \\
  \left[-\tilde{S}(y_j,y_i)\right]_{i,j=1}^m  & \left[\tilde{I}(y_i,y_j)\right]_{i,j=1}^m  \\
\end{array}\right]\nonumber
\end{equation}
where
\begin{itemize}
    \item
$\tilde{D}(y_i,y_j)=\dfrac{1}{h_{N-1}}\;(y_i-y_j)\left\langle
d(y_i)d(y_j)\right\rangle_{\T_{2N-2}^{(1)}(f)}$ \item
$\tilde{I}(y_i,y_j)=h_{N}\;(y_i-y_j)\left\langle
\dfrac{1}{d(y_i)d(y_j)}\right\rangle_{\T_{2N+2}^{(1)}(f)}$
\item $\tilde{S}(y_i,y_j)=\left\{%
\begin{array}{ll}
\dfrac{1}{y_i-y_j}\;\left\langle
\dfrac{d(y_i)}{d(y_j)}\right\rangle_{\T_{2N}^{(1)}(f)}, & i\neq j \\
\dfrac{d}{dx}\biggl|_{x=y_i}\;\left\langle
\dfrac{d(x)}{d(y_i)}\right\rangle_{\T_{2N}^{(1)}(f)}, & i=j
\end{array}%
\right.$
\end{itemize}

 It remains to take the residues.
 Rewrite once again the Pfaffian as a sum over permutations. Taking into account the definition  of the Pfaffian it is not hard to see that
after taking $\underset{y_1=z_1}{\mbox{Res}},\ldots
\underset{y_m=z_m}{\mbox{Res}}$ the two point functions
$\tilde{D}$, $\tilde{I}$, and $\tilde{S}$ turn into $D$, $S$ and
$I$.
\end{proof}
\begin{prop}
The correlation functions of the discrete symplectic ensemble are
given by the formulas of Proposition \ref{PPropositionmpointfor1}
with averages over $\triangle^{(1)}$ replaced by similar averages
over $\triangle^{(4)}$.
\end{prop}
\begin{cor}\label{COROLLARYMDISC}The $m$-point correlation function for the discrete
orthogonal ensemble $\T_{2N}^{(1)}(f) $ is given by the quaternion
determinant $Tdet$ of the $2\times 2$ matrix valued kernel
$\sigma(z_i,z_j)$, $i,j=\overline{1,m}$,
\begin{equation}
\varrho_m(z_1,\ldots ,z_m)=Tdet\;\sigma(z_i,z_j)_{i,j=1}^m
\end{equation}
where
\begin{equation}
\sigma(z_i,z_j)=\left(%
\begin{array}{cc}
  -S(z_i,z_j) & I(z_i,z_j) \\
  -D(z_i,z_j) & -S(z_j,z_i) \\
\end{array}%
\right)
\end{equation}
the two-point functions $D$, $I$, $S$ are given by
\begin{itemize}
    \item
$D(z_i,z_j)=\sum\limits_{l=0}^{N-1}\dfrac{p_{2l+1}(z_i)p_{2l}(z_j)-p_{2l}(z_i)p_{2l+1}(z_j)}{h_l}$
\item
$I(z_i,z_j)=f(z_i)f(z_j)\;\biggl[\sum\limits_{l=0}^{N-1}\dfrac{q_{2l+1}(z_i)q_{2l}(z_j)-q_{2l}(z_i)q_{2l+1}(z_j)}{h_l}+\epsilon(z_i,z_j)\biggr]$
\item
$S(z_i,z_j)=f(z_j)\;\biggl[\sum\limits_{l=0}^{N-1}\dfrac{p_{2l+1}(z_i)q_{2l}(z_j)-p_{2l}(z_i)q_{2l+1}(z_j)}{h_l}\biggr]
$
\end{itemize}
and the functions $q_k(z)$ are defined by
\begin{equation}\label{functionsq}
\underset{\zeta=z}{\mbox{Res}}\;h_k(\zeta)=f(z)q_k(z)
\end{equation}
\end{cor}
\begin{rem}
With the specific choice $\epsilon(x,y)=\epsilon(x-y)$ on a one
dimensional lattice,
\begin{equation}
\epsilon(x,y)=\left\{%
\begin{array}{ll}
    -1/2, & x>y \\
0, & x=y \\
1/2, & x<y \\\end{array}%
\right.
\end{equation}
the formula for the $m$-point correlation function is reduced to
that of Nagao and Forrester \cite{nagao}.
\end{rem}
\begin{proof}
Compute the two point functions $D$, $I$, $S$ in Proposition
\ref{PPropositionmpointfor1}. We use the formulas which express
averages of two characteristic polynomials as the
Christoffel-Darbyoux type kernels constructed from the skew
orthogonal polynomials and their Cauchy type transforms,
Proposition \ref{PAveragesAsChristoffelDarby1}. Taking into
account equation (\ref{functionsq}) we immediately obtain the
expressions for $D$, $I$, $S$ stated in the Corollary. Thus we
have
\begin{equation}
\varrho_m(z_1,\ldots ,z_m)=(-)^{\frac{m(m-1)}{2}}\mbox{Pf} \left[
\begin{array}{cc}
  \left[D(z_i,z_j)\right]_{i,j=1}^m & \left[S(z_i,z_j)\right]_{i,j=1}^m  \\
 & \\
  \left[-S(z_j,z_i)\right]_{i,j=1}^m  & \left[I(z_i,z_j)\right]_{i,j=1}^m  \\
\end{array}\right]\nonumber
\end{equation}
It remains to rewrite the righthand side of this expression as the
quaternion determinant of $2\times 2$ matrix valued kernel. This
is achieved in two steps. The first step is to observe that
\begin{equation}
\PF\;\bigl[W(x_1,\ldots,x_m; y_1,\ldots, y_m\arrowvert
x_1,\ldots,x_m; y_1,\ldots,
y_m)\biggr]=(-)^{\frac{m(m-1)}{2}}\PF\biggl[w(z_i,z_j)\biggr]_{i,j=1}^m
\end{equation}
where
\begin{equation}
w(z_i,z_j)=\biggl[\begin{array}{cc}
  W(x_i,x_j) & W(x_i,y_j) \\
  W(y_i,x_j) & W(y_i,y_j) \\
\end{array}\biggr]\nonumber
\end{equation}
The second step is to apply the formula
\begin{equation}
\PF\left[ZQ\right]=\mbox{Tdet}\;Q,\;\;\; Z=\left(%
\begin{array}{cccc}
  J & 0 & \ldots & 0 \\
  0 & J & \ldots & 0 \\
  \vdots & & \ddots &  \\
  0 & 0 & \ldots & J \\
\end{array}%
\right)\nonumber
\end{equation}
where $Q$ is a $2\times 2$ block antisymmetric matrix, and $J=\left(%
\begin{array}{cc}
  0 & 1 \\
  -1 & 0 \\
\end{array}%
\right)$
\end{proof}
\begin{rem}
The statement similar to Corollary \ref{COROLLARYMDISC} holds for
the discrete symplectic ensembles as well. The proof is just the
same.
\end{rem}
\section{Continuous Limit}\label{SECTIONCONTINUOUS}
\subsection{Exact formula for unitary ($\beta=2$) ensembles}
Let $\mu$ be a positive measure on $\mathbb{R} $ with
finite moments: $\int_\mathbb{R}|x|^n\mu(dx)<\infty$ for any $n\ge
0$, such that
\begin{equation}
C^{(\beta)}_{N}:=\frac{1}{N!}\int\limits_{\mathbb{R}^N}\prod_{1\le
i<j\le N}|x_i-x_j|^\beta \mu(dx_1)\cdots\mu(dx_N)\ne 0\nonumber
\end{equation}
for any $N\ge 1$ and some (equivalently, all) $\beta>0$. Denote by
$\triangle^{(\beta)}_{N}$ the probability measure on
$\mathbb{R}^N$ given by
\begin{equation}
p_{N}^{(\beta)}(dx_1,\dots ,dx_N)=\frac 1{N!C^{(\beta)}_{N}}
\prod_{1\le i<j\le N}|x_i-x_j|^\beta\,
\mu(dx_1)\otimes\cdots\otimes\mu(dx_N).\nonumber
\end{equation}
For any symmetric function $g:\mathbb{R}^N\to \mathbb{C}$ set
\begin{equation}
\langle g
\rangle_{\triangle_N^{(\beta)}}:=\int\limits_{\mathbb{R}^N}
g(x_1,\dots,x_N)\,p^{(\beta)}_{N}(dx_1,\dots, dx_N)\nonumber
\end{equation}
provided that the integral converges. For $\zeta\in\mathbb{C}$ set
\begin{equation}
D(\zeta)=\prod_{i\ge 1}(\zeta-x_i).\nonumber
\end{equation}
The number of factors in such products will always be clear from
the context. Recall that for finite sets $A=\{a_1,\dots,a_p\}$ and
$B=\{b_1,\dots,b_q\}$ we use the notation $\prod (A;B)$ to denote
the product of pairwise differences
\begin{equation}
\prod (A;B)= \prod_{i=1}^p\prod_{j=1}^q(a_i-b_j),\nonumber
\end{equation}
and $V(A)$ stands for the Vandermonde determinant $\prod_{1\le
i<j\le p}(a_i-a_j)$ (note that we use an ordering of $a_i$'s
here).
\begin{thm}\label{CONTTHEOREMB2}(Case $\beta=2$) For any integers $N\ge 1$
and $S>1-N$, and finite sets of mutually distinct complex numbers
\begin{equation}
\begin{split}
\alpha^-=\{\alpha^-_1,\dots,\alpha^-_{m_1}\},\qquad
\alpha^+=\{\alpha^+_1,\dots,\alpha^-_{k_1}\},\\
\beta^-=\{\beta_1^-,\dots,\beta^-_{m_2}\},\qquad
\beta^+=\{\beta_1^+,\dots,\beta^+_{k_2}\},\nonumber
\end{split}
\end{equation}
with $|\alpha^-|-|\alpha^+|=|\beta^-|-|\beta^+|=S$, such that
\begin{equation}
\alpha^-\cap\alpha^+=\varnothing,\quad
\beta^-\cap\beta^+=\varnothing,\quad \alpha^+\cap
\mathbb{R}=\varnothing,\quad \beta^+\cap
\mathbb{R}=\varnothing,\nonumber
\end{equation}
one has
\begin{equation}\label{MainContB2}
\begin{split}
\left\langle\dfrac{\prod_{i=1}^{m_1}D(\alpha_i^-)\prod_{i=1}^{m_2}
D(\beta_i^-)}{\prod_{j=1}^{k_1}D(\alpha_j^+)\prod_{j=1}^{k_2}D(\beta_j^+)}
\right\rangle_{\triangle_N^{(2)}}=
(-1)^{\frac{1}{2}\left[\left(|\alpha^-|+|\beta^-|\right)^2
+|\beta^-|-|\alpha^-|\right]}\;\\
\times\dfrac{C_{N+S}^{(2)}}{C_N^{(2)}}\;\dfrac{\prod(\alpha^-;\alpha^+)\prod(\beta^-;\beta^+)}{V(\alpha^-)V(\alpha^+)V(\beta^-)V(\beta^+)}
\; \det\left[W_N^{(2)}(\alpha^-,\beta^+\vert
\beta^-,\alpha^+)\right].
\end{split}
\end{equation}
Here $W_N^{(2)}(\alpha^-,\beta^+\vert \beta^-,\alpha^+)$ is a
matrix with rows parameterized by elements of $\alpha^-$ and
$\beta^+$, columns parameterized by elements of $\beta^-$ and
$\alpha^+$, and with matrix elements
\begin{equation}
\begin{split}
 &W_N^{(2)}(\alpha_i^-,\beta_j^-)=\dfrac{C^{(2)}_{N+S-1}}{C^{(2)}_{N+S}}\;
  \left\langle
    D(\alpha_i^-)D(\beta_j^-)\right\rangle_{\triangle_{N+S-1}^{(2)}}\\
&W_N^{(2)}(\alpha_i^-,\alpha_j^+)=\dfrac{1}{\alpha_i^--\alpha_j^+}\;
\left\langle\dfrac{D(\alpha_i^-)}{D(\alpha_j^+)}\right\rangle_{\triangle_{N+S}^{(2)}}
\\
&W_N^{(2)}(\beta_i^+,\beta_j^-)=\dfrac{1}{\beta_i^+-\beta_j^-}\;
\left\langle\dfrac{D(\beta_j^-)}{D(\beta_i^+)}\right\rangle_{\triangle_{N+S}^{(2)}}
\\
&W_N^{(2)}(\beta_i^+,\beta_j^+)=\dfrac{C^{(2)}_{N+S+1}}{C^{(2)}_{N+S}}\;
\left\langle\dfrac{1}{D(\beta_i^+)D(\beta_j^+)}\right\rangle_{\triangle_{N+S+1}^{(2)}}
\end{split}\nonumber
\end{equation}
\end{thm}
\begin{proof}
 Since all the moments of $\mu$ are finite, and we are
averaging functions of at most polynomial growth, it suffices to
prove the theorem for the measures with compact support. Clearly,
formula (\ref{MainContB2}) remains intact under the operation of
taking weak limits of measures $\mu$ having the same compact
support. On the other hand, formula (\ref{MainContB2}) was already
proved in Theorem \ref{THEOREMAvergesCharacteristicPolynomials}
for all measures $\mu$ supported by a (large enough) finite set of
points. Since any compactly supported probability measure can be
weakly approximated by finitely supported ones, the proof is
complete.
\end{proof}
\subsection{Exact formulae for orthogonal $\beta=1$ and symplectic $\beta=4$
ensembles}
\begin{thm}\label{CONTTHEOREMB1B4}(Cases $\beta=1$ and $\beta=4$) (i) For any
integers $N\ge 1$ and $S>1-N$, and finite sets of mutually
distinct complex numbers
\begin{equation}\label{ALPHABETA}
\alpha=\{\alpha_1,\dots,\alpha_{k}\},\qquad
\beta=\{\beta_1,\dots,\beta_{m}\},\qquad k-m=2S,
\end{equation}
such that $\beta\cap\mathbb{R}=\varnothing$, one has
\begin{equation}\label{MaincontB1}
\biggl\langle\dfrac{\prod_{i=1}^k D(\alpha_i)}{\prod_{i=1}^m
D(\beta_i)}\biggr\rangle_{\triangle_{2N}^{(1)}}
=\dfrac{C_{2N+2S}^{(1)}}{C_{2N}^{(1)}}\;
\dfrac{\prod(\alpha;\beta)}{V(\alpha)V(\beta)}\PFAFF\;
\biggl[W_N^{(1)}(\alpha,\beta\arrowvert\alpha,\beta)\biggr]
\end{equation}
where $W_N^{(1)}$ is a skew-symmetric $(k+m)\times(k+m)$ matrix
with rows and columns parameterized by elements of $\alpha$ and
$\beta$, and with matrix elements given by
\begin{equation}
\begin{split}
&W^{(1)}_N(\alpha_i,\alpha_j)
=\frac{C_{2N+2S-2}^{(1)}}{C_{2N+2S}^{(1)}}\;(\alpha_i-\alpha_j)\left\langle
D(\alpha_i)D(\alpha_j)\right\rangle_{\T_{2N+2S-2}^{(1)}}
\\
&W^{(1)}_N(\alpha_i,\beta_j)=\dfrac{1}{\alpha_i-\beta_j}\left\langle
\dfrac{D(\alpha_i)}{D(\beta_j)}\right\rangle_{\T_{2N+2S}^{(1)}}
\\
&W^{(1)}_N(\beta_i,\beta_j)=\frac{C_{2N+2S+2}^{(1)}}{C_{2N+2S}^{(1)}}\;
(\beta_i-\beta_j)\left\langle
\dfrac{1}{D(\beta_i)D(\beta_j)}\right\rangle_{\T_{2N+2S+2}^{(1)}}
\end{split}\nonumber
\end{equation}
(ii) For any integers $N\ge 1$ and $S>1-N$, and finite sets of
mutually distinct complex numbers (\ref{ALPHABETA}) such that
$\beta\cap \mathbb{R}=\varnothing$, one has
\begin{equation}\label{MainContB4}
\biggl\langle\dfrac{\prod_{i=1}^k D^2(\alpha_i)}{\prod_{i=1}^m
D^2(\beta_i)}\biggr\rangle_{\T_{N}^{(4)}}
=\dfrac{C_{N+S}^{(4)}}{C_{N}^{(4)}}\;
\dfrac{\prod(\alpha;\beta)}{V(\alpha)V(\beta)}\PFAFF\;
\biggl[W_N^{(4)}(\alpha,\beta\arrowvert\alpha,\beta)\biggr]
\end{equation}
where $W_N^{(4)}$ is a skew-symmetric matrix with rows and columns
parameterized by elements of $\alpha$ and $\beta$, and with matrix
elements given by
\begin{equation}
\begin{split}
&W^{(4)}_N(\alpha_i,\alpha_j)=\frac{C^{(4)}_{N+S-1}}{C^{(4)}_{N+S}}\;
(\alpha_i-\alpha_j)\left\langle
D^2(\alpha_i)D^2(\alpha_j)\right\rangle_{\T_{N+S-1}^{(4)}}
\\
&W^{(4)}_N(\alpha_i,\beta_j)=\dfrac{1}{\alpha_i-\beta_j}\left\langle
\dfrac{D^2(\alpha_i)}{D^2(\beta_j)}\right\rangle_{\T_{N+S}^{(4)}}
\\
&W^{(4)}_N(\beta_i,\beta_j)=\frac{C_{N+S+1}^{(4)}}{C_{N+S}^{(4)}}\;
(\beta_i-\beta_j)\left\langle \dfrac{1}{D^2(\beta_i)D^2(\beta_j)}
\right\rangle_{\T_{N+S+1}^{(4)}}
\end{split}\nonumber
\end{equation}
\end{thm}
\begin{proof} Arguing as in the proof of Theorem \ref{CONTTHEOREMB2} we see that it
suffices to prove (\ref{MaincontB1}) and (\ref{MainContB4}) for
compactly supported measures $\mu$ with continuous density with
respect to the Lebesgue measure:
\begin{equation}
\mu(dx)=F(x)dx, \quad F\in C(\mathbb{R}), \quad
\operatorname{supp}(F)\subset [-A,A],\quad A\in
\mathbb{R}_+.\nonumber
\end{equation}
Let us consider the case $\beta=1$ first. The idea is to reduce
the statement to Theorem
\ref{THEOREMAVERAGESCHARACTERISTICPOLYNOMIALSORTHOGON}. Split the
segment $[-A,A]$ into $M>2N$ intervals of length $2A/M$:
\begin{equation}
[-A,A]=I_1\cup\dots \cup I_M,\qquad
I_k=\left[-A+\tfrac{2(k-1)A}{M}\,,-A+\tfrac{2kA}{M}\right],\nonumber
\end{equation}
and take a discrete set $\X$ with $2M$ points such that each of
the intervals $I_k$ contains exactly two points of $\X$. One may
take for example
\begin{equation}
\X=\left\{-A+\tfrac A{2M}\,, -A+\tfrac {3A}{2M}\,; -A+\tfrac
{5A}{2M}\,, -A+\tfrac {7A}{2M}\,;\dots;A-\tfrac {3A}{2M}\,,
A-\tfrac {A}{2M}\right\}.\nonumber
\end{equation}
Then for any choice of $2N$ intervals $I_{k_j}$, $j=1,\dots,2N$,
with $k_1<k_2<\dots<k_{2N}$, there exists a unique way to choose a
point of $\X$ in each of these intervals so that the chosen $2N$
points of $\X$ form a point configuration from
$Conf_{2N}^{(1)}(\X)$. Conversely, any element of
$Conf_{2N}^{(1)}(\X)$ produces $2N$ intervals $I_{k_j}$ which
contain the points of the configuration. Thus, for any continuous
function $g:[-A,A]^{2N}\to \mathbb{C}$,
\begin{equation}
\left(\tfrac M{2A}\right)^{2N}\sum_{(x_1<\dots<x_{2N})\in
Conf_{2N}^{(1)}(\X)}g(x_1,\dots,x_{2N})\nonumber
\end{equation}
can be viewed as a Riemannian sum for the integral
\begin{equation}
\int\limits_{-A<x_1<\dots<x_{2N}<A}g(x_1,\dots,x_{2N})dx_1\cdots
dx_{2N}.\nonumber
\end{equation}
Taking $M$ to $\infty$ brings the diameter of the partition
$[-A,A]=I_1\cup\dots\cup I_{M}$ to zero, and formula
(\ref{MaincontB1}) directly follows from Theorem
\ref{THEOREMAVERAGESCHARACTERISTICPOLYNOMIALSORTHOGON}. The proof
of (i) is complete.

Let consider the case $\beta=4$ now. We want to derive
(\ref{MainContB4}) from Theorem \ref{B4DISCRETEMAINTHEOREM}. Take
$M>2N$ and
\begin{equation}
\X=\left\{-A+\frac{(2k-1)A}M\right\}_{k=1}^M\subset [-A,A].
\nonumber
\end{equation}
Then for any continuous function $g:[-A,A]^{N}\to \mathbb{C}$,
\begin{equation}
\begin{split}
\sum_{X=({}^lx_1<x_1<\dots<{}^lx_{N}<x_{N})\in
Conf_{2N}^{(4)}(\X)}g(x_1,\dots,x_{N})\,|V(X)| =\left(\tfrac
M{2A}\right)^{N}\qquad\qquad\qquad\\
\times\underset{-A+\frac {A}{M}<x_1<\dots<x_{N}}{\sum_{
x_i\in\X}}g(x_1,\dots,x_{N})\,\prod_{1\le i<j\le N}
(x_i-x_j)^2\left(x_i-x_j-\tfrac {2A}M\right)\left(x_i-x_j+\tfrac
{2A}M\right)\\=\int\limits_{-A<x_1<\dots<x_{N}<A}g(x_1,\dots,x_{N})\prod_{1\le
i<j\le N} (x_i-x_j)^4 \,dx_1\cdots dx_N+O\left(M^{-1}\right).
\end{split}\nonumber
\end{equation}
Thus, Theorem \ref{B4DISCRETEMAINTHEOREM} implies
(\ref{MainContB4}).
\end{proof}
\begin{rem}The conditions imposed on the measure $\mu$ at
the beginning of this section can be relaxed. Namely, instead of
finiteness of all the moments it suffices to require the
convergence of the averages involved in (\ref{MaincontB1}) and
(\ref{MainContB4}), and instead of requiring all $C_N^{(\beta)}$
to be nonzero it also suffices to require that only of the
constants involved in these formulas. This allows $\mu$ to have
finite support.
\end{rem}
\begin{rem}
The statements of Theorems \ref{CONTTHEOREMB2} and
\ref{CONTTHEOREMB1B4} admit an interpretation in terms of random
matrices. Indeed, the measures $\T_N^{(2)}$, $\T_{2N}^{(1)}$, and
$\T_{N}^{(4)}$ are radial parts (i.e., projections onto
(different) eigenvalues) of the measures
\begin{equation}
P(dH)=\operatorname{const}\cdot \exp(-Q(H))\,dH,\nonumber
\end{equation}
where $Q(x)$ is an even degree polynomial with positive highest
coefficient, and $H$ belongs to the linear space of $N\times N$
Hermitian, $2N\times 2N$ real symmetric, or $N\times N$ quaternion
real Hermitian matrices, respectively. This corresponds to our
measure $\mu$ being equal $\mu(dx)=\exp(-Q(x))dx$ for $\beta=1$
and $2$, and $\mu(dx)=\exp(-2Q(x))dx$ for $\beta=4$. In the cases
$\beta=1,2$ the characteristic polynomial of the random matrix
$H$, $\det(\zeta-H)$, is exactly our product-function $D(\zeta)$,
while for $\beta=4$ we have $\det(\zeta-H)=D^2(\zeta)$. Thus, in
Theorems 1,2 we deal with average values of products and ratios of
characteristic polynomials with respect to the measures on
matrices introduced above. In the case $Q(x)=\const x^2$ these
matrix probability spaces are called {\it Gaussian Unitary
($\beta=2$), Orthogonal ($\beta=1$), and Symplectic ($\beta=4$)
Ensembles} (GUE, GOE, GSE, for short).
\end{rem}
\begin{rem}
Formulas (\ref{MainContB2}), (\ref{MaincontB1}),
(\ref{MainContB4}) require that the total number of factors in the
left-hand side is even (we count factors in both the numerator and
the denominator). These formulas can be easily extended to the
situation when the total number of factors is odd. In order to do
that, one needs to take the even case and send one of the
parameters $\alpha_i^\pm,\beta_j^\pm$ in (\ref{MainContB2}) and
$\alpha_i,\beta_j$ in (\ref{MaincontB1}), (\ref{MainContB4}) to
$\infty$. Given that $D(\zeta)\sim\zeta^N$ in $\T_N^{(\beta)}$,
the limiting formulas are readily obtained. After the limit
transition, some of the two-point averages in the matrix elements
of $W_N^{(\beta)}$ in the right-hand side turn into one-point
averages of the form $\langle D(\gamma)\rangle, \langle
1/D(\gamma)\rangle$ in the case $\beta=1,2$, or $\langle
D^2(\gamma)\rangle$, $\langle 1/D^2(\gamma)\rangle$ in the case
$\beta=4$.
\end{rem}
\subsection{Correlation functions}\label{SectionCorrCont}
Theorems \ref{CONTTHEOREMB2} and \ref{CONTTHEOREMB1B4} allow to
compute the correlation functions of $\T_N^{(\beta)}$ for
$\beta=1,2,4$ and reproduce fundamental results of the Random
Matrix Theory.

Define the {\it $n$th correlation measure} of $\T_N^{(\beta)}$ by
$$ \rho_{n,N}^{(\beta)}(dx_1,\dots,dx_n)=\frac{N!}{(N-n)!}
\int\limits_{x_{n+1},x_{n+2},\dots,x_N}p_N^{(\beta)}
(dx_1,\dots,dx_N).
$$
For a function $f(z)$ of a complex variable $\zeta$, which is
continuous in both half-planes $\Img \zeta>0$ and $\Img \zeta<0$
up to the real axis, we will denote by $[f(\zeta)]_{\zeta=x}$,
$x\in \mathbb{R}$, the difference of the limit values of
$f(\zeta)$ as $\zeta\to x$  from bottom and from top divided by
$2\pi i$:
$$
[f(\zeta)]_x=\frac 1{2\pi i}\,(f(x-i0)-f(x+i0)).
$$
\begin{thm}\label{CONTTHEOREMCORRELATIONFUNCTIONS}
 Take any $n\ge 1$ and assume that near $n$
points $x_1,\dots,x_n\in\mathbb{R}$ the measure $\mu$ is
absolutely continuous with respect to the Lebesgue measure, and
its density their is uniformly H\"older continuous. Then the $n$th
correlation measure of $\T_N^{(\beta)}$ (or $\T_{2N}^{(\beta)}$
for $\beta=1$) has a continuous density near $(x_1,\dots,x_n)$
which is given by \\
$\bullet$ For $\beta=2$
$$
\rho_{n,N}^{(2)}(x_1,\dots,x_n)=\det\;[K^{(2)}_N(x_i,x_j)]_{i,j=1}^n
$$
where for $x\ne y$ the kernel is given by
$$
K^{(2)}_N(x,y)=
\frac{1}{x-y}\left[\left\langle\frac{D(x)}{D(\zeta)}\right\rangle_{\T_N^{(2)}}
\right]_{\zeta=y},
$$
and for $x=y$ the kernel is defined by continuity. \\
$\bullet$ For $\beta=1$
$$
\rho_{n,2N}^{(1)}(x_1,\dots,x_n)=\PFAFF\;[K^{(1)}_{2N}(x_i,x_j)]_{i,j=1}^n
$$
where the skew-symmetric $2\times 2$ matrix kernel for $x\ne y$ is
given by
$$
 K_{2N}^{(1)}(x,y)=\\
 \bmatrix
\frac{C_{2N-2}^{(1)}}{C_{2N}^{(1)}}\,(x-y)\left\langle
D(x)D(y)\right\rangle_{\T_{2N-2}^{(1)}}&
\frac{1}{x-y}\left[\left\langle
\frac{D(x)}{D(\zeta)}\right\rangle_{\T_{2N}^{(1)}}\right]_{\zeta=y}
\\
-\frac{1}{x-y}\left[\left\langle
\frac{D(x)}{D(\zeta)}\right\rangle_{\T_{2N}^{(1)}}\right]_{\zeta=y}&
\frac{C_{2N+2}^{(1)}}{C_{2N}^{(1)}}\,(x-y) \left[\left\langle
\frac{1}{D(\zeta)D(\eta)}\right\rangle_{\T_{2N+2}^{(1)}}\right]_{\zeta=x,\,\eta=y}
\endbmatrix
$$
and for $x=y$ the kernel is defined by continuity.\\
 $\bullet$ For
$\beta=4$
$$
\rho_{n,N}^{(4)}(x_1,\dots,x_n)=\PFAFF\;[K^{(4)}_{N}(x_i,x_j)]_{i,j=1}^n
$$
where the skew-symmetric $2\times 2$ matrix kernel for $x\ne y$ is
given by
$$
 K_{N}^{(4)}(x,y)=\\
\bmatrix
 \frac{C^{(4)}_{N-1}}{2C^{(4)}_{N}}\; (x-y)\left\langle
D^2(x)D^2(y)\right\rangle_{\T_{N+1}^{(4)}}&
\frac{1}{2(x-y)}\left[\left\langle
\frac{D^2(x)}{D^2(\zeta)}\right\rangle_{\T_{N}^{(4)}}\right]_{\zeta=y}\\
-\frac{1}{2(x-y)}\left[\left\langle
\frac{D^2(x)}{D^2(\zeta)}\right\rangle_{\T_{N}^{(4)}}\right]_{\zeta=y}&
\frac{C_{N+1}^{(4)}}{2C_{N}^{(4)}}\; (x-y)\left[\left\langle
\frac{1}{D^2(\zeta)D^2(\eta)}
\right\rangle_{\T_{N-1}^{(4)}}\right]_{\zeta=x,\,\eta=y}
\endbmatrix
$$
and for $x=y$ the kernel is defined by continuity.
\end{thm}
\begin{proof} The Sokhotski-Plemelj formula implies that
$$
\rho_{n,N}^{(\beta)}(x_1,\dots,x_n)=\left[\frac{\partial^n}{\partial
u_1\cdots\partial u_n}\Biggl|_{u=v}\left\langle\frac{D(u_1)\cdots
D(u_n)}{D(v_1)\cdots
D(v_n)}\right\rangle_{\T_N^{(\beta)}}\right]_{v_1=x_1,\dots,v_n=x_n}
$$
$$
\rho_{n,N}^{(\beta)}(x_1,\dots,x_n)=\frac{1}{2^n}\left[\frac{\partial^n}{\partial
u_1\cdots\partial
u_n}\Biggl|_{u=v}\left\langle\frac{D^2(u_1)\cdots
D^2(u_n)}{D^2(v_1)\cdots
D^2(v_n)}\right\rangle_{\T_N^{(\beta)}}\right]_{v_1=x_1,\dots,v_n=x_n}
$$
We use the first formula for $\beta=1,2$ and the second one for
$\beta=4$, and then proceed similarly to Propositions
\ref{KernelAsAnaveragesOftwocharacteristicpolynomials} and
\ref{PPropositionmpointfor1}.
\end{proof}
\begin{rem} If instead of asking for the measure $\mu$ to
have nice density near $x_1,\dots,x_n$, we require that $\mu$ is
purely atomic near these points, then the formulas of Theorem
\ref{CONTTHEOREMCORRELATIONFUNCTIONS} will continue to hold if we
understand the symbol $[f(\zeta)]_x$ as the residue of the
meromorphic function $f(\zeta)$ at the point $x$. The proof is
very similar.
\end{rem}
The one- and two-point averages
$$
\left\langle D(x)\right\rangle,\quad \left\langle \frac
1{D(x)}\right\rangle,\qquad \left\langle
D(x)D(y)\right\rangle,\quad \left\langle
\frac{D(x)}{D(y)}\right\rangle,\quad \left\langle \frac
1{D(x)D(y)}\right\rangle
$$
for $\beta=1,2$, and similar expressions with $D^2$ instead of $D$
for $\beta=4$, can all be evaluated in terms of (skew)-orthogonal
polynomials associated with the measure $\mu$.

For $\beta=2$ one needs the usual orthogonal polynomials
associated with $\mu$. The computation has been done by Strahov
and Fyodorov \cite{strahov}, see also \cite{baik}.

For $\beta=1$ and $\beta=4$ the skew-orthogonal polynomials
associated to the skew-symmetric inner products
$$
\langle f,g\rangle_1 =\frac
12\iint_{\mathbb{R}^2}\operatorname{sgn}(x-y) f(x)g(y) \mu(dx)\mu(dy),\\
\langle f,g\rangle_4= \frac 12\int_{\mathbb{R}}
(f(x)g'(x)-f'(x)g(x))\mu(dx)
$$
need to be constructed. The expression of the one- and two-point
functions in terms of these polynomials are exactly the same as in
$\beta=2$ case  modulo the change of the inner product. Since the
proofs are essentially the same, we omit them here.

If one substitutes the expressions in terms of the
(skew)-orthogonal polynomials into the formulas for the
correlation kernels in Theorem
\ref{CONTTHEOREMCORRELATIONFUNCTIONS}, one recovers the classical
formulas of Random Matrix Theory, see e.g. \cite{tracy}.
\section{Asymptotics }\label{SectionAsymptotics}
This Section presents the asymptotic analysis of kernels and
correlation functions of characteristic polynomials. The three
cases are considered: the case of the Gaussian Unitary ensemble
(GUE), the case of the Gaussian Orthogonal Ensemble (GOE), and the
case of the Gaussian Symplectic Ensemble (GSE).
\subsection{Statement of Asymptotic
Results}\label{SectionStatementofAsymptoticResults}
\subsubsection{The Gaussian Unitary Ensemble} In the notation of Section \ref{SECTIONCONTINUOUS}, we consider the case of $\beta=2$,
$\mu(dx)=e^{-x^2}dx$. The computations of correlation functions
for GUE involve monic orthogonal polynomials defined by the
following  inner product:
$$
\left<f,g\right>_2=\int\limits_{-\infty}^{+\infty}dx
e^{-x^2}f(x)g(x),
$$
and the relevant family of (monic)  orthogonal polynomials
$\{\pi_n\}$, $n=0,1,2,\ldots $, has the property
$$
\left<\pi_l,\pi_k\right>_2=h_k^{(2)}\delta_{kl}.
$$
\begin{rem}The inner products as well as the averages
over ensembles are denoted by the same symbol $\left<\right>$
throughout the paper. These notations should not lead to a
confusion.
\end{rem}
In Section \ref{SECTIONCONTINUOUS} we have found that averages of
products
 and ratios of characteristic polynomials can be expressed in
 terms of kernels. In the case of GUE these kernels are
\begin{itemize}
  \item  $
W_{N,I}^{(2)}(\zeta,\eta)=\dfrac{1}{h_N^{(2)}}\;
(\zeta-\eta)\left<D(\zeta)D(\eta)\right>_{\triangle_{N}^{(2)}}
\;\;\; \zeta\in\;\mathbb{C},\eta\in\;\mathbb{C}.$
  \item $
W_{N,II}^{(2)}(\zeta,\eta)= \dfrac{1}{\zeta-
\eta}\;\biggl<\dfrac{D(\zeta)}{D(\eta)
}\biggr>_{\triangle_{N}^{(2)}},\;\;\;
\zeta\in\;\mathbb{C},\eta\in\;\mathbb{C}\backslash\mathbb{R}. $
  \item
  $W_{N,III}^{(2)}(\zeta,\eta)=
  h^{(2)}_{N-1}(\zeta-\eta)\left<\dfrac{1}{D(\zeta)D(\eta)}\right>_{\triangle_{N}^{(2)}}
\; \zeta,\eta\in\;\mathbb{C}\backslash\mathbb{R} $
\end{itemize}
The following statement was proved in Ref. \cite{strahov}.
\begin{thm}\label{TEOREMA0}(Scaling limits of kernels at the origin of the
spectrum). The following limit relations hold:
\begin{equation}
\begin{split}
\underset{N\rightarrow\infty}{\lim}\dfrac{1}{\sqrt{2N}}&
W^{(2)}_{N,I}(\dfrac{\zeta}{\sqrt{2N}},\dfrac{\eta}{\sqrt{2N}})=
\mathbb{S}_{GUE}^{I}(\zeta,\eta),\;\;\;\;
\underset{N\rightarrow\infty}{\lim}\dfrac{1}{\sqrt{2N}}W^{(2)}_{N,II}(\dfrac{\zeta}{\sqrt{2N}},\dfrac{\eta}{\sqrt{2N}})=
\mathbb{S}_{GUE}^{II}(\zeta,\eta),\\
&\underset{N\rightarrow\infty}{\lim}\dfrac{1}{\sqrt{2N}}W^{(2)}_{N,III}(\dfrac{\zeta}{\sqrt{2N}},\dfrac{\eta}{\sqrt{2N}})=
\mathbb{S}_{GUE}^{III}(\zeta,\eta).
\end{split}\nonumber
\end{equation}
where
\begin{itemize}
  \item  $
\mathbb{S}_{GUE}^{I}(\zeta,\eta)=
\dfrac{1}{\pi}\biggl[\dfrac{\sin(\zeta-\eta)}{\zeta-\eta}\biggr],
\;\;\; \zeta\in\;\mathbb{C},\;\;\eta\in\;\mathbb{C}.$
  \item $\mathbb{S}_{GUE}^{II}(\zeta,\eta)=
  \begin{cases}
    \dfrac{e^{i(\eta-\zeta)}}{\zeta-\eta} & \Img \eta >0, \\
    \dfrac{e^{-i(\eta-\zeta)}}{\zeta-\eta} & \Img \eta<0.
  \end{cases}
,\;\;\;
\zeta\in\;\mathbb{C},\;\;\eta\in\;\mathbb{C}\backslash\mathbb{R}.$
  \item
  $\mathbb{S}_{GUE}^{III}(\zeta,\eta)=2\pi i
  \begin{cases}
    \quad\dfrac{e^{i(\zeta-\eta)}}{\zeta-\eta} & \Img \zeta >0, \Img \eta <0 \\
    -\dfrac{e^{-i(\zeta-\eta)}}{\zeta-\eta} & \Img \zeta <0, \Img \eta >0 \\
    \quad 0 & \text{otherwise}.
  \end{cases}
,\;\;\;\; \zeta, \eta\in\;\mathbb{C}\backslash\mathbb{R}.$
\end{itemize}
\end{thm}
\begin{thm} For any integers $N\ge 1$ and $S>1-N$, and finite sets
of mutually distinct complex numbers
\begin{equation}
\begin{split}
\alpha^-=\{\alpha^-_1,\dots,\alpha^-_{m_1}\},\qquad
\alpha^+=\{\alpha^+_1,\dots,\alpha^-_{k_1}\},\\
\beta^-=\{\beta_1^-,\dots,\beta^-_{m_2}\},\qquad
\beta^+=\{\beta_1^+,\dots,\beta^+_{k_2}\},\nonumber
\end{split}
\end{equation}
with $|\alpha^-|-|\alpha^+|=|\beta^-|-|\beta^+|=S$ and
$$
\alpha^-\cap\alpha^+=\varnothing,\quad \alpha^+\cap
\mathbb{R}=\beta^+\cap \mathbb{R}=\varnothing,\quad
\beta^-\cap\beta^+=\varnothing,\quad
$$
one has
\begin{equation}
\begin{split}
&\underset{N\rightarrow\infty}{\lim}\left\{\dfrac{1}{\mathrm{T}^{(2)}(N)}
\left<\dfrac{\prod_{i=1}^{m_1}D(\tfrac{\alpha_i^{-}}{\sqrt{2N}})\prod_{i=1}^{m_2}
D(\tfrac{\beta_i^{-}}{\sqrt{2N}})}{\prod_{j=1}^{k_1}D(\tfrac{\alpha_i^{+}}{\sqrt{2N}})\prod_{j=1}^{k_2}
D(\tfrac{\beta_j^{+}}{\sqrt{2N}})}\right>_{\triangle_N^{(2)}}
\right\}\\
&=(-1)^{\frac{(|\beta^-|+|\alpha^-|)^2+(|\beta^-|-|\alpha^-|)}{2}}
\dfrac{\prod(\alpha^-;\alpha^+)\prod(\beta^-;\beta^+)}{V(\alpha^-)V(\alpha^+)V(\beta^-)V(\beta^+)}
\det\left[\mathbb{S}_{GUE}(\alpha^-,\beta^+\vert
\beta^-,\alpha^+)\right],
\end{split}\nonumber
\end{equation}
where
$$
\dfrac{1}{\mathrm{T}^{(2)}(N)}=\dfrac{C_N^{(2)}}{C_{N+S}^{(2)}}\;
(2N)^{\tfrac{S^2}{2}},
$$
$\mathbb{S}_{GUE}(\alpha^-,\beta^+\vert \beta^-,\alpha^+)$ is a
matrix with rows parameterized by elements of $\alpha^{-}$ and
$\beta^{+}$, columns parameterized by $\beta^{-},\alpha^{+}$, and
with matrix elements:
\begin{itemize}
  \item  $
\mathbb{S}_{GUE}(\alpha_i^{-},\beta_j^{-})=\mathbb{S}_{GUE}^{I}(\alpha_i^{-},\beta_j^{-})$
  \item $\mathbb{S}_{GUE}(\alpha_i^{-},\alpha_j^{+})=\mathbb{S}_{GUE}^{II}(\alpha_i^{-},\alpha_j^{+})
  $
  \item $\mathbb{S}_{GUE}(\beta_i^{+},\beta_j^{-})=\mathbb{S}_{GUE}^{II}(\beta_j^{-},\beta_i^{+})
  $
  \item
  $\mathbb{S}_{GUE}(\beta_i^{+},\alpha_j^{+})=\mathbb{S}_{GUE}^{III}(\beta_i^{+},\alpha_j^{+})
  $
\end{itemize}
\end{thm}
The constant $C_{N}^{(2)}$ can be explicitly computed, see Mehta
\cite{mehta}:
\begin{equation}
C_{N}^{(2)}=\pi^{\tfrac{N}{2}}2^{-\tfrac{N(N-1)}{2}}\cdot\prod\limits_{j=1}^{N}j!
\nonumber
\end{equation}
\subsubsection{The Gaussian Orthogonal Ensemble}
Let us now consider the case $\beta=1$,
$\mu(dx)=e^{-\frac{x^2}2}dx$. The computations of correlation
functions for GOE involve skew orthogonal polynomials. They are
defined by the following skew symmetric inner product:
\begin{equation}
\left<f,g\right>_1=\dfrac{1}{2}\int\limits_{-\infty}^{+\infty}dx
e^{-\frac{x^2}{2}}\int\limits_{-\infty}^{+\infty}dy
e^{-\frac{y^2}{2}} \sgn(y-x)f(x)g(y).\nonumber
\end{equation}
Specifically, the relevant family of (monic) skew orthogonal
polynomials $\{p_n\}$, $n=0,1,2,\ldots $, is given by
\begin{equation}
\left<p_{2k},p_{2l+1}\right>_1=-\left<p_{2l+1},p_{2k}\right>_1=h_k^{(1)}\delta_{kl}.\nonumber
\end{equation}
and by the condition that all other skew symmetric products
between these polynomials are zeros.

 The averages of products
 and ratios of characteristic polynomials can be expressed in
 terms of kernels. In the case of GOE these kernels are
\begin{itemize}
  \item  $
W_{N,I}^{(1)}(\zeta,\eta)=\dfrac{1}{h_N^{(1)}}\;
(\zeta-\eta)\left<d(\zeta)d(\eta)\right>_{\triangle_{2N}^{(1)}}
\;\;\; \zeta\in\;\mathbb{C},\eta\in\;\mathbb{C}.$
  \item $
W_{N,II}^{(1)}(\zeta,\eta)= \dfrac{1}{\zeta-
\eta}\;\biggl<\dfrac{d(\zeta)}{d(\eta)
}\biggr>_{\triangle_{2N}^{(1)}},\;\;\;
\zeta\in\;\mathbb{C},\eta\in\;\mathbb{C}\backslash\mathbb{R}. $
  \item
  $W_{N,III}^{(1)}(\zeta,\eta)=
  h^{(1)}_{N-1}(\zeta-\eta)\left<\dfrac{1}{d(\zeta)d(\eta)}\right>_{\triangle_{2N}^{(1)}}
\;\;\; \zeta,\eta\in\;\mathbb{C}\backslash\mathbb{R}. $
\end{itemize}
The kernels $W_{N,I}^{(1)}(\zeta,\eta)$,
$W_{N,II}^{(1)}(\zeta,\eta)$, and $W_{N,III}^{(1)}(\zeta,\eta)$
admit representations as Christoffel-Darboux type sums. Namely,
\begin{equation}\label{WNIASCAUCHYDARBOUX}
W_{N,I}^{(1)}(\zeta,\eta)=\sum\limits_{i=0}^N
\dfrac{p_{2i+1}(\zeta)p_{2i}(\eta)-p_{2i}(\zeta)p_{2i+1}(\eta)}{h_i^{(1)}},
\end{equation}
\begin{equation}\label{WNIIASCAUCHYDARBOUX}
W_{N,II}^{(1)}(\zeta,\eta)= \dfrac{1}{\zeta-\eta}+
\sum\limits_{i=0}^{N-1}
\dfrac{p_{2i+1}(\zeta)h_{2i}(\eta)-p_{2i}(\zeta)h_{2i+1}(\eta)}{h_i^{(1)}},
\end{equation}
\begin{equation}\label{WNIIIASCAUCHYDARBOUX}
 W_{N,III}^{(1)}(\zeta,\eta)=\sum\limits_{i=0}^{N-2}\dfrac{
 h_{2i+1}(\zeta)h_{2i}(\eta)-h_{2i}(\zeta)h_{2i+1}(\eta)}{h_i^{(1)}}
 +\left<R_{\zeta},R_{\eta}\right>_1.
\end{equation}
Here $R_{\zeta}\equiv (\zeta-x)^{-1}$, and the functions $h_k$ are
the skew symmetric products between skew  orthogonal polynomials
and $R_{\zeta}$:
\begin{equation}\label{DefinitionOfhk}
h_k(\eta)=\left<p_k,R_{\eta}\right>_1,\;\;\;\eta\in\;\mathbb{C}\backslash\mathbb{R}.
\end{equation}
\begin{thm}\label{TEOREMA1}(Scaling limits of kernels at the origin of the
spectrum). The following limit relations hold:
\begin{equation}
\begin{split}
\underset{N\rightarrow\infty}{\lim}\dfrac{1}{2N}&
W^{(1)}_{N,I}(\dfrac{\zeta}{\sqrt{2N}},\dfrac{\eta}{\sqrt{2N}})=
\mathbb{S}_{GOE}^{I}(\zeta,\eta),\;\;\;\;
\underset{N\rightarrow\infty}{\lim}\dfrac{1}{\sqrt{2N}}W^{(1)}_{N,II}(\dfrac{\zeta}{\sqrt{2N}},\dfrac{\eta}{\sqrt{2N}})=
\mathbb{S}_{GOE}^{II}(\zeta,\eta),\\
&\underset{N\rightarrow\infty}{\lim}W^{(1)}_{N,III}(\dfrac{\zeta}{\sqrt{2N}},\dfrac{\eta}{\sqrt{2N}})=
\mathbb{S}_{GOE}^{III}(\zeta,\eta).
\end{split}\nonumber
\end{equation}
where
\begin{itemize}
  \item  $
\mathbb{S}_{GOE}^{I}(\zeta,\eta)=
-\dfrac{1}{\pi}\dfrac{d}{d\zeta}\biggl[\dfrac{\sin(\zeta-\eta)}{\zeta-\eta}\biggr],
\;\;\; \zeta\in\;\mathbb{C},\;\;\eta\in\;\mathbb{C}.$
  \item $\mathbb{S}_{GOE}^{II}(\zeta,\eta)=
  \begin{cases}
    \dfrac{e^{i(\eta-\zeta)}}{\zeta-\eta} & \Img \eta >0, \\
    \dfrac{e^{-i(\eta-\zeta)}}{\zeta-\eta} & \Img \eta<0.
  \end{cases}
,\;\;\;
\zeta\in\;\mathbb{C},\;\;\eta\in\;\mathbb{C}\backslash\mathbb{R}.$
  \item
  $\mathbb{S}_{GOE}^{III}(\zeta,\eta)=2\pi i
  \begin{cases}
    \quad\int\limits_{1}^{+\infty}\dfrac{e^{i(\zeta-\eta)t}dt}{t} & \Img \zeta >0, \Img \eta <0 \\
    -\int\limits_{1}^{+\infty}\dfrac{e^{-i(\zeta-\eta)t}dt}{t} & \Img \zeta <0, \Img \eta >0 \\
    \quad 0 & \text{otherwise}.
  \end{cases}
,\;\zeta, \eta\in\;\mathbb{C}\backslash\mathbb{R}.$
\end{itemize}
\end{thm}
\begin{thm}\label{THEOREMSCALINGB1}For any integers $N\geq 1$ and $S>1-N$, and finite sets
of mutually distinct complex numbers
$$
\alpha=\{\alpha_1,\ldots,\alpha_k\},\;\;\;\beta=\{\beta_1,\ldots
,\beta_m \}, \;\; k-m=2S
$$
such that
$$\beta\cap \mathbb{R}=\varnothing
$$
 one has
\begin{equation}
\begin{split}
\underset{N\rightarrow\infty}{\lim}&
\left\{\dfrac{1}{\mathrm{T}^{(1)}(N)}
\left<\dfrac{\prod_{i=1}^{k}D(\tfrac{\alpha_i}{\sqrt{2N}})}{\prod_{i=1}^{m}D(\tfrac{\beta_i}{\sqrt{2N}})}
\right>_{\triangle_{2N}^{(1)}}
\right\}=\dfrac{\prod(\alpha;\beta)}{V(\alpha)V(\beta)}\PF\;\biggl[\mathbb{S}_{GOE}(\alpha,\beta\vert
\alpha,\beta)\biggr],
\end{split}\nonumber
\end{equation}
where
$$
\dfrac{1}{\mathrm{T}^{(1)}(N)}=\dfrac{C_{2N}^{(1)}}{C_{2N+2S}^{(1)}}\;
(2N)^{\tfrac{km}{2}-\tfrac{k(k+1)}{4}-\tfrac{m(m-1)}{4}},
$$
$\mathbb{S}_{GOE}(\alpha,\beta\vert \alpha,\beta)$ is a skew
symmetric $(k+m)\times (k+m)$ matrix with rows and columns
parameterized by elements of $\alpha$ and $\beta$, and with matrix
elements:
\begin{itemize}
  \item  $
\mathbb{S}_{GOE}(\alpha_i,\alpha_j)=\mathbb{S}_{GOE}^{I}(\alpha_i,\alpha_j)$
  \item $\mathbb{S}_{GOE}(\alpha_i,\beta_j)=\mathbb{S}_{GOE}^{II}(\alpha_i,\beta_j)
  $
  \item $\mathbb{S}_{GOE}(\beta_i,\beta_j)=\mathbb{S}_{GOE}^{III}(\beta_i,\beta_j)
  $
\end{itemize}

\end{thm}
The constants $C_{2N}^{(1)}$ can be explicitly computed, see Mehta
\cite{mehta}:
$$
(2N)!C_{2N}^{(1)}=(2\pi)^{N}\prod\limits_{j=0}^{2N-1}\frac{\Gamma
(\tfrac{3}{2}+\tfrac{j}{2})}{\Gamma(\tfrac{3}{2})}
$$
\subsubsection{The Gaussian Symplectic  Ensemble} Finally, let us take $\beta=4$, $\mu(dx)=e^{-x^2}dx$.
Given two functions $f, g$ the skew symmetric inner product for
the Gaussian Symplectic Ensemble (GSE) is defined by
\begin{equation}
\left<f,g\right>_4=\dfrac{1}{2}\int\limits_{-\infty}^{+\infty} dx
e^{-x^2}\left(f(x)g'(x)-f'(x)g(x)\right).\nonumber
\end{equation}
The family of monic skew orthogonal polynomials associated with
GSE is introduced by
\begin{equation}
\left<p_{2k},p_{2l+1}\right>_4=-\left<p_{2l+1},p_{2k}\right>_4=h_{k}^{(4)}\delta_{kl}.
\nonumber
\end{equation}
All other brackets are zeros. In the case of GSE the kernels that
determine averages of characteristic polynomials are:
\begin{itemize}
  \item  $
W_{N,I}^{(4)}(\zeta,\eta)=\dfrac{1}{h_{N}^{(4)}}\;
(\zeta-\eta)\left<D^2(\zeta)D^2(\eta)\right>_{\triangle_{N}^{(4)}},
\;\zeta\in\;\mathbb{C},\;\eta\in\;\mathbb{C}.$
  \item $
W_{N,II}^{(4)}(\zeta,\eta)= \dfrac{1}{\zeta-
\eta}\;\biggl<\dfrac{D^2(\zeta)}{D^2(\eta)
}\biggr>_{\triangle_{N}^{(4)}},\;
\zeta\in\;\mathbb{C},\;\eta\in\;\mathbb{C}\backslash\mathbb{R}. $
  \item
  $W_{N,III}^{(4)}(\zeta,\eta)=
  h_{N-1}^{(4)}(\zeta-\eta)\left<\dfrac{1}{D^2(\zeta)D^2(\eta)}\right>_{\triangle_{N}^{(4)}},
\;
\zeta\in\;\mathbb{C}\backslash\mathbb{R},\eta\in\;\mathbb{C}\backslash\mathbb{R}.
$
\end{itemize}
Similar to the case of $GOE$ the following expressions were
obtained:
\begin{equation}\label{4WNIASCAUCHYDARBOUX}
W_{N,I}^{(4)}(\zeta,\eta)=\dfrac{1}{h_{N}^{(4)}}\sum\limits_{i=0}^N
\dfrac{p_{2i+1}(\zeta)p_{2i}(\eta)-p_{2i}(\zeta)p_{2i+1}(\eta)}{h_i^{(4)}}
\end{equation}
\begin{equation}\label{4WNIIASCAUCHYDARBOUX}
W_{N,II}^{(4)}(\zeta,\eta)= \dfrac{1}{\zeta-\eta}+
\sum\limits_{i=0}^{N-1}
\dfrac{p_{2i+1}(\zeta)h_{2i}(\eta)-p_{2i}(\zeta)h_{2i+1}(\eta)}{h_i^{(4)}},
\end{equation}
\begin{equation}\label{4WNIIIASCAUCHYDARBOUX}
 W_{N,III}^{(4)}(\zeta,\eta)=h_{N-1}^{(4)}\biggl[\sum\limits_{i=0}^{N-2}\dfrac{
 h_{2i+1}(\zeta)h_{2i}(\eta)-h_{2i}(\zeta)h_{2i+1}(\eta)}{h_i^{(4)}}
 +\left<R_{\zeta},R_{\eta}\right>_4\biggr]
\end{equation}
\begin{thm}\label{THEOREMA2}(Scaling limits of kernels at the origin of the
spectrum). The following limit relations hold
\begin{equation}
\begin{split}
\underset{N\rightarrow\infty}{\lim}&
W^{(4)}_{N,I}(\dfrac{\zeta}{\sqrt{2N}},\dfrac{\eta}{\sqrt{2N}})=
\mathbb{S}_{GSE}^{I}(\zeta,\eta),\;\;\;\;
\underset{N\rightarrow\infty}{\lim}\dfrac{1}{\sqrt{2N}}W^{(4)}_{N,II}(\dfrac{\zeta}{\sqrt{2N}},\dfrac{\eta}{\sqrt{2N}})=
\mathbb{S}_{GSE}^{II}(\zeta,\eta),\\
&\underset{N\rightarrow\infty}{\lim}\dfrac{1}{4N}W^{(4)}_{N,III}(\dfrac{\zeta}{2\sqrt{N}},\dfrac{\eta}{2\sqrt{N}})=
\mathbb{S}_{GSE}^{III}(\zeta,\eta).
\end{split}\nonumber
\end{equation}
where
\begin{itemize}
  \item  $
\mathbb{S}_{GSE}^{I}(\zeta,\eta)=
\dfrac{1}{\pi}\int\limits_{0}^{1}dx\dfrac{\sin (\zeta-\eta) x}{x},
\;\;\; \zeta\in\;\mathbb{C},\;\;\eta\in\;\mathbb{C}.$
  \item $\mathbb{S}_{GSE}^{II}(\zeta,\eta)=
  \begin{cases}
    \dfrac{e^{i(\eta-\zeta)}}{\zeta-\eta} & \Img \eta >0, \\
    \dfrac{e^{-i(\eta-\zeta)}}{\zeta-\eta} & \Img \eta<0.
  \end{cases}
,\;
\zeta\in\;\mathbb{C},\;\;\eta\in\;\mathbb{C}\backslash\mathbb{R}.$
  \item
  $\mathbb{S}_{GSE}^{III}(\zeta,\eta)=2\pi i
  \begin{cases}
    -\dfrac{d}{d\zeta}\biggl[\dfrac{e^{-i(\zeta-\eta)}}{\zeta-\eta}\biggr] & \Img \zeta <0, \Img \eta >0 \\
     \quad\dfrac{d}{d\zeta}\biggl[\dfrac{e^{i(\zeta-\eta)}}{\zeta-\eta}\biggr] & \Img \zeta >0, \Img \eta <0 \\
    \quad 0 & \text{otherwise}.
  \end{cases}
,\;\zeta,\eta\in\;\mathbb{C}\backslash\mathbb{R}. $
\end{itemize}

\end{thm}
\begin{thm}\label{THEOREMSCALINGB4} For any integers $N\geq 1$ and $S>1-N$, and finite
sets of mutually distinct complex numbers
$$
\alpha=\{\alpha_1,\ldots,\alpha_k\},\;\;\;\beta=\{\beta_1,\ldots
,\beta_m \}, \;\; k-m=2S
$$
such that
$$\beta\cap \mathbb{R}=\varnothing
$$
one has
\begin{equation}
\begin{split}
\underset{N\rightarrow\infty}{\lim}&
\left\{\dfrac{1}{\mathrm{T}^{(4)}(N)}
\left<\dfrac{\prod_{i=1}^{k}d(\tfrac{\alpha_i}{\sqrt{2N}})}{\prod_{i=1}^{m}d(\tfrac{\beta_i}{\sqrt{2N}})}
\right>_{\triangle_{N}^{(4)}(e^{-x^2/2})}
\right\}=\dfrac{\prod(\alpha;\beta)}{V(\alpha)V(\beta)}\PFAFF\;\biggl[\mathbb{S}_{GSE}(\alpha,\beta\vert
\alpha,\beta)\biggr],
\end{split}\nonumber
\end{equation}
where
$$
\dfrac{1}{\mathrm{T}^{(4)}(N)}=\dfrac{C_{N}^{(4)}}{C_{N+S}^{(4)}}\;
(2N)^{\tfrac{km}{2}-\tfrac{k(k-1)}{4}-\tfrac{m(m+1)}{4}},
$$
$\mathbb{S}_{GSE}(\alpha,\beta\vert \alpha,\beta)$ is a skew
symmetric $(k+m)\times (k+m)$ matrix with rows and columns
parameterized by elements of $\alpha$ and $\beta$, and with matrix
elements:
\begin{itemize}
  \item  $
\mathbb{S}_{GSE}(\alpha_i,\alpha_j)=\mathbb{S}_{GSE}^{I}(\alpha_i,\alpha_j)$
  \item $\mathbb{S}_{GSE}(\alpha_i,\beta_j)=\mathbb{S}_{GSE}^{II}(\alpha_i,\beta_j)
  $
  \item $\mathbb{S}_{GSE}(\beta_i,\beta_j)=\mathbb{S}_{GSE}^{III}(\beta_i,\beta_j)
  $
\end{itemize}
\end{thm}
The constants $C_{N}^{(4)}$ can be explicitly computed, see Mehta
\cite{mehta}:
$$
N!C_N^{(4)}=\frac{(2\pi)^{N(N+1/2)}}{2^{N(N+1/2)}}\left[\prod\limits_{j=0}^{N-1}(2j+2)!\right]
$$
The following Sections include the proofs of Theorems
\ref{TEOREMA1} and \ref{THEOREMA2}.
\begin{rem}\label{RemarkAsymptoticsCorrelationFunctions}
As we have seen in Sections \ref{SectionCorrFunctionsDiscrete},
\ref{SectionCorrCont} the correlation functions of GUE, GOE, GSE
can be expressed through 2-point averages of characteristic
polynomials, see Theorem \ref{CONTTHEOREMCORRELATIONFUNCTIONS}. If
we insert the asymptotic expressions for $\mathbb{S}_{GUE}$,
$\mathbb{S}_{GOE}$, $\mathbb{S}_{GSE}$ to the formulas of Theorem
\ref{CONTTHEOREMCORRELATIONFUNCTIONS} we obtain the following
asymptotic formulas for the correlation kernels:
\begin{equation}
\underset{N\rightarrow\infty}{\lim}\dfrac{1}{\sqrt{2N}}K_N^{(2)}(\dfrac{x}{\sqrt{2N}},\dfrac{y}{\sqrt{2N}})
=\dfrac{\sin (x-y)}{\pi(x-y)}\nonumber
\end{equation}
\begin{equation}
\underset{N\rightarrow\infty}{\lim}\left(%
\begin{array}{cc}
 \frac{1}{\sqrt{2N}}  & 0\\
  0 & 1 \\
\end{array}%
\right)
K_N^{(1)}(\dfrac{x}{\sqrt{2N}},\dfrac{y}{\sqrt{2N}})\left(%
\begin{array}{cc}
 \frac{1}{\sqrt{2N}}  & 0\\
  0 & 1 \\
  \end{array}%
\right)
=\dfrac{1}{\pi}\left(%
\begin{array}{cc}
  -\frac{d}{dx}\frac{\sin(x-y)}{x-y} & \frac{\sin(x-y)}{x-y} \\
  -\frac{\sin(x-y)}{x-y} & -\int\limits_{1}^{+\infty}\frac{dt\;\sin(x-y)t}{t} \\
\end{array}%
\right)\nonumber
\end{equation}
\begin{equation}
\underset{N\rightarrow\infty}{\lim}\left(%
\begin{array}{cc}
 \frac{1}{\sqrt{2N}}  & 0\\
  0 & 1 \\
\end{array}%
\right)
K_N^{(4)}(\dfrac{x}{\sqrt{2N}},\dfrac{y}{\sqrt{2N}})\left(%
\begin{array}{cc}
 \frac{1}{\sqrt{2N}}  & 0\\
  0 & 1 \\
  \end{array}%
\right)
=\dfrac{1}{2\pi}\left(%
\begin{array}{cc}
  \int\limits_{0}^1\frac{dt\;\sin(x-y)t}{t} & \frac{\sin(x-y)}{x-y} \\
  -\frac{\sin(x-y)}{x-y} & -\frac{d}{dx}\frac{\sin(x-y)}{x-y} \\
\end{array}%
\right)\nonumber
\end{equation}
The expression in the righthand sides are well known\footnote{The
correlation functions in $\beta=1,4$ cases are usually written as
quaternion determinants while we use pfaffian representations. The
relation between them is explained in Section
\ref{SectionCorrFunctionsDiscrete}.}, see e. g. Forrester
\cite{forrester0}, Chapter 5. To derive these expressions we
observe that
\begin{equation}
\left[\mathbb{S}^{II}_{GUE}(x,\eta)\right]_{\eta=y}=\left[\mathbb{S}^{II}_{GOE}(x,\eta)\right]_{\eta=y}=
\left[\mathbb{S}^{II}_{GSE}(x,\eta)\right]_{\eta=y}=\dfrac{\sin(x-y)}{\pi(x-y)}\nonumber
\end{equation}
\begin{equation}
\left[\mathbb{S}^{III}_{GOE}(\zeta,\eta)\right]_{\zeta=x,\eta=y}=-\dfrac{1}{\pi}\int\limits_{1}^{+\infty}
\dfrac{dt\;\sin(x-y)t}{t}\nonumber
\end{equation}
\begin{equation}
\left[\mathbb{S}^{III}_{GSE}(\zeta,\eta)\right]_{\zeta=x,\eta=y}=-\dfrac{1}{\pi}\dfrac{d}{dx}\dfrac{\sin(x-y)}{x-y}
\nonumber
\end{equation}
\end{rem}
\subsection{Summation formulae
for kernels}
\subsubsection{Summation formulae for kernels in the
GOE case}
\begin{prop}\label{PropositionCDSum}
The Christoffel-Darboux type sum constructed from the skew
orthogonal polynomials $\{ p_n\}$, $n=0,1,2,\ldots $, is
representable in terms of the orthonormal functions associated
with the Hermite polynomials:
\begin{equation}\label{CauchyDarbouxSum}
\begin{split}
\sum\limits_{i=0}^{N-1} \;
& \dfrac{p_{2i}(x)p_{2i+1}(y)-p_{2i}(x)p_{2i+1}(y)}{h_i}=\\
&
e^{\frac{x^2+y^2}{2}}\dfrac{c_N}{c_{N-1}}\biggl[\frac{d}{dx}\biggl(\dfrac{\psi_N(x)\psi_{N-1}(y)
-\psi_{N-1}(x)\psi_N(y)}{x-y}\biggr)+\psi_{N-1}(y)\psi_N(x)\biggr],
\end{split}
\end{equation}
where the functions $\psi_n(x)$ are orthonormal functions
associated with the Hermite polynomials. Namely,
\begin{equation}\label{psi}
\psi_n(x)= \dfrac{\pi_n(x)e^{-\frac{x^2}{2}}}{c_n}
=\dfrac{H_n(x)e^{-\frac{x^2}{2}}}{2^nc_n},\;\;\;
\int\psi_n(x)\psi_m(x)dx= \delta_{nm}.
\end{equation}
Here $H_n(x)$ are the Hermite polynomials, $\pi_n$ are the monic
orthogonal polynomials defined by the weight $w(x)=e^{-x^2}$, and
the coefficients $c_n$ are their norms,
\begin{equation}\label{Asc_n}
c_n=\pi^{1/4}2^{-n/2}\sqrt{n!}.
\end{equation}
\end{prop}
This result is well-known and can be found in the literature on
 Random Matrix Theory (see, for example, Forrester, Honner and
Nagao \cite{forrester}, Adler, Forrester, Nagao, and van Moerbeke
\cite{adler}, Widom \cite{widom}).

Proposition \ref{PropositionCDSum} lead us to the representations
for kernels which are suitable for the asymptotic analysis.
\begin{prop}
The kernels $W_{N,I}^{(1)}(\zeta,\eta)$,
$W_{N,II}^{(1)}(\zeta,\eta)$, and $W_{N,III}^{(1)}(\zeta,\eta)$
are representable in terms of the orthonomal functions $\psi_n(x)$
defined by equation (\ref{psi}), and by their Cauchy type
transforms $\Psi_n(\zeta)$,
\begin{equation}\label{CauchyTransformPsi}
\Psi_n(\zeta)=\int\dfrac{dx\; e^{-x^2/2}\psi_n(x)}{x-\zeta};\;\;\;
\zeta\in \mathbb{C}\backslash \mathbb{R}.
\end{equation}
Namely,
\begin{equation}\label{WNIEXACT}
W_{N,I}^{(1)}(\zeta,\eta)=-\dfrac{c_{N+1}}{c_{N}}\;e^{\frac{\zeta^2+\eta^2}{2}}
\biggl[\frac{d}{d\zeta}\biggl(\dfrac{\psi_{N+1}(\zeta)\psi_{N}(\eta)
-\psi_{N}(\zeta)\psi_{N+1}(\eta)}{\zeta-\eta}\biggr)+\psi_{N}(\eta)\psi_{N+1}(\zeta)\biggr].
\end{equation}
\begin{equation}\label{ExactWII}
\begin{split}
W_{N,II}^{(1)}(\zeta,\eta)=&
\dfrac{c_N}{c_{N-1}}\;e^{\zeta^2/2}\biggl[\dfrac{\Psi_N(\eta)\psi_{N-1}(\zeta)
-\Psi_{N-1}(\eta)\psi_{N}(\zeta)}{\zeta-\eta}\\
&+ \dfrac{1}{2}\psi_{N-1}(\zeta)\iint \dfrac{dxdy\;\psi_N(x)
e^{-y^2/2}\sgn(y-x)}{\eta-y}\biggr].
\end{split}
\end{equation}
\begin{equation}\label{ExactWNIII}
\begin{split}
W_{N,III}^{(1)}&(\zeta,\eta)=\dfrac{1}{2}\dfrac{c_{N-1}}{c_{N-2}}\;\biggl[
\Psi_{N-1}(\zeta)\iint\dfrac{dxdy\; e^{-y^2/2}\sgn(y-x)\psi_{N-2}(x)}{(y-\eta)(x-\zeta)}\\
&-\Psi_{N-2}(\zeta)\iint\dfrac{dxdy\; e^{-y^2/2}\sgn(y-x)\psi_{N-1}(x)}{(y-\eta)(x-\zeta)}\\
&-1/2\iint\dfrac{dxdy\;
e^{-y^2/2}\sgn(y-x)\psi_{N-1}(x)}{(y-\zeta)} \iint\dfrac{dxdy\;
e^{-y^2/2}\sgn(y-x)\psi_{N-2}(x)}{(y-\eta)}\biggr].
\end{split}
\end{equation}
\end{prop}
\begin{proof}
Equation (\ref{WNIEXACT}) is obtained immediately from equations
(\ref{CauchyDarbouxSum}) and (\ref{WNIASCAUCHYDARBOUX}). Let us
derive formula (\ref{ExactWII}). We start from the representation
of the second kernel as the Christoffel-Darboux type sum, equation
(\ref{WNIIASCAUCHYDARBOUX}). In this representation we rewrite the
functions $h_{2i}(\eta)$, $h_{2i+1}(\eta)$ explicitly as integral
transformations (equation (\ref{DefinitionOfhk}) ) of the monic
skew orthogonal polynomials $p_{2i}$ and $p_{2i+1}$. Then we apply
formula (\ref{CauchyDarbouxSum}) and obtain
\begin{equation}\label{WNIITECH}
\begin{split}
W_{N,II}^{(1)}&(\zeta,\eta)=\dfrac{1}{\zeta-\eta}
+\dfrac{1}{2}\;e^{\zeta^2/2}\dfrac{c_N}{c_{N-1}}\biggl[\\
&\iint\dfrac{dxdy e^{-y^2/2} \sgn(y-x)}{\eta-y}
\dfrac{d}{dx}\left(\dfrac{\psi_N(x)\psi_{N-1}(\zeta)
-\psi_{N-1}(x)\psi_N(\zeta)}{x-\zeta}\right)\\
&+ \psi_{N-1}(\zeta)\; \iint\dfrac{dxdy\psi_N(x)
e^{-y^2/2}\sgn(y-x)}{\eta-y}\biggr].
\end{split}
\end{equation}
The first term in the brackets is simplified if we perform the
integration by parts. After that we use
\begin{equation}
\dfrac{1}{(y-\zeta)(y-\eta)}=
\dfrac{1}{\zeta-\eta}\left(\dfrac{1}{y-\zeta}-\dfrac{1}{y-\eta}\right),
\nonumber
\end{equation}
 and decompose the resulting integral into two terms.
In each term the integrals are the Cauchy type transforms $\Psi_n$
of functions $\psi_n$. Now we apply the well-known formula
\begin{equation}
\Psi_N(\zeta)\psi_{N-1}(\zeta)-\Psi_{N-1}(\zeta)\psi_{N}(\zeta)=
\dfrac{c_{N-1}}{c_N}\;e^{-\zeta^2/2}. \nonumber
\end{equation}
This formula follows, for example, from the fact that the
determinant of the solution of the Riemann-Hilbert problem for
orthogonal polynomials is identically 1, see Deift \cite{deift},
page 44. Then one term (the first term in equation
(\ref{WNIITECH})) is cancelled with $\frac{1}{\zeta-\eta}$ and
equation (\ref{ExactWII}) is obtained. Equation (\ref{ExactWNIII})
is derived from equations (\ref{WNIIIASCAUCHYDARBOUX}) and
(\ref{CauchyDarbouxSum}) by the same procedure.
\end{proof}
It is convenient to introduce two functions,
\begin{equation}
I_N(\eta)=-\iint\dfrac{dxdy\;e^{-y^2/2}\sgn(y-x)\psi_{N-1}(x)}{y-\eta},
\nonumber
\end{equation}
and
\begin{equation}
E_N(\zeta,\eta)=-\iint\dfrac{dxdy\;e^{-y^2/2}\sgn(y-x)\psi_{N-1}(x)}{(y-\eta)(x-\zeta)}.
\nonumber
\end{equation}
Then the kernels $W_{N,II}^{(1)}(\zeta,\eta)$ and
$W_{N,III}^{(1)}(\zeta,\eta)$ can be rewritten as follows,
\begin{equation}\label{ExactWIIBrief}
\begin{split}
W_{N,II}^{(1)}(\zeta,\eta)=&
\dfrac{c_N}{c_{N-1}}\;e^{\zeta^2/2}\biggl[\dfrac{\Psi_N(\eta)\psi_{N-1}(\zeta)
-\Psi_{N-1}(\eta)\psi_{N}(\zeta)}{\zeta-\eta}+
\dfrac{1}{2}\psi_{N-1}(\zeta)I_N(\eta)\biggr],
\end{split}
\end{equation}
and
\begin{equation}\label{ExactWNIIIbrief}
\begin{split}
W_{N,III}^{(1)}&(\zeta,\eta)=\dfrac{1}{2}\dfrac{c_{N-1}}{c_{N-2}}\;\biggl[
E_{N-1}(\zeta,\eta)\Psi_{N-2}(\zeta)-E_{N-2}(\zeta,\eta)\Psi_{N-1}(\zeta)
-\dfrac{1}{2}I_{N-1}(\zeta)I_{N-2}(\eta)\biggr].
\end{split}
\end{equation}
\subsubsection{Summation formulae for kernels in the GSE case}
Given two functions, $f(x)$ and $g(x)$, introduce new skew
symmetric brackets $(,)$  by
\begin{equation}
(f,g)=\dfrac{1}{2}\int dx (f(x)g'(x)-f'(x)g(x))
\end{equation}
It is easy to check that
\begin{equation}\label{brackets}
\left<f,g\right>_4=(f(x)e^{-x^2/2},g(x)e^{-x^2/2}).
\end{equation}
Orthonormal functions associated with the skew orthogonal
polynomials are defined by
\begin{equation}
\phi_{2i}(x)=\dfrac{1}{\sqrt{h_i^{(4)}}}\;e^{-\tfrac{x^2}{2}}p_{2i}(x),\;\;
\phi_{2i+1}(x)=\dfrac{1}{\sqrt{h_i^{(4)}}}\;e^{-\tfrac{x^2}{2}}p_{2i+1}(x)
\nonumber
\end{equation}
Relation (\ref{brackets}) implies that
\begin{equation}
\left(\phi_{2i},\phi_{2j+1}\right)\equiv\dfrac{1}{2}\int dx
\left(\phi_{2i}(x)\phi_{2j+1}'(x)-\phi_{2i+1}(x)\phi_{2j}'(x)\right)=\delta_{ij},
\nonumber
\end{equation}
and all other possible brackets are zeros. The following formula
is well known, see Forrester, Nagao and Honner\cite{forrester},
Widom \cite{widom}, and Adler, Forrester, Nagao, and van Moerbeke
\cite{adler}
\begin{equation}\label{BasicSummationFormula}
\begin{split}
\sum\limits_{i=0}^{N-1}\biggl[\phi_{2i}(x)&\phi_{2i+1}'(y)-
\phi_{2i+1}(x)\phi_{2i}'(y)\biggr]=\\
&\dfrac{c_{2N}}{c_{2N-1}}\left[
\dfrac{\psi_{2N}(x)\psi_{2N-1}(y)-\psi_{2N-1}(x)\psi_{2N}(y)}{x-y}+
\psi_{2N}(y)\int\limits_{-\infty}^{x}\psi_{2N-1}(t)dt\right],
\end{split}
\end{equation}
where the functions $\psi_n(x)$ are orthonormal functions
associated with the Hermite polynomials (see equation
(\ref{psi})).
\begin{prop}\label{ProposititionWNI}
The summation formula for the first kernel,
$W_{N,I}^{(4)}(\zeta,\eta)$, is
\begin{equation}\label{WI4}
\begin{split}
 W_{N,I}^{(4)}(\zeta,\eta)&=\dfrac{ c_{2N+2}}{
c_{2N+1}}\;e^{\tfrac{\zeta^2+\eta^2}{2}}\\
& \times\biggl[
\int\limits_{\eta}^{\zeta}dx\;\dfrac{\psi_{2N+2}(\eta)\psi_{2N+1}(x)-\psi_{2N+1}(\eta)\psi_{2N+2}(x)}{\eta-x}+
\int\limits_{\eta}^{\zeta}\psi_{2N}(x)dx\int\limits_{-\infty}^{\eta}\psi_{2N-1}(t)dt\biggr]
\end{split}
\end{equation}
\end{prop}
\begin{proof}
Application of basic summation formula
(\ref{BasicSummationFormula}) together with expression
(\ref{4WNIASCAUCHYDARBOUX}) for the kernel.
\end{proof}
\begin{prop}\label{Propositionhi}
It is possible to rewrite the functions $h_{2i}(\eta)$,
$h_{2i+1}(\eta)$ as follows
\begin{equation}\label{h2i}
h_{2i}(\eta)=\sqrt{h_{i}^{(4)}}\int\dfrac{dx}{x-\eta}\;e^{-x^2/2}\phi_{2i}'(x)
\end{equation}
\begin{equation}\label{h2i+1}
h_{2i+1}(\eta)=\sqrt{h_{i}^{(4)}}\int\dfrac{dx}{x-\eta}\;e^{-x^2/2}\phi_{2i+1}'(x)
\end{equation}
\end{prop}
\begin{proof}
Apply relation (\ref{brackets}) and integrate by parts.
\end{proof}
\begin{prop}
The second kernel, $W_{N,II}^{(4)}(\zeta,\eta)$, can be rewritten
as follows
\begin{equation}\label{SeconExact4}
\begin{split}
W_{N,II}^{(4)}(\zeta,\eta)=e^{\zeta^2/2}\dfrac{c_{2N}}{c_{2N-1}}\biggl[&
\dfrac{\Psi_{2N}(\eta)\psi_{2N-1}(\zeta)
-\Psi_{2N-1}(\eta)\psi_{2N}(\zeta)}{\zeta-\eta}+
\Psi_{2N}(\eta)\int\limits_{-\infty}^{\zeta}\psi_{2N-1}(t)dt\biggr].
\end{split}
\end{equation}
where $\Psi_N$ was introduced by equation
(\ref{CauchyTransformPsi})
\end{prop}
\begin{proof}
We insert the expressions for the functions $h_{2i}(\eta)$,
$h_{2i+1}(\eta)$ obtained in Proposition \ref{Propositionhi} into
the formula for the kernel $W_{N,II}^{(4)}(\zeta,\eta)$, equation
(\ref{4WNIIASCAUCHYDARBOUX}). After that we use
\begin{equation}
\dfrac{1}{(y-\zeta)(y-\eta)}=
\dfrac{1}{\zeta-\eta}\left(\dfrac{1}{y-\zeta}-\dfrac{1}{y-\eta}\right),
\nonumber
\end{equation}
 and decompose the first resulting integral into two parts.
Then we apply the well-known formula
\begin{equation}\label{WELLKNOWNFORMULA}
\Psi_{2N}(\zeta)\psi_{2N-1}(\zeta)-\Psi_{2N-1}(\zeta)\psi_{2N}(\zeta)=
\dfrac{c_{2N-1}}{c_{2N}}\;e^{-\zeta^2/2}.
\end{equation}
and derive equation (\ref{SeconExact4}).
\end{proof}
It is worth noting that the first term in the brackets of equation
(\ref{SeconExact4}) is exactly the same as in the case of GOE, see
(\ref{ExactWIIBrief}).
\begin{prop}
Let $\zeta\in \mathbb{C}\backslash \mathbb{R}$, $\eta\in
\mathbb{C}\backslash \mathbb{R}$. The summation formula for the
third kernel, $W_{N,III}^{(4)}(\zeta,\eta)$, is
\begin{equation}\label{WIIIFINAL}
\begin{split}
W_{N,III}^{(4)}(\zeta,\eta)=\dfrac{c_{2N-2}}{c_{2N-3}}
\biggl[\Psi_{2N-3}(\eta)F_{2N-2}&(\eta,\zeta)-\Psi_{2N-2}(\eta)F_{2N-3}(\eta,\zeta)\\
&- \Psi_{2N-2}(\eta)\Psi_{2N-3}(\zeta)\biggr],
\end{split}
\end{equation}
where we have introduced:
\begin{equation}\label{FEXACT}
F_N(\eta,\zeta)=\int\dfrac{dx\;\psi_N(x)}{\eta-x}\dfrac{d}{dx}\dfrac{e^{-x^2/2}}{\zeta-x}.
\end{equation}
\end{prop}
\begin{proof}
Equation (\ref{4WNIIIASCAUCHYDARBOUX}) shows that we need to
compute the following kernel:
\begin{equation}\label{SNIII}
S_N^{III}(\zeta,\eta)=\sum\limits_{i=0}^{N-2}\dfrac{
 h_{2i+1}(\zeta)h_{2i}(\eta)-h_{2i}(\zeta)h_{2i+1}(\eta)}{h_i^{(4)}}
\end{equation}
We rewrite the functions $h_{2i+1}(\zeta)$, $h_{2i}(\zeta)$,
$h_{2i+1}(\eta)$, $h_{2i}(\eta)$ explicitly using equation
(\ref{h2i}) and (\ref{h2i+1}),
\begin{equation}
h_{2i}(\zeta)=\sqrt{h_{i}^{(4)}}\int
dx\;\phi_{2i}(x)\dfrac{d}{dx}\dfrac{e^{-x^2/2}}{\zeta-x},\;\;\;\;h_{2i+1}(\zeta)=\sqrt{h_{i}^{(4)}}\int
dx\;\phi_{2i+1}(x)\dfrac{d}{dx}\dfrac{e^{-x^2/2}}{\zeta-x};\nonumber
\end{equation}
\begin{equation}
h_{2i}(\eta)=\sqrt{h_{i}^{(4)}}\int\dfrac{dx}{x-\eta}\;e^{-x^2/2}\phi_{2i}'(x),\;\;\;\;
h_{2i+1}(\eta)=\sqrt{h_{i}^{(4)}}\int\dfrac{dx}{x-\eta}\;e^{-x^2/2}\phi_{2i+1}'(x).
\nonumber
\end{equation}
Insert the above expressions to equation (\ref{SNIII}), and obtain
\begin{equation}
S_N^{III}(\zeta,\eta)=\sum\limits_{i=0}^{N-2}\iint
dxdy\left[\dfrac{e^{-y^2/2}}{\eta-y}\dfrac{d}{dx}\dfrac{e^{-x^2/2}}{\zeta-x}\right]
\left(\phi'_{2i+1}(y)\phi_{2i}(x)-\phi_{2i}'(y)\phi_{2i+1}(x)\right)
\nonumber
\end{equation}
Now we apply  basic summation formula
(\ref{BasicSummationFormula}) and find
\begin{equation}
\begin{split}
S_N^{III}(\zeta,\eta)=\dfrac{c_{2N-2}}{c_{2N-3}}\biggl\{\iint
dxdy\biggl[\dfrac{e^{-y^2/2}}{\eta-y}\dfrac{d}{dx}\dfrac{e^{-x^2/2}}{\zeta-x}\biggr]
\biggl[\dfrac{\psi_{2N-2}(x)\psi_{2N-3}(y)-\psi_{2N-3}(x)\psi_{2N-2}(y)}{x-y}\biggr]\\
+\iint
dxdy\biggl[\dfrac{e^{-y^2/2}}{\eta-y}\dfrac{d}{dx}\dfrac{e^{-x^2/2}}{\zeta-x}\biggr]
\biggl[\psi_{2N-2}(y)\int\limits_{-\infty}^x\psi_{2N-3}(t)dt\biggr]\biggr\}\\
\equiv\dfrac{c_{2N-2}}{c_{2N-3}}\biggl[
I_1(\zeta,\eta)+I_2(\zeta,\eta)\biggr]
\end{split}\nonumber
\end{equation}
Rewrite $I_1(\zeta,\eta)$  as a repeated integral
\begin{equation}\label{I1}
I_1(\zeta,\eta)=\int
dx\;\dfrac{d}{dx}\dfrac{e^{-x^2/2}}{\zeta-x}\;\int
dy\;\dfrac{e^{-y^2/2}}{\eta-y}\;\dfrac{\psi_{2N-2}(x)\psi_{2N-3}(y)-\psi_{2N-3}(x)\psi_{2N-2}(y)}{x-y}
\end{equation}
Apply identity
\begin{equation}
\dfrac{1}{(\eta-y)(x-y)}=\dfrac{1}{\eta-x}\left(\dfrac{1}{x-y}-\dfrac{1}{\eta-y}\right)\nonumber
\end{equation}
in the inner integral of equation (\ref{I1}). Then we obtain,
using equation (\ref{WELLKNOWNFORMULA}),
\begin{equation}
\begin{split}
\int
dy\;\dfrac{e^{-y^2/2}}{\eta-y}\;\dfrac{\psi_{2N-2}(x)\psi_{2N-3}(y)-\psi_{2N-3}(x)\psi_{2N-2}(y)}{x-y}
\\
=\dfrac{1}{\eta-x}\int
dy\;e^{-y^2/2}\;\dfrac{\psi_{2N-2}(x)\psi_{2N-3}(y)-\psi_{2N-3}(x)\psi_{2N-2}(y)}{x-y}
\\
-\dfrac{1}{\eta-x}\int
dy\;e^{-y^2/2}\;\dfrac{\psi_{2N-2}(x)\psi_{2N-3}(y)-
\psi_{2N-3}(x)\psi_{2N-2}(y)}{\eta-y} \\
=\dfrac{1}{\eta-x}e^{-x^2/2}+\dfrac{1}{\eta-x}\left(\psi_{2N-2}(x)\Psi_{2N-3}(\eta)-
\psi_{2N-3}(x)\Psi_{2N-2}(\eta)\right)
\end{split}\nonumber
\end{equation}
Therefore,
\begin{equation}
\begin{split}
I_1(\zeta,\eta)&=\int
dx\;\dfrac{d}{dx}\dfrac{e^{-x^2/2}}{\zeta-x}\;\dfrac{1}{\eta-x}e^{-x^2/2}+\\
&+\int
dx\;\dfrac{d}{dx}\dfrac{e^{-x^2/2}}{\zeta-x}\dfrac{1}{\eta-x}\left(\psi_{2N-2}(x)\Psi_{2N-3}(\eta)-
\psi_{2N-3}(x)\Psi_{2N-2}(\eta)\right)
\end{split}\nonumber
\end{equation}
or
\begin{equation}
I_1(\zeta,\eta)=\left<\dfrac{1}{\eta-x},\dfrac{1}{\zeta-x}\right>+
\Psi_{2N-3}(\eta)F_{2N-2}(\eta,\zeta)-\Psi_{2N-2}(\eta)F_{2N-3}(\eta,\zeta)\nonumber
\end{equation}
We simplify $I_2(\zeta,\eta)$ by  partial integration, obtain an
expression for $S_N^{III}(\zeta,\eta)$, insert this to  formula
 (\ref{4WNIIIASCAUCHYDARBOUX}), and
obtain (\ref{WIIIFINAL}).
\end{proof}
\subsection{Asymptotics of functions associated with Hermite
polynomials}
\subsubsection{Asymptotics of
$\psi_N(x)$}\label{SectionAsympsiN} The large $N$ asymptotics of
the functions $\psi_N(x)$ in different regions of the real line
can be found in Szeg\"o \cite{Szego}, formula (8.22.14). However,
the asymtotic expression near the edges of the spectrum is valid
only in  infinitesimally small neighborhoods (as
$N\rightarrow\infty$) of the edges. In what follows we need
uniform asymptotics of functions $\psi_N(x)$ on the entire real
line. Theorem 2.2 in Deift, Kriecherbauer, McLaughlin and Zhou
\cite{deift2} provides the needed asymptotics of orthogonal
polynomials with respect to the weights $w(x)dx=e^{-Q(x)}dx$ on
the real line, where $Q(x)=\sum_{k=0}^{2m}q_kx^k$, $q_{2m}>0$ (see
also Deift, Kriecherbauer, McLaughlin and Zhou \cite{deift3}). The
expressions in Theorem 2.2 are valid in the entire complex plane.
We adopt this result to the particular case of $Q(x)=x^2$.

The parity of functions $\psi_N(x)$ defined by equation
(\ref{psi}) corresponds to the parity of $N$, i. e. if $N$ is
even, $\psi_N(x)$ is an even function, and if $N$ is odd,
$\psi_N(x)$ is an odd function. Thus it is enough to consider the
asymptotics of functions $\psi_N(x)$ on $\mathbb{R}_+$. We
decompose $\mathbb{R}_+$ into three regions: $B_{\delta}$,
$C_{\delta}$ and $A_{\delta}$, see Fig. 4.
\begin{picture}(400,200) (-60,-80)
\put (-80,50) {\line(1,0) {300}} \put (-80,50) {\line(0,1) {5}}
\put (-80,50) {\line(0,-1) {5}} \put (-80, 35) {0} \put (10,50)
{\line(0,1) {5}} \put (10,50) {\line(0,-1) {5}} \put (5, 35)
{$1-\delta$}\put (50,50) {\line(0,1) {5}} \put (50,50)
{\line(0,-1) {5}} \put (50, 35) {$1$} \put (90,50) {\line(0,1)
{5}} \put (90,50) {\line(0,-1) {5}} \put (85, 35) {$1+\delta$}
\put (-40, 60) {$B_{\delta}$} \put (50, 60) {$C_{\delta}$} \put
(130, 60) {$A_{\delta}$} \put (-100,-10) {\textbf{Fig}. 4.
Different asymptotic regions for $\psi_N(\sqrt{2N} x)$}
\end{picture}

\begin{thm}(Plancherel-Rotach asymptotics of $\psi_N(\sqrt{2N} x)$
on the real line.) There exists a $\delta_0$ such that for all
$0<\delta \leq \delta_0$ the following asymptotic formulae hold:\\
(i) For $x\in A_{\delta}$\\
\begin{equation}
\begin{split}
\psi_N(\sqrt{2N}
x)&=\dfrac{1}{2^{3/4}\pi^{1/2}N^{1/4}}\biggl[\left(\dfrac{x-1}{x+1}\right)^{1/4}+
\left(\dfrac{x+1}{x-1}\right)^{1/4}\biggr]
\\
& \times \exp\biggl[-\dfrac{\pi
N}{2}\left[x\sqrt{x^2-1}-\ln(x+\sqrt{x^2-1})\right]\biggr]\left(1+
\mathcal{O}(\dfrac{1}{N})\right).
\end{split}
\end{equation}
(ii) For $x\in B_{\delta}$\\
\begin{equation}\label{asymptoticsPsiBDELTA}
\begin{split}
\psi_N(\sqrt{2N}
x)=&\dfrac{2^{1/4}}{\pi^{1/2}N^{1/4}}\dfrac{1}{(1-x^2)^{1/4}}
\biggl\{\cos\left(Nx\sqrt{1-x^2}+(N+1/2)\arcsin
x-\dfrac{N\pi}{2}\right)\\
&\times\;\left(1+\mathcal{O}(\dfrac{1}{N})\right)+\sin\left(Nx\sqrt{1-x^2}+(N-1/2)\arcsin
x-\dfrac{N\pi}{2}\right)\;\mathcal{O}(\dfrac{1}{N})\biggr\}.
\end{split}
\end{equation}
(iii) For $x\in C_{\delta}$\\
\begin{equation}
\begin{split}
\psi_N(\sqrt{2N} x)=&\dfrac{1}{(2N)^{1/4}} \biggl\{
\dfrac{(x+1)^{1/4}}{(x-1)^{1/4}}\left(f_N(x)\right)^{1/4}
\Airy(f_N(x))\left(1+\mathcal{O}(\dfrac{1}{N})\right)-
\\
&\dfrac{(x-1)^{1/4}}{(x+1)^{1/4}}\left(f_N(x)\right)^{-1/4} \Airy
'(f_N(x))\left(1+\mathcal{O}(\dfrac{1}{N})\right) \biggr\}.
\end{split}
\end{equation}
where the function $f_N(x)$ is given by
\begin{equation}
f_N(x)=
\left(\dfrac{3N}{2}\right)^{2/3}\left[x\sqrt{x^2-1}-\ln(x+\sqrt{x^2-1})\right]^{2/3},
\end{equation}
and $\Airy$ denotes the Airy function determined as the solution
of
\begin{equation}
\Airy ''(x)=x\Airy(x),
\end{equation}
satisfying
\begin{equation}
\underset{x\rightarrow \infty}{\lim} \Airy(x)\sqrt{4\pi} x^{1/4}
e^{\frac{2}{3}x^{3/2}}=1.
\end{equation}
All the error terms are uniform for $\delta\in$ compact subsets of
$(0,\delta_0]$ and for $x\in X_{\delta}$, where $X=A,B,C$.
\end{thm}
\begin{rem}
When $|x-1|<\epsilon$, $\epsilon$ is a small parameter,
$f_N(x)=(x-1)(\phi_N(x))^{2/3}$, where $\phi_N(x)$ is a strictly
positive function at $x=1$. Therefore the function
$\psi_N(\sqrt{2N} x)$ is finite at $x=1$.
\end{rem}
\begin{cor}\label{UniformBoundForPsi}
$\underset{x\in \mathbb{R}}{\sup}\vert \psi_N(x)\vert\leq
\const\cdot N^{-\frac{1}{12}}.$
\end{cor}
\begin{proof}
This is evident from the asymptotic expressions above  for
$\psi_N(x)$, and from the asymptotic expansion of the Airy
function and its derivative.
\end{proof}
\begin{prop}\label{PropositionAsymOfPsiIntheBalk}
Assume that $x\in \left[-M,M\right]$; $M\rightarrow \infty$ as
$N\rightarrow \infty$, $\dfrac{M}{\sqrt{N}}\rightarrow 0$ as
$N\rightarrow \infty$. Then
\begin{equation}
\psi_N(x)=\dfrac{2^{1/4}}{\pi^{1/2}
N^{1/4}}\left[\cos\left(x\sqrt{2N+1}-\dfrac{\pi N}{2}\right)+
\mathcal{O}\left(\dfrac{M^3}{\sqrt{N}}\right) \right].
\end{equation}
The error term is uniform for $x\in \left[-M,M\right]$.
\end{prop}
\begin{proof} This formula follows directly from equation
(\ref{asymptoticsPsiBDELTA}), where we replace $x$ by
$\dfrac{x}{\sqrt{2N}}$ and expand in powers of
$\dfrac{x}{\sqrt{2N}}$ the fraction
$\dfrac{1}{\left(1-\frac{x^2}{2N}\right)^{1/4}}$, and the
expression inside the cosine.
\end{proof}
\subsubsection{ Asymptotics of $\Psi_N(\zeta)$} Recall that the
Cauchy type transform $\Psi_N(\zeta)$ of the function $\psi_N(x)$
was defined by equation (\ref{CauchyTransformPsi}).
\begin{prop}\label{PropositionLargePsiAs1} Assume that $\zeta\in\; \mathbb{C}\backslash
\mathbb{R}$,  $M\rightarrow \infty$ as $N\rightarrow \infty$, $\ln
N\ll M\ll \sqrt{N}$ as $N\rightarrow \infty$. Then the following
asymptotic expression for $\Psi_N(\zeta)$ holds:
\begin{equation}\label{LargePsiAs1}
\Psi_N(\zeta)=\dfrac{2^{1/4}}{\pi^{1/2} N^{1/4}}
\int_{-M}^{M}\dfrac{dx\;e^{-x^2/2}}{x-\zeta}\;\cos\left[x\sqrt{2N+1}-\dfrac{\pi
N}{2}\right]+ \mathcal{O}\left(\dfrac{M^3\ln M }{N^{3/4}}\right).
\end{equation}
The error term is uniform for all $\zeta$ taken from a compact
subset of $\mathbb{C}\backslash \mathbb{R}$.
\end{prop}
\begin{proof}
We decompose the integral in the definition of the function
$\Psi_N(\zeta)$ (equation (\ref{CauchyTransformPsi})) into three
parts:
\begin{equation}\label{DecompositionToThree}
\int\dfrac{dx\; e^{-x^2/2}\psi_n(x)}{x-\zeta}=
\int\limits_{-M}^{M}\dfrac{dx\; e^{-x^2/2}\psi_n(x)}{x-\zeta}+
\int\limits_{M}^{+\infty}\dfrac{dx\; e^{-x^2/2}\psi_n(x)}{x-\zeta}
+\int\limits_{-\infty}^{M}\dfrac{dx\;
e^{-x^2/2}\psi_n(x)}{x-\zeta}.
\end{equation}
By Corollary \ref{UniformBoundForPsi} $\psi_N(x)$ is uniformly
bounded by $\const\cdot N^{-1/12}$. This enables us to estimate
the second and the third integrals in the expression above.
Namely,
\begin{equation}
\begin{split}
\biggl|\int\limits_{M}^{+\infty}\dfrac{dx\;
e^{-x^2/2}\psi_n(x)}{x-\zeta}\biggr|&\leq \const\cdot
N^{-1/12}\int_M^{+\infty}dx e^{-x^2/2}x\\
&=\const\cdot N^{-1/12} e^{-M^2/2}.
\end{split}\nonumber
\end{equation}
This shows that the second and third integrals in
(\ref{DecompositionToThree}) are small. To obtain formula
(\ref{LargePsiAs1}) we insert the asymptotic formula for
$\psi_N(x)$ (equation (\ref{asymptoticsPsiBDELTA})) into the first
integral in (\ref{DecompositionToThree}), and note that
\begin{equation}
\int\limits_{-M}^{M}\dfrac{dx\;
e^{-x^2/2}}{\left|x-\zeta\right|}\leq
\int\limits_{-M}^M\dfrac{dx}{\vert
x-\zeta\vert}=\mathcal{O}\left(\ln M \right).
\end{equation}\nonumber
\end{proof}
\begin{prop}\label{PropositionAsymptoticsofLargePsiBalk}
Under the same assumptions as in Proposition
\ref{PropositionLargePsiAs1} the asymptotic formula for the
 Cauchy type transform $\Psi_N(\zeta)$ of $\psi_N(x)$ is
 \begin{equation}\label{LargePsiAs2}
 \Psi_N(\dfrac{\zeta}{\sqrt{2N}})=\dfrac{2^{1/4}\pi^{1/2}}{N^{1/4}}
  \begin{cases}
    ie^{i(\zeta-\frac{\pi N}{2})}, & \Img \zeta>0, \\
      -ie^{-i(\zeta-\frac{\pi N}{2})}, & \Img \zeta<0
  \end{cases}
+\mathcal{O}\left(\dfrac{M^3\ln M}{N^{3/4}} \right).
\end{equation}
The error term is uniform for all $\zeta$ taken from a compact
subset of $\mathbb{C}\backslash \mathbb{R}$.
\end{prop}
\begin{proof}
We replace $x$ by $\dfrac{x}{\sqrt{2N}}$ in the integral in
equation (\ref{LargePsiAs1}). The obtained integral is finite as
$N\rightarrow \infty$ and has the following limiting value
\begin{equation}
\int\limits_{-\infty}^{+\infty}\dfrac{dx}{x-\zeta}\cos\left[x-\dfrac{\pi
N }{2}\right].\nonumber
\end{equation}
We compute this integral by the standard residue calculations and
get formula (\ref{LargePsiAs2}).
\end{proof}
\subsubsection{Asymptotics of $I_N(\dfrac{\eta}{\sqrt{2N}})$} The
aim of this subsection is to prove the following Proposition:
\begin{prop}\label{PropositionBoundOfI}Let $\eta$ be taken from a compact
subset $K$ of $\mathbb{C}\backslash \mathbb{R}$. Then the
following estimate is valid
\begin{equation}
\underset{\eta\in\; K
}{\sup}|I_N(\dfrac{\eta}{\sqrt{2N}})|\leq\const\cdot
N^{-1/2}\nonumber
\end{equation}
\end{prop}
\begin{proof}
The function $I_N(\eta)$ can be rewritten as
\begin{equation}
I_N(\eta)=\int_{-\infty}^{+\infty}dx \psi_N(x)\phi_{\eta}(x),
\nonumber
\end{equation}
where
\begin{equation}
\phi_{\eta}(x)=\int_{-\infty}^{+\infty}dy
e^{-y^2/2}\sgn(y-x)\dfrac{1}{\eta-y}. \nonumber
\end{equation}
Let $M\rightarrow \infty$ as $ N\rightarrow\infty$,
$\dfrac{M}{\sqrt{2N}}\rightarrow 0$ as $N\rightarrow\infty$. Then
the integral in the expression for $I_N(\eta)$ can be restricted
to the domain $[-M,M]$, with an exponentially small error:
\begin{equation}\label{OgranichenieI}
I_N(\eta)=\int_{-M}^{M}dx
\psi_N(x)\phi_{\eta}(x)+\mathcal{O}(e^{-M}).
\end{equation}
In order to show this we note that the integral can be rewritten
as follows
\begin{equation}
I_N(\eta)=\int_{-\infty}^{+\infty}dy
e^{-y^2/2}\dfrac{1}{\eta-y}\int_{-\infty}^{+\infty}\sgn(y-x)\psi_N(x)dx.
\nonumber
\end{equation}
We define
\begin{equation}
\epsilon\psi_N(y)=\int_{-\infty}^{+\infty}\sgn(y-x)\psi_N(x)dx.
\nonumber
\end{equation}
 Decompose the integral $I_N(\eta)$ into three parts:
\begin{equation}\label{SplitI}
\begin{split}
I_N(\eta)&=\int_{-M}^{M}dy
e^{-y^2/2}\dfrac{1}{\eta-y}\epsilon\psi_N(y)+\int_{-\infty}^{-M}dy
e^{-y^2/2}\dfrac{1}{\eta-y}\epsilon\psi_N(y)\\
&+\int_{M}^{+\infty}dy
e^{-y^2/2}\dfrac{1}{\eta-y}\epsilon\psi_N(y).
\end{split}
\end{equation}
The global estimate for $\epsilon\psi_N(y)$ is known (see, Nagao
and Wadati \cite{nagao}, equation (A.1)):
\begin{equation}
\underset{y\in \mathbb{R}
}{\sup}\left|\epsilon\psi_N(y)\right|\leq \mbox{const}\cdot
N^{-1/4}\nonumber
\end{equation}
We use this estimate to evaluate  the second and third integrals
in equation (\ref{SplitI}). For example, for the third integral we
obtain :
\begin{equation}
\begin{split}
\biggl|\int_{M}^{+\infty}dy
e^{-y^2/2}\dfrac{1}{\eta-y}\epsilon\psi_N(y)\biggr|\leq
\dfrac{\const}{N^{1/4}}\int_M^{+\infty}e^{-y^2/2}ydy=\mathcal{O}(e^{-M}).
\end{split}\nonumber
\end{equation}
The same estimate is valid for the second integral in equation
(\ref{SplitI}). By equation (\ref{OgranichenieI}) the problem of
evaluation of $I_N(\dfrac{\eta}{\sqrt{2N}})$ is reduced to the
evaluation of the following integral:
\begin{equation}
I_{N,M}(\eta)=\int_{-M}^{M}dx \psi_N(x)\phi_{\eta}(x),\nonumber
\end{equation}
We change the variables of the integration, $x=t/\sqrt{2N}$,
$y=s/\sqrt{2N}$, and find
\begin{equation}
I_{N,M}(\dfrac{\eta}{\sqrt{2N}})=\dfrac{1}{\sqrt{2N}}\;\int\limits_{-M\sqrt{2N}}^{M\sqrt{2N}}dt\psi_N(\dfrac{t}{\sqrt{2N}})
\int\limits_{-\infty}^{+\infty}ds
e^{-s^2/4N}\sgn(s-t)\dfrac{1}{\eta-s}.\nonumber
\end{equation}
The last integral can be rewritten as
\begin{equation}
I_{N,M}(\dfrac{\eta}{\sqrt{2N}})=\dfrac{1}{\sqrt{2N}}\;\int\limits_{-\infty}^{+\infty}ds
e^{-s^2/4N}\dfrac{1}{\eta-s}\int\limits_{-M\sqrt{2N}}^{M\sqrt{2N}}dt\psi_N(\dfrac{t}{\sqrt{2N}})\sgn(s-t).
\nonumber
\end{equation}
Now we can insert the asymptotic expression for
$\psi_N\left(\dfrac{t}{(2N)^{1/2}}\right)$ into the second
integral and obtain
\begin{equation}
\int\limits_{-M\sqrt{2N}}^{M\sqrt{2N}}dt\psi_N(\dfrac{t}{\sqrt{2N}})\sgn(s-t)=
\dfrac{2^{1/4}}{\pi^{1/2}N^{1/4}}\int\limits_{-M\sqrt{2N}}^{M\sqrt{2N}}dt
\sgn(s-t)\biggl(\cos\left[t-\dfrac{\pi
N}{2}\right]+\mathcal{O}\left(\dfrac{M^3}{\sqrt{2N}}\right)\biggr).
\nonumber
\end{equation}
This shows that the integral
\begin{equation}
\int\limits_{-M\sqrt{2N}}^{M\sqrt{2N}}dt\psi_N(\dfrac{t}{\sqrt{2N}})\sgn(s-t)\nonumber
\end{equation}
is bounded by  $\dfrac{\const\cdot M^4}{N^{1/4}}$. The integral $
\int\limits_{-\infty}^{+\infty}ds e^{-s^2/4N}\dfrac{1}{|\eta-s|} $
is bounded by $\const\cdot\log N$ uniformly for all $\eta\in K$,
where $K$ is a compact subset of $\mathbb{C}\backslash
\mathbb{R}$. Choosing $M=M(N)$ appropriately (such that
$\frac{M^4(N)\log N}{N^{1/4}}\leq 1$ as $N\rightarrow \infty$) we
prove the Proposition.
\end{proof}
\subsubsection{Estimate of
$\int\limits_{M}^{+\infty}\dfrac{\psi_N(x)\;dx}{x}$}
\begin{prop}\label{PropositionEstpsixx}
Assume that $M\ll\sqrt{N}$. Then the following estimate is valid:
\begin{equation}
\biggl|\int\limits_{M}^{+\infty}\dfrac{\psi_N(x)\;dx}{x}\biggr|\leq\Max\left[\dfrac{\const}{M^2
N^{1/12} }, \dfrac{\const M}{N^{3/4}}\right].\nonumber
\end{equation}
where the constants do not depend on $N,M$.
\end{prop}
\begin{proof}
We use representation of $\psi_N(x)$ in terms of the Hermite
polynomial $H_N(x)$ (see equation (\ref{psi})), the relation
$H'_N(x)=2NH_{N-1}(x)$, and take the integral
$\int\limits_{M}^{+\infty}\dfrac{\psi_N(x)\;dx}{x}$ by parts. As a
result we obtain
\begin{equation}
\int\limits_{M}^{+\infty}\dfrac{\psi_N(x)\;dx}{x}=\dfrac{c_{N+1}}{(N+1)c_N}
\left[-\dfrac{\psi_{N+1}(M)}{M^2}+\int\limits_{M}^{+\infty}\psi_{N+1}(x)dx
+\int\limits_{M}^{+\infty}\dfrac{\psi_{N+1}(x)\;dx}{x^2}\right].\nonumber
\end{equation}
Since $|\psi_{N+1}(M)|\leq\const\cdot N^{-1/12}$ (Corollary
\ref{UniformBoundForPsi}) the first term in the brackets is
estimated by $\const\cdot N^{-1/12}M^{-2}$. As for the second term
we note that
\begin{equation}
\begin{split}
\biggl|\int\limits_{M}^{+\infty}\psi_{N+1}(x)dx\biggr|
&\leq\biggl|\int\limits_{0}^{+\infty}\psi_{N+1}(x)dx\biggr|+
\biggl|\int\limits_{0}^{M}\psi_{N+1}(x)dx\biggr|\\
&\leq \const\cdot N^{-1/4}+\const\cdot N^{-1/4}M\leq\const\cdot
N^{-1/4}M,
\end{split}\nonumber
\end{equation}
where we have used the estimate proved in Nagao an Wadati
\cite{nagao}, equation (A.1):
\begin{equation}
\biggl|\int\limits_{0}^{+\infty}\psi_{N}(x)dx\biggr|\leq
\const\cdot N^{-1/4}.\nonumber
\end{equation}
The third integral is evaluated as follows:
\begin{equation}
\biggl|\int\limits_{M}^{+\infty}\dfrac{\psi_{N+1}(x)\;dx}{x^2}\biggr|
\leq
\dfrac{1}{M^2}\int\limits_M^{+\infty}\left|\psi_{N+1}(x)\right|dx\leq
\dfrac{\const \sqrt{2N}}{N^{1/12}M^{2}}.\nonumber
\end{equation}
Noting that the factor $\frac{c_{N+1}}{(N+1)c_N}\sim
\frac{1}{\sqrt{2N}}$, as $N\rightarrow\infty$, we obtain the
estimate in the statement of the Proposition.
\end{proof}
\subsubsection{Asymptotics of $E_N(\zeta,\eta)$}

\begin{prop}\label{PropositionAsymptoticsOfE}
Take $\zeta$, $\eta\in\mathbb{C}\backslash \mathbb{R}$. Then
\begin{equation}\label{EAs}
\begin{split}
E_N\left(\dfrac{\zeta}{\sqrt{2N}},\dfrac{\eta}{\sqrt{2N}}\right)=&-\dfrac{2^{1/4}}{\pi^{1/2}N^{1/4}}
\int\limits_{\mathbb{R}} \dfrac{dy\;e^{-y^2/4N}}{\zeta-y}
\int\limits_{\mathbb{R}}\dfrac{dx\cos(x-\dfrac{\pi
N}{2})\sgn(y-x)}{\eta-x}\\
&+\mathcal{O}\left(\Max\left[\dfrac{M^3\log^2N  }{N^{3/4}},
\dfrac{\log N}{M^2N^{1/12}}\right]\right)
\end{split}
\end{equation}
as $N\rightarrow\infty$. The error term,
$\mathcal{O}\left(\Max\left[\dfrac{M^3\log^2N  }{N^{3/4}},
\dfrac{\log N}{M^2N^{1/12}}\right]\right)$, is uniform for
$\zeta,\eta$ taken from a compact subset  of $\mathbb{C}\backslash
\mathbb{R}$.
\end{prop}
\begin{rem}
 The  error term
$\mathcal{O}\left(\Max\left[\frac{M^3\log^2N }{N^{3/4}},
\frac{\log N}{M^2N^{1/12}}\right]\right)$ in  Proposition
\ref{PropositionAsymptoticsOfE} is negligible in comparison with
the main term,
\begin{equation}
 -\frac{2^{1/4}}{\pi^{1/2}N^{1/4}}\int\limits_{\mathbb{R}} \dfrac{dy\;e^{-y^2/4N}}{\zeta-y}
\int\limits_{\mathbb{R}}\dfrac{dx\cos(x-\dfrac{\pi
N}{2})\sgn(y-x)}{\eta-x} \nonumber
\end{equation}
 if
\begin{equation}\label{conditions}
    \frac{M^3\log^2N}{N^{3/4}}\leq\frac{1}{N^{1/4}},\;\;\;
    \frac{\log N}{M^2 N^{1/12}}\leq\frac{1}{N^{1/4}}
\end{equation}
Once $M=M(N)$ is chosen such that $N^{1/12}(\log N)^{1/2}\ll M \ll
N^{1/6}(\log N)^{-2/3}$ the conditions above (equation
(\ref{conditions})) are satisfied.
\end{rem}
\begin{proof}
It is convenient to rewrite
$E_N\left(\dfrac{\zeta}{\sqrt{2N}},\dfrac{\eta}{\sqrt{2N}}\right)$
as
\begin{equation}\label{E}
E_N\left(\dfrac{\zeta}{\sqrt{2N}},\dfrac{\eta}{\sqrt{2N}}\right)=
\int
\frac{e^{-y^2/4N}g_N(\frac{y}{\sqrt{2N}},\frac{\eta}{\sqrt{2N}})\;dy}{\zeta-y},
\end{equation}
where
\begin{equation}\label{gN}
g_N(\dfrac{y}{\sqrt{2N}},\dfrac{\eta}{\sqrt{2N}})=\int\frac{\sgn(y-x)\psi_N(\frac{x}{\sqrt{2N}})\;dx}{x-\eta}.
\end{equation}
Equation (\ref{EAs}) means that the function $\psi_{N-1}(x)$ in
expression (\ref{E}) can be replaced by its asymptotics in the
center of the spectrum if we are interested in the scaling limit
of $E_N(\zeta,\eta)$. In order to prove that such replacement is
allowed we proceed as follows. First we will show that the
integration over $y$ in equation (\ref{E}) can be resticted to a
domain on which the asymptotics of $\psi_{N-1}(x)$ is given by
Proposition \ref{PropositionAsymOfPsiIntheBalk}. Second we will
observe that the integration over $x$ in equation (\ref{gN}) can
be restricted to the same domain as well.

Let $K$ denote a compact subset of $\mathbb{C}\backslash
\mathbb{R}$. The function $g_N(y,\dfrac{\eta}{\sqrt{2N}})$ has the
following bound which is uniform in $y\in \mathbb{R}$ and in
$\eta\in K$,
\begin{equation}\label{UniformBoundg}
\underset{\eta\in\; \mathbb{K}
}{\sup}\biggl\{g_N(y,\frac{\eta}{\sqrt{2N}})\biggr\}\leq
\const\cdot N^{\alpha},\;\;\;\; \alpha-\mbox{real positive.}
\end{equation}
In order to prove this inequality we rewrite the integral in
equation (\ref{gN}) as a sum of three integrals:
\begin{equation}\label{gN1}
\begin{split}
g_N(y,\dfrac{\eta}{\sqrt{2N}})=&\int\limits_{-\sqrt{2N}(1+\epsilon)}^{\sqrt{2N}(1+\epsilon)}
\frac{\sgn(y-x)\psi_N(x)\;dx}{x-\frac{\eta}{\sqrt{2N}}}+\int\limits_{\sqrt{2N}(1+\epsilon)}^{+\infty}
\frac{\sgn(y-x)\psi_N(x)\;dx}{x-\frac{\eta}{\sqrt{2N}}}\\
&+ \int\limits_{-\infty}^{-\sqrt{2N}(1+\epsilon)}
\frac{\sgn(y-x)\psi_N(x)\;dx}{x-\frac{\eta}{\sqrt{2N}}},
\end{split}
\end{equation}
where $\epsilon>0$. The following estimate  is evident from the
asymptotic expressions for the function $\psi_N(x)$ (see Section
\ref{SectionAsympsiN}),
\begin{equation}
|\psi_N(x)|< e^{-\const \sqrt{N} |x|},\;\;\dfrac{|x|}{\sqrt{2N}}>
1+\epsilon,\; \epsilon >0.\nonumber
\end{equation}
Using this estimate we obtain that the second and the third
integral in equation (\ref{gN1}) are exponentially small. As for
the first integral in equation (\ref{gN1}) it is clearly
polynomially bounded since $|x-\frac{\eta}{\sqrt{2N}}|\geq
\frac{\const}{\sqrt{2N}}$, ($\Img \eta\neq 0$), and
$\underset{x\in \mathbb{R}}{\sup}|\psi_N(x)|\leq
\frac{\const}{N^{1/12}}$ (see Corollary \ref{UniformBoundForPsi}).
Thus inequality (\ref{UniformBoundg}) is proved.

Let $M\rightarrow \infty$ as $N\rightarrow\infty$, but
$\dfrac{M}{\sqrt{N}}\rightarrow 0$ as $N\rightarrow\infty$. Then
we claim that
\begin{equation}
E_N\left(\dfrac{\zeta}{\sqrt{2N}},\dfrac{\eta}{\sqrt{2N}}\right)=\int\limits_{-M}^{M}
\frac{e^{-y^2/2}g_N(y,\frac{\eta}{\sqrt{2N}})\;dy}{\frac{\zeta}{\sqrt{2N}}-y}+\mathcal{O}(N^{\alpha}e^{-M^2/2}),
\nonumber
\end{equation}
where the error term, $\mathcal{O}(N^{\alpha}e^{-M^2/2})$, is
uniform for $\zeta,\eta$ taken from a compact subset of
$\mathbb{C}\backslash \mathbb{R}$. Indeed,  let us take $M$ as
described above and rewrite
$E_N\left(\dfrac{\zeta}{\sqrt{2N}},\dfrac{\eta}{\sqrt{2N}}\right)$
as
\begin{equation}
\begin{split}
E_N\left(\dfrac{\zeta}{\sqrt{2N}},\dfrac{\eta}{\sqrt{2N}}\right)=&\int\limits_{-M}^{M}
\frac{e^{-y^2/2}g_N(y,\frac{\eta}{\sqrt{2N}})\;dy}{\frac{\zeta}{\sqrt{2N}}-y}+
\int\limits_{M}^{+\infty}
\frac{e^{-y^2/2}g_N(y,\frac{\eta}{\sqrt{2N}})\;dy}{\frac{\zeta}{\sqrt{2N}}-y}\\
&+\int\limits_{-\infty}^{-M}
\frac{e^{-y^2/2}g_N(y,\frac{\eta}{\sqrt{2N}})\;dy}{\frac{\zeta}{\sqrt{2N}}-y}.
\end{split}\nonumber
\end{equation}
and estimate the second and the third integrals. Namely,
\begin{equation}
\begin{split}
\biggl|\int\limits_{M}^{+\infty} \frac{e^{-y^2/2}g_N(y,
\frac{\eta}{\sqrt{2N}})\;dy}{\frac{\zeta}{\sqrt{2N}}-y}\biggr|
&\leq \biggl|\int\limits_{M}^{+\infty}
e^{-y^2/2}g_N(y,\frac{\eta}{\sqrt{2N}})y\;dy\biggr| \\
&\leq\const\cdot N^{\alpha}\int\limits_{M}^{+\infty}
e^{-y^2/2}y\;dy=\const\cdot N^{\alpha}\cdot e^{-M^2/2}.
\end{split}\nonumber
\end{equation}
where equation (\ref{UniformBoundg}) was used. This shows that the
integral over $y$ in the expression for
$E_N(\frac{\zeta}{\sqrt{2N}}, \frac{\eta}{\sqrt{2N}})$ can be
restricted to the domain $[-M,M]$ with a small error term.

Now assume that $y$ is taken from the interval $[-M,M]$, and
$M=M(N)$ is chosen such that $M\rightarrow \infty$ as
$N\rightarrow\infty$, but $\dfrac{M}{\sqrt{N}}\rightarrow 0$ as
$N\rightarrow\infty$. Then
\begin{equation}\label{gnlimiting}
\int\limits_{-\infty}^{+\infty}\dfrac{\sgn(y-x)\psi_N(x)\;dx}{x-\frac{\eta}{\sqrt{2N}}}
=\int\limits_{-M}^{M}\dfrac{\sgn(y-x)\psi_N(x)\;dx}{x-\frac{\eta}{\sqrt{2N}}}
+\mathcal{O}\biggl\{\Max[M N^{-3/4},M^{-2}N^{-1/12}]\biggr\},
\end{equation}
as $N\rightarrow\infty$. The error term,
$\mathcal{O}\biggl\{\Max[M N^{-3/4},M^{-2}N^{-1/12}]\biggr\}$, is
uniform on $y\in \mathbb{R}$, $\eta\in\; \mathbb{C}\backslash
\mathbb{R}$, where $\eta$ is taken from a compact subset of
$\mathbb{C}\backslash \mathbb{R}$ . In order to prove
(\ref{gnlimiting})  we decompose the integral in the left-hand
side of equation (\ref{gnlimiting}) into three integrals on
intervals $[-M,M]$, $(-\infty,-M)$, and $(M,+\infty)$. The
integrals on intervals $(-\infty,-M)$ and $(M,+\infty)$ are
evaluated using Proposition \ref{PropositionEstpsixx}.

To summarize we have proved that
\begin{equation}
\begin{split}
E_N(\dfrac{\zeta}{\sqrt{2N}},\dfrac{\eta}{\sqrt{2N}})=&
\int\limits_{-M}^{M}\dfrac{e^{-y^2/2}dy}{\frac{\zeta}{\sqrt{2N}}-y}
\left\{\int\limits_{-M}^{M}\dfrac{\sgn(y-x)\psi_N(x)dx}{x-\frac{\eta}{\sqrt{2N}}}+
\mathcal{O}\left[\Max[MN^{-3/4}, M^{-2}N^{-1/12}]\right]\right\}\\
&+ \mathcal{O}(N^{\alpha}e^{-M^2/2}).
\end{split}\nonumber
\end{equation}
 The error term $\mathcal{O}(N^{\alpha}e^{-M^2/2})$ is exponentially
 small. Integrating the error term $\mathcal{O}\left[\Max[\frac{M}{N^{3/4}},
 \frac{1}{M^{2}N^{1/12}}]\right]$
 we obtain:
 \begin{equation}
 \begin{split}
&\biggl|\int\limits_{-M}^{M}\dfrac{e^{-y^2/2}dy}{\frac{\zeta}{\sqrt{2N}}-y}\;\mathcal{O}\left[\Max[MN^{-3/4},
M^{-2}N^{-1/12}]\right]\biggr|=\\
&=\biggl|\int\limits_{-M\sqrt{2N}}^{M\sqrt{2N}}\dfrac{e^{-y^2/4N}dy}{\zeta-y}\;\mathcal{O}\left[\Max[MN^{-3/4},
M^{-2}N^{-1/12}]\right]\biggr|\leq\\
&\leq \mathcal{O}\left[\Max[M\log NN^{-3/4}, M^{-2}\log
NN^{-1/12}]\right].
\end{split}\nonumber
\end{equation}
Another error term is due to the replacement of $\psi_N(x)$ by its
asymptotics taken from Proposition
\ref{PropositionAsymOfPsiIntheBalk}. This error term is evaluated
as follows:
\begin{equation}
\biggl|\int\limits_{-M\sqrt{2N}}^{M\sqrt{2N}}\dfrac{e^{-y^2/4N}dy}{\zeta-y}
\int\limits_{-M\sqrt{2N}}^{M\sqrt{2N}}\dfrac{dx}{x-\eta}
\;\mathcal{O}\left[\dfrac{M^3}{N^{3/4}}\right]\biggr|\leq
\mathcal{O}\left[\dfrac{M^3(\log N)^2}{N^{3/4}}\right].\nonumber
\end{equation}
This completes the proof.
\end{proof}
\subsubsection{Asymptotics of $F_N(\zeta,\eta)$}
\begin{prop}\label{F}
Let $\zeta,\eta$ be chosen from a compact subset of
$\mathbb{C}\backslash \mathbb{R}$. Then
\begin{equation}\label{EqF}
\begin{split}
F_N(\dfrac{\eta}{2\sqrt{N}},\dfrac{\zeta}{2\sqrt{N}})=
2N\biggl[\dfrac{2^{1/4}}{\pi^{1/2}N^{1/4}}&\int\limits_{-M}^{M}dt\;\cos(t-\dfrac{\pi
N}{2})\left[\dfrac{1}{\eta-t}\dfrac{d}{dt}\dfrac{e^{-t^2/4N}}{\zeta-t}\right]\\
&+
\mathcal{O}\left(Max\left[\dfrac{M^4}{N^{3/4}},\dfrac{1}{N^{1/12}M^2}\right]\right)\biggr]
\end{split}
\end{equation}
where $M=M(N)$ is chosen such that $M(N)\rightarrow\infty$ as
$N\rightarrow\infty$, but $\dfrac{M(N)}{\sqrt{N}}\rightarrow 0$ as
$N\rightarrow\infty$.
\end{prop}
\begin{proof}
Recall that $F_N(\zeta,\eta)$ was defined by an integral in
equation (\ref{FEXACT}). Change the variable of integration in
this integral, and decompose it into three integrals. The first
one runs from $-\infty$ to $-M$, the second one runs from $-M$ to
$M$, and the third one runs from $M$ to $+\infty$. In the second
integral replace the function $\psi_N(x)$ by its asymptotics in
the bulk of the spectrum (see Proposition
\ref{PropositionAsymOfPsiIntheBalk}). The integration of the error
term gives $\mathcal{O}(\dfrac{M^4}{N^{3/4}})$. The first and the
third integrals are of order
$\mathcal{O}(\dfrac{1}{M^2N^{1/12}})$. It can be seen rewriting
the expression
$\dfrac{1}{\eta-t}\dfrac{d}{dt}\dfrac{e^{-t^2/4N}}{\zeta-t}$
explicitly, and using the fact that $\underset{t\in
R}{\sup}\left|\psi_N(t)\right|\leq\const\cdot N^{-1/12}$
(Corollary \ref{UniformBoundForPsi}).
\end{proof}
\begin{rem}
The error term,
$\mathcal{O}\left(Max\left[\dfrac{M^4}{N^{3/4}},\dfrac{1}{N^{1/12}M^2}\right]\right)$,
is negligible in comparison with the first term in the brackets of
equation (\ref{EqF}), if the following conditions are satisfied:
\begin{equation}
\dfrac{M^4}{N^{3/4}}\ll\dfrac{1}{N^{1/4}},\;\;\dfrac{1}{N^{1/12}M^2}\ll\dfrac{1}{N^{1/4}}.
\nonumber
\end{equation}
The first condition implies that $M\ll N^{1/8}$, and the second
equation implies that $N^{1/12}\ll M$. Therefore if $M(N)$ is
chosen such that $N^{1/12}\ll M\ll N^{1/8}$ the error term is
negligible in comparison with the main term.
\end{rem}
\subsection{Asymptotics of the kernels: GOE case}
\subsubsection{Asymptotics of
$W_{N,I}^{(1)}(\dfrac{\zeta}{\sqrt{2N}},\dfrac{\eta}{\sqrt{2N}})$}
\begin{thm}(Scaling limit of the first kernel) For $\zeta\in \mathbb{C}$, $\eta\in \mathbb{C}$,
\begin{equation}
\underset{N\rightarrow\infty}{\lim}
\biggl\{\dfrac{1}{2N}W_{N,I}^{(1)}(\dfrac{\zeta}{\sqrt{2N}},
\dfrac{\eta}{\sqrt{2N}})\biggr\}= -\dfrac{1}{\pi}\dfrac{d}{d\zeta}
\left[\dfrac{\sin(\zeta-\eta)}{\zeta-\eta}\right].\nonumber
\end{equation}
\end{thm}
\begin{proof}
Use formula (\ref{WNIEXACT}), the following relation which follows
immediately from the properties of the Hermite polynomials,
\begin{equation}
\dfrac{d}{dx}\psi_n(x)=n\dfrac{c_{n-1}}{c_n}\psi_{n-1}(x)-x\psi_n(x)
\nonumber
\end{equation}
together with the asymptotic formula for the functions
$\psi_n(x)$, Proposition
\ref{PropositionAsymptoticsofLargePsiBalk}.
\end{proof}
\subsubsection{Asymptotics of
$W_{N,II}^{(1)}(\dfrac{\zeta}{\sqrt{2N}},\dfrac{\eta}{\sqrt{2N}})$}
\begin{prop}\label{proposititionIITechnical}
Take $\zeta,\eta\in\mathbb{C}\backslash \mathbb{R}$. Then
\begin{equation}
\underset{N\rightarrow\infty}{\lim}\biggl\{\sqrt{2N}\;\dfrac{
\Psi_N(\frac{\eta}{\sqrt{2N}})\psi_{N-1}(\frac{\zeta}{\sqrt{2N}})
-\Psi_{N-1}(\frac{\eta}{\sqrt{2N}})\psi_{N}(\frac{\zeta}{\sqrt{2N}})}{\zeta-\eta}\biggr\}
=
  \begin{cases}
    \dfrac{2}{\zeta-\eta}\;e^{i(\eta-\zeta)}, & \Img\eta >0, \\
   \dfrac{2}{\zeta-\eta}\;e^{-i(\eta-\zeta)}, & \Img\eta <0 \\
  \end{cases}\nonumber
\end{equation}
\end{prop}
\begin{proof}
Apply Propositions \ref{PropositionAsymOfPsiIntheBalk} and
\ref{PropositionAsymptoticsofLargePsiBalk}.
\end{proof}
\begin{thm}(Scaling limit of the second kernel)
Let $\zeta\in \mathbb{C}$, $\eta\in \mathbb{C}\backslash
\mathbb{R}$. Then
\begin{equation}
\underset{N\rightarrow\infty}{\lim}\dfrac{1}{\sqrt{2N}}
W_{N,II}^{(1)}(\dfrac{\zeta}{\sqrt{2N}},\dfrac{\eta}{\sqrt{2N}}) =
  \begin{cases}
    \dfrac{e^{i(\eta-\zeta)}}{\zeta-\eta} & \Img \eta >0, \\
    \dfrac{e^{-i(\eta-\zeta)}}{\zeta-\eta} & \Img \eta<0.
  \end{cases}\nonumber
\end{equation}
\end{thm}
\begin{proof}
This follows immediately from equation (\ref{ExactWIIBrief}),
Propositions \ref{proposititionIITechnical},
\ref{PropositionBoundOfI}, and from the fact that
$\dfrac{c_N}{c_{N-1}}\sim\sqrt{\dfrac{N}{2}}$ as
$N\rightarrow\infty$.
\end{proof}
\subsubsection{Asymptotics of
$W_{N,III}^{(1)}(\dfrac{\zeta}{\sqrt{2N}},\dfrac{\eta}{\sqrt{2N}})$}
\begin{thm}(Scaling limit of the third kernel)
Take $\zeta\in \mathbb{C}\backslash \mathbb{R}$, $\eta\in
\mathbb{C}\backslash \mathbb{R}$. Then
\begin{equation}\label{WIIIA}
\underset{N\rightarrow\infty}{\lim}W_{N,III}^{(1)}(\dfrac{\zeta}{\sqrt{2N}},
\dfrac{\eta}{\sqrt{2N}})=2\pi i
  \begin{cases}
    \quad\int\limits_{1}^{+\infty}\dfrac{e^{i(\zeta-\eta)t}dt}{t} & \Img \zeta >0, \Img \eta <0 \\
    -\int\limits_{1}^{+\infty}\dfrac{e^{-i(\zeta-\eta)t}dt}{t} & \Img \zeta <0, \Img \eta >0 \\
    \quad 0 & \text{otherwise}.
  \end{cases}
,\;\;\;\;\;
\zeta\in\;\mathbb{C}\backslash\mathbb{R},\;\;\;\eta\in\;\mathbb{C}\backslash\mathbb{R}.
\end{equation}
\end{thm}
\begin{proof}
Recall that the kernel $W_{N,III}^{(1)}(\zeta,\eta)$ is given by
equation (\ref{ExactWNIIIbrief}). In order to determine the large
$N$ asymptotics of $W_{N,III}^{(1)}(\zeta,\eta)$ consider the
expression:
\begin{equation}
A_N(\zeta,\eta)\equiv
E_{N-1}(\zeta,\eta)\Psi_{N-2}(\zeta)-E_{N-2}(\zeta,\eta)\Psi_{N-1}(\zeta).
\nonumber
\end{equation}
The asymptotics for $E_N(\zeta,\eta)$ (equation (\ref{EAs})), and
the asymptotics for $\Psi_{N}(\zeta)$ (equation
(\ref{LargePsiAs2})) give us the leading term of
$A_N(\zeta,\eta)$:
\begin{equation}\label{AN}
A_N(\zeta,\eta)\simeq\left\{%
\begin{array}{ll}
    \sqrt{\dfrac{2}{N}}\int\limits_{\mathbb{R}}\dfrac{dy\;e^{-y^2/4N}}{\eta-y}\int\limits_{\mathbb{R}}\dfrac{e^{i(\zeta-x)}\sgn(y-x)}{\zeta-x}, & \Img \zeta>0 \\
    -\sqrt{\dfrac{2}{N}}\int\limits_{\mathbb{R}}\dfrac{dy\;e^{-y^2/4N}}{\eta-y}\int\limits_{\mathbb{R}}\dfrac{e^{-i(\zeta-x)}\sgn(y-x)}{\zeta-x}, & \Img \zeta<0
\end{array}%
\right.
\end{equation}
In what follows  the  integral representation for $\sgn(y-x)$ will
play a role,
\begin{equation}
\sgn(y-x)=\dfrac{1}{i\pi}\int\limits_{\mathbb{R}}\dfrac{e^{i(y-x)t}dt}{t}.
\nonumber
\end{equation}
If we insert this expression to equation (\ref{AN}), and perform
the computations formally (i. e. changing the order of
integrations, and replacing $e^{-y^2/4N}$ by $1$) we obtain:
\begin{equation}\label{ANA}
A_N(\dfrac{\zeta}{\sqrt{2N}},\dfrac{\eta}{\sqrt{2N}})\simeq 2\pi i
\begin{cases}
    \quad\dfrac{4}{\sqrt{2N}}\int\limits_{1}^{+\infty}\dfrac{e^{i(\zeta-\eta)t}dt}{t} & \Img \zeta >0, \Img \eta <0 \\
    -\dfrac{4}{\sqrt{2N}}\int\limits_{1}^{+\infty}\dfrac{e^{-i(\zeta-\eta)t}dt}{t} & \Img \zeta <0, \Img \eta >0 \\
    \quad 0 & \text{otherwise}.
  \end{cases}
\end{equation}
Since the factor
$\frac{1}{2}\frac{c_{N-1}}{c_{N-2}}\simeq\frac{\sqrt{N}}{2\sqrt{2}}$,
as $N\rightarrow\infty$, and the term
$I_{N-1}(\frac{\zeta}{\sqrt{2N}})I_{N-2}(\frac{\zeta}{\sqrt{2N}})$
is negligible in the large $N$ limit in comparison with
$A_N(\frac{\zeta}{\sqrt{2N}},\frac{\eta}{\sqrt{2N}})$ (see
Proposition \ref{PropositionBoundOfI}), formula (\ref{WIIIA})
follows.

However these  computations involve the integrals which are not
absolutely convergent, so formula (\ref{ANA}) must be justified.
In order to prove this formula rigorously we observe that the
following holds:
\begin{equation}\label{IntegralOtSGN}
\int\limits_{\mathbb{R}}\dfrac{e^{i(\zeta-x)}\sgn(y-x)}{\zeta-x}=2\int\limits_{1}^{+\infty}\dfrac{e^{-i(y-\zeta)t}\;dt}{t},\;\;\;\Img\zeta>0
\end{equation}
Indeed, the integral in the left-hand side of equation
(\ref{IntegralOtSGN}) is equal to
$2\int_{y}^{+\infty}\tfrac{dx\;e^{i(\zeta-x)}}{x-\zeta}$ (since
$\int_{-\infty}^{+\infty}\tfrac{dx\;e^{i(\zeta-x)}}{x-\zeta}=0$,
when $\Img\zeta>0$). To check that
$\int_{y}^{+\infty}\tfrac{dx\;e^{i(\zeta-x)}}{x-\zeta}$
 is
equal to
$\int\limits_{1}^{+\infty}\tfrac{e^{-i(y-\zeta)t}\;dt}{t}$ we
first observe that the derivatives with respect to $y$ are equal
to each other. Second, it is evident that the first integral tends
to zero as $y\rightarrow +\infty$. The fact that the second
integral tends to zero when $y\rightarrow +\infty$ can be proved
by partial integration. Thus it remains to compute the large $N$
limit of the expression
\begin{equation}\label{2Integrala}
\int\limits_{\mathbb{R}}\dfrac{dy\;e^{-y^2/4N}}{\eta-y}\int\limits_{1}^{+\infty}\dfrac{e^{-i(y-\zeta)t}dt}{t},\;\;\;
\Img \zeta>0.
\end{equation}
The integrand, as a function in $(y,t)$, decays exponentially as
$\mbox{max}(|y|,|t|)\rightarrow\infty$ Therefore we are allowed to
change the order of integration, and expression (\ref{2Integrala})
is equal to
\begin{equation}\label{2IntNaoborot}
\int\limits_{1}^{+\infty}\dfrac{e^{i\zeta
t}dt}{t}\int\limits_{\mathbb{R}}\dfrac{dy\;e^{-iyt-y^2/4N}}{\eta-y},\;\;\;
\Img \zeta>0.
\end{equation}
The large $N$ limit of the inner integral can be computed as
follows. Take $M=M(N)$ such that $M(N)\rightarrow\infty$ as
$N\rightarrow\infty$, but $M^2(N)/N\rightarrow 0$ as
$N\rightarrow\infty$. Decompose the inner integral into three
integrals: the first one runs from $-\infty$ to $-M$, the second
one runs from $-M$ to $M$, and the third one runs from $M$ to
$+\infty$. Then the first and the third integrals are of order
$\mathcal{O}(\frac{1}{M})$. As for the second integral, it
asymptotically equals to $(2\pi i)e^{-i\eta t}$, when
$\Img\eta<0$, and to zero, when $\Img\eta>0$. It can be observed
considering the semicircle contour $C_M\cup [-M,M]$, where $C_M$
is a semicircle from the point $M$ to the point $-M$ passing
through lower half of the complex plane. The Jordan Lemma implies
that
\begin{equation}
\biggl|\int_{C_M}dz\dfrac{e^{-z^2/4N}e^{-izt}}{\eta-z}\biggr|\leq\dfrac{\const\;e^{-M^2/4N}}{M\cdot
t}
\end{equation}
Since $t\geq 1$ we obtain that
\begin{equation}
\int_{C_M}dz\dfrac{e^{-z^2/4N}e^{-izt}}{\eta-z}=\mathcal{O}\left(\dfrac{1}{M}\right)
\end{equation}
where the estimate is uniform for $t\in[1,+\infty)$. Thus the
inner integral in equation (\ref{2IntNaoborot}) is determined by
the residue at $\eta$ of the integrand (uniformly for
$t\in[1,+\infty)$ with error of order
$\mathcal{O}\left(\dfrac{1}{M}\right)$). Replacing the inner
integral by this residue we again arrive to formula (\ref{ANA}).
\end{proof}

\subsection{Asymptotics of the kernels: GSE case}
\subsubsection{Asymptotic of the first kernel}
\begin{thm}(Scaling limit of the first kernel)
For complex $\zeta, \eta$
\begin{equation}\label{WNIAS}
 \underset{N\rightarrow\infty}{\lim}\;
W_{N,I}^{(4)}(\dfrac{\zeta}{2\sqrt{N}},\dfrac{\eta}{2\sqrt{N}})=\dfrac{1}{\pi}\int\limits_0^{1}
dt\;\dfrac{\sin(\zeta-\eta) t}{t}
\end{equation}
\end{thm}
\begin{proof}
Propositions \ref{ProposititionWNI} gives exact formula for the
kernel  $W_{N,I}^{(4)}$ (equation (\ref{WI4})). The second term in
the brackets in equation (\ref{WI4}) is the product of two
integrals. The first integral is of order $\mathcal{O}(N^{-3/4})$,
since $\left|\psi_{2N}(\frac{u}{2\sqrt{N}})\right|\leq\const\cdot
N^{-1/4}$ uniformly for $u$ chosen from a compact real interval.
(Note that the asymptotics of functions $\psi_N$ in a real
neighborhood of zero can be extended to a complex neighborhood of
zero , see Szeg\"o \cite{Szego}; Deift, Kriecherbauer, McLaughlin,
Venakides and Zhou \cite{deift2}, such that the bound
$\left|\psi_{2N}(\frac{u}{2\sqrt{N}})\right|\leq\const\cdot
N^{-1/4}$ remains valid). The second integral is bounded by
$\const\cdot N^{-1/4}$ (see Nagao and Wadati \cite{nagao},
equation (A.1)). Thus the second term in the brackets in equation
(\ref{WI4}) (where $\zeta$ is replaced by
$\dfrac{\zeta}{2\sqrt{N}}$, and $\eta$ is replaced by
$\dfrac{\eta}{2\sqrt{N}}$) is of order $\mathcal{O}(N^{-1})$. As
for the first term in the brackets of equation (\ref{WI4}) it is
of order $\mathcal{O}(N^{-1/2})$. Indeed, this term is an integral
over a compact interval, so the functions $\psi_{2N+2}$,
$\psi_{2N+1}$ can be replaced by their asymptotic expressions (
Proposition \ref{PropositionAsymOfPsiIntheBalk}) in the bulk of
the spectrum. Thus we observe that the second term in the brackets
in equation (\ref{WI4}) is negligible with respect to the first
term, and formula (\ref{WNIAS}) is obtained.
\end{proof}
\subsubsection{Asymptotics of the second kernel}
\begin{thm}(Scaling limit of the second kernel)
Let $\zeta\in \mathbb{C}$. Then
\begin{equation}
\underset{N\rightarrow\infty}{\lim}\dfrac{1}{2\sqrt{N}}
W_{N,II}^{(4)}(\dfrac{\zeta}{2\sqrt{N}},\dfrac{\eta}{2\sqrt{N}}) =
  \begin{cases}
    \dfrac{e^{i(\eta-\zeta)}}{\zeta-\eta} & \Img \eta >0, \\
    \dfrac{e^{-i(\eta-\zeta)}}{\zeta-\eta} & \Img \eta<0.
  \end{cases}
\end{equation}
\end{thm}
\begin{proof}
The exact expression for the second kernel in the case of GSE is
essentially the same as in the case of GOE. Thus the asymptotics
is obtained by the same method.
\end{proof}
\subsubsection{Asymptotics of the third kernel}
\begin{thm}(Scaling limit of the third kernel)
Let $\zeta$, $\eta$ are chosen from a compact subset of
$\mathbb{C}\backslash \mathbb{R}$. Then
\begin{equation}\label{FINALASYMWIII}
\begin{split}
\underset{N\rightarrow\infty}{\lim}\left[\dfrac{1}{4N}W_{N,III}^{(4)}(\dfrac{\zeta}{2\sqrt{N}},
\dfrac{\eta}{2\sqrt{N}})\right]=2\pi i
  \begin{cases}
    e^{i\eta} \dfrac{d}{d\zeta}\left(\dfrac{e^{-i\zeta}}{\eta-\zeta}\right) & \Img \zeta <0, \Img \eta >0 \\
   - e^{-i\eta} \dfrac{d}{d\zeta}\left(\dfrac{e^{i\zeta}}{\eta-\zeta}\right) & \Img \zeta >0, \Img \eta <0 \\
    \quad 0 & \text{otherwise}.
  \end{cases}
.
\end{split}
\end{equation}
\end{thm}
\begin{proof}
Consider equation (\ref{WIIIFINAL}). Replace the functions
$\Psi_{2N-2}$, $\Psi_{2N-3}$, $F_{2N-2}$, $F_{2N-3}$ by their
asymptotics given by equations (\ref{LargePsiAs2}) and
(\ref{EqF}). The first two terms in the brackets of equation
(\ref{WIIIFINAL}) are then represented by an integral which can be
directly computed by residue calculations. As for the third term
in the brackets of equation (\ref{WIIIFINAL}) it is asymptotically
negligible in comparison with two first terms.
\end{proof}
\begin{rem} The asymptotics of  the kernels
$W_{N,I}^{(2)}(\alpha,\beta)$, $W_{N,II}^{(2)}(\alpha,\beta)$,
$W_{N,III}^{(2)}(\alpha,\beta)$ is determined in Ref.
\cite{strahov}. Note that this asymptotics can be obtained
exploiting the asymptotic expressions for $\psi_N$ ( Proposition
\ref{PropositionAsymOfPsiIntheBalk}) and for $\Psi_N$ (Proposition
\ref{PropositionAsymptoticsofLargePsiBalk}).
\end{rem}
\section{Appendix. Some useful facts from the Linear Algebra}
\subsection{The formula for the Cauchy determinant} Let
$A=(a_1,\ldots ,a_k)$, $B=(b_1,\ldots ,b_k)$ be two
non-intersecting sets. Then the following formula is valid:
\begin{equation}\label{CauchyDeterminant}
\mbox{det}\left(\dfrac{1}{a_i-b_j}\right)_{i,j=1}^k=
(-)^{\frac{k(k-1)}{2}}\;\dfrac{V(A)V(B)}{\prod(A;B)}
\end{equation}
\subsection{Determinant of the block matrix}
If the matrix $H$ has the following block structure,
\begin{equation}
H=\left(
\begin{array}{cc}
  0 & A \\
  B & 0 \\
\end{array}
\right)
\end{equation}
with square matrices $A$ and $B$, then
\begin{equation}\label{determinantoftheblockmatrix}
\mbox{det}\; H=(-)^{\vert A\vert\cdot\vert B\vert}\;\mbox{det}\;
A\cdot\mbox{det}\; B.
\end{equation}
\subsection{Minors of the inverse matrix}
Assume that two $N\times N$ matrices $A$ and $B$ are inverse to
each other,
\begin{equation}
A=B^{-1}.
\end{equation}
Then an arbitrary minor of the matrix $A$ corresponding to the
rows $\alpha_1,\ldots ,\alpha_r$ and columns $\beta_1,\ldots
,\beta_r$ can be expressed as follows:
\begin{equation}
\mbox{det}\; A(\alpha_1,\ldots ,\alpha_r\vert\beta_1,\ldots
,\beta_r)=\qquad\qquad\qquad\qquad\qquad
\end{equation}
\begin{equation}
\qquad
=(-)^{\sum\limits_{i=1}^{r}\alpha_i+\beta_i}\;\dfrac{\mbox{det}\;B\left(1,\ldots
,\check{\beta_1},\ldots,\check{\beta_r},...,N\vert 1,\ldots
,\check{\alpha_1},\ldots,\check{\alpha_r},...,N\right)}{\mbox{det}\;
B}\nonumber
\end{equation}
Here $\check{k}$ denotes the $k^{\mbox{th}}$ row (or column) which
is removed from the matrix.
\subsection{The expansion of the minors of the matrix $I+A$}
Let $\check{\alpha}=(\check{\alpha_1},\ldots,\check{\alpha_r})$,
$\check{\beta}=\left(\check{\beta_1},\ldots,\check{\beta_r}\right)$
denote removed columns or rows. Assume that
$\check{\alpha}\cap\check{\beta}$ is empty. Then the following
expansion of the minor of the matrix $I+A$ can be obtained:
\begin{equation}
\mbox{det}\left(I+A\right)\left(1,\ldots
,\check{\alpha_1},\ldots,\check{\alpha_r},...,N\vert 1,\ldots
,\check{\beta_1},\ldots,\check{\beta_r},...,N\right)=
\end{equation}
\begin{equation}
(-)^{\sum\limits_{i=1}^{r}\alpha_i+\beta_i+r}
\sum\limits_{X}\mbox{det}\; A\left(\beta_1,\ldots
,\beta_r,X\vert\alpha_1,\ldots ,\alpha_r,X\right)\nonumber
\end{equation}
Here the sum is over all subsets $X$ of the set $[1,\ldots ,N]$,
which do not intersect $\alpha$ and $\beta$.
\subsection{The minors of the matrix $K$}
Applying two previous expressions it is easy to relate the minors
of the matrix $K$  defined  in terms of the $L$ matrix of the
$L$-ensemble as
\begin{equation}
K=I-\dfrac{I}{I+L}
\end{equation}
with minors of the matrix $L$. Namely we have
\begin{equation}\label{KMinor}
\mbox{det}\; K(\alpha\vert\beta)=
\sum\limits_{X}\dfrac{\mbox{det}\;L(\alpha,X\vert\beta,X)}{\mbox{det}\;(I+L)}
\end{equation}
where $\alpha=(\alpha_1,\ldots ,\alpha_r)$, $\beta=(\beta_1,\ldots
,\beta_r)$, $\alpha_i\neq \beta_j$. Here again the sum is over all
subsets $X$ of the set $[1,\ldots ,N]$.
\subsection{Definition of Pfaffian}\label{ASectionDefinitionOFPfaffian}
 The Pfaffian of a $2N\times
2N$ antisymmetric matrix $A=\|A_{jk}\|_{j,k=1}^{2N}$ is defined as
\begin{equation}
\PF\;A=\sum_{\substack{ \sigma=(i_1,\ldots ,i_{2N})\in S_{2N}\\
i_1<i_2,\ldots ,i_{2N-1}<i_{2N}\\
i_1<i_3<\ldots <i_{2N-1}}} \mbox{sgn}(\sigma)\;A_{i_1i_2}\ldots
A_{i_{2N-1}i_{2N}}
\end{equation}
\subsection{Pfaffian of the block matrix}
The following formula is valid
\begin{equation}
\PF\bigg[\begin{array}{cc}
  0 & A \\
  -A^{\dag} & 0 \\
\end{array}\biggr]=\mbox{det}\; A\;(-)^{\frac{|A|(|A|-1)}{2}}
\end{equation}
\subsection{The expansion of Pfaffians of submatrices of the
matrix $J+A$} Let A be a $2N\times 2N$ antisymmetric matrix,
\begin{equation}\label{AppendixA}
\begin{split}
A[1'&,1'',\ldots ,N',N''\vert 1',1'',\ldots ,N',N'']=\\
&\left[%
\begin{array}{ccccc}
 0 & A(1',1'') & \ldots & A(1',N') & A(1',N'') \\
  -A(1',1'') & 0 &  & A(1'',N') & A(1'',N'')\\
  \vdots &  &  &  &  \\
  -A(1',N') & -A(1',N'') &  & 0 & A(N',N'') \\
  -A(1',N'') & -A(1'',N'') &  & -A(N',N'') & 0 \\
\end{array}%
\right]
\end{split}
\end{equation}
and $J$ be the antisymmetric matrix of format $2N\times 2N$
defined by
\begin{equation}
J=\left[%
\begin{array}{ccccccc}
  0 & 1 & 0 & 0 & \ldots & 0 & 0 \\
  -1 & 0 & 0 & 0 &  & 0 & 0 \\
  0 & 0 & 0 & 1 &  & 0 & 0 \\
  0 & 0 & -1 & 0 &  & 0 & 0 \\
  \vdots &  &  &  &  &  & \\
  0 & 0 & 0 & 0 & \ldots & 0 & 1 \\
  0 & 0 & 0 & 0 & \ldots & -1 & 0 \\
\end{array}%
\right]
\end{equation}
then the following expansion is valid ($2\leq 2k\leq N$):
\begin{equation}\label{AppendixPffafianJA}
\begin{split}
&\PF\;(J+A)\biggl[1',2',\ldots , 2k'\arrowvert 1',2',\ldots ,2k'\biggr]= \\
&\sum_{\substack{X=(x_1',x_1'',\ldots ,x_d',x_d'')\\
(x_1,\ldots x_d)\subset (2k+1,\ldots
,N)}}\PF\;A\biggl[1'',2'',\ldots ,2k'',X\arrowvert 1'',2'',\ldots
,2k'',X\biggr]
\end{split}
\end{equation}
\subsection{The Pfaffian of a submatrix of the inverse matrix}
With $A$ defined by equation (\ref{AppendixA}) consider the
inverse matrix $B=A^{-1}$. For $2\leq 2m\leq N$ we find
\begin{equation}\label{AppendixPfaffianOfSubmatrixOfTheInverse}
\begin{split}
&\PF\;B\biggl[1',2',\ldots , 2m'\arrowvert 1',2',\ldots
,2m'\biggr]=
\dfrac{1}{\PF\;A}\\
&\times\;\PF\;A\biggl[ 1'',\ldots ,2m'',2m+1',2m+1'',\ldots ,N',N''\arrowvert\\
& \qquad\qquad\qquad\qquad 1'',\ldots ,2m'',2m+1',2m+1'',\ldots
,N',N''\biggr]
\end{split}
\end{equation}

\newpage
\setlength{\unitlength}{1cm}
\begin{picture}(5,10) (5,5)
\put (5,5){\line(1,0){15}} \put (5,5){\line(0,1){5}} \put
(20,5){\line(0,1){5}} \put (5,10){\line(1,0){15}}
\put(12.5,10.5){$\X$}
\put(5.1,8){\line(1,0){7.4}} \put(5.1,8){\line(0,-1){3}}
\put(12.6,8){\line(1,0){7.2}} \put(19.9,8){\line(0,-1){3}}
\put(8,8.5){$\X_-$}\put(17,8.5){$\X_+$}
\put(12.5,8){\line(0,-1){3}} \put(12.6,8){\line(0,-1){3}}
\put(5.2,7){\line(1,0){5.2}} \put(5.2,7){\line(0,-1){2}}
\put(10.5,7){\line(1,0){1.9}} \put(12.4,7){\line(0,-1){2}}
\put(10.4,7){\line(0,-1){2}} \put(10.5,7){\line(0,-1){2}}
\put(8,7.2){$\X_-\setminus \X_0$} \put(11.5,7.2){$\X_0$}
\end{picture}
\\
\begin{center}
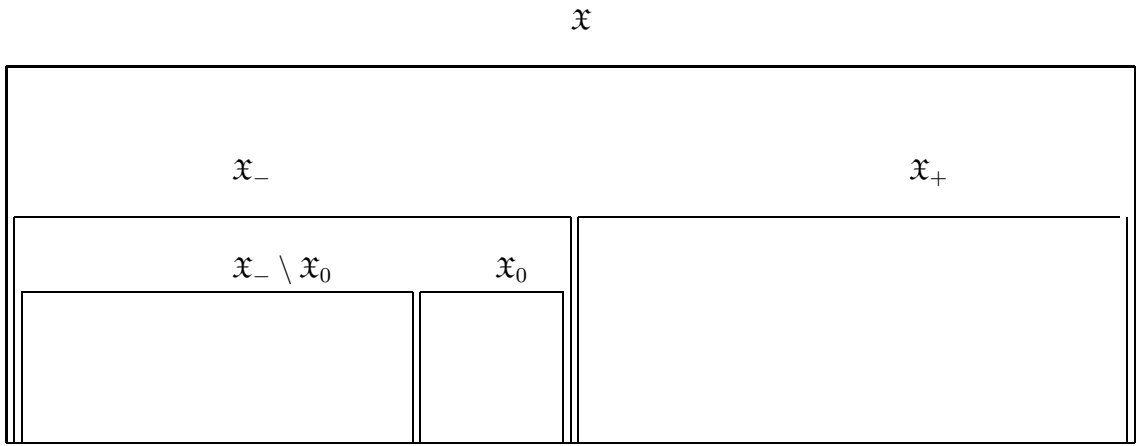
$\qquad\qquad$ Fig 1. The decomposition of the discrete set $\X$
\end{center}
\newpage
\begin{picture}(15,15) (7,1)
\put (6,6) {\circle*{0.3}} \put (7,6) {\circle*{0.3}}
\put(8,6){\circle{0.3}}
 \put (9,6) {\circle*{0.3}}
 \put
(10,6){\circle{0.3}} \put(11,6){\circle{0.3}}
 \put (12,6){\circle*{0.3}}
 \put (13,6) {\circle{0.3}}
  \put (14,6){\circle{0.3}}
\put (15,6){\circle{0.3}} \put (16,6) {\circle{0.3}}
\put(17,6){\circle{0.3}}
 \put(18,6){\circle{0.3}}
 \put (19,6){\circle{0.3}}
 \put (20,6) {\circle*{0.3}}
\put (21,6) {\circle{0.3}}
\put (5.5,5.5) {\line(1,0){8}}
 \put (13.5,5.5) {\line(0,1){0.5}}
 \put (5.5,5.5) {\line(0,1){0.5}}
\put (13.6,5.5) {\line(1,0){8}}
 \put (13.6,5.5) {\line(0,1){0.5}}
 \put (21.6,5.5) {\line(0,1){0.5}}
\put (6,6) {\line(0,1){5}} \put (6,6) {\line(1,1){3}}
 \put (7,6)
{\line(-1,5){1}} \put (7,6) {\line(2,3){2}}
\put(9,6){\line(0,1){3}}
 \put (9,6) {\line(-3,5){3}}
 \put (12,6) {\line(1,2) {2.4}}
 \put (20,6) {\line (-2,3) {1.8}}
\put (12,6) {\line(-1,1){3}}
\put (10.5,5.5) {\line(0,1){0.5}}
\put(8, 5.1) {$\frak{X}_-\setminus\frak{X}_0$}
\put(11.6,5.1){$\frak{X}_0$} \put(17.6,5.1){$\frak{X}_+$}
\put(6.8, 11) {$X_-^I$} \put(9.3, 8.9) {$X_-$} \put(14.8, 11)
{$X_-^{II}$}\put(18,8.9){$X_+$}
\end{picture}
\\

\begin{center}
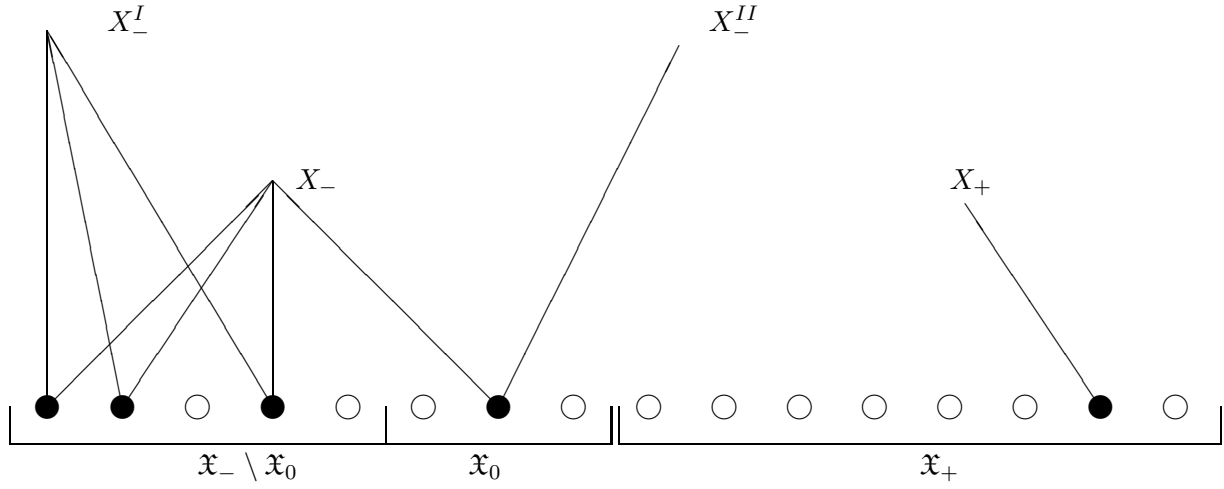
 Fig 2. Unbalanced particle-particle configurations.
The configuration $X$ consists of four negative particles and one
positive particle. The set $\frak{X}_0$ is chosen such that
$\arrowvert\frak{X}_0\arrowvert=\arrowvert
X_-\arrowvert-\arrowvert X_+\arrowvert$.
\end{center}
\newpage
\begin{picture}(15,15) (7,1)
\put (6,6) {\circle*{0.3}} \put (7,6) {\circle*{0.3}}
\put(8,6){\circle{0.3}}
 \put (9,6) {\circle*{0.3}}
 \put
(10,6){\circle{0.3}} \put(11,6){\circle{0.3}}
 \put (12,6){\circle*{0.3}}
 \put (13,6) {\circle{0.3}}
  \put (14,6){\circle{0.3}}
\put (15,6){\circle{0.3}} \put (16,6) {\circle{0.3}}
\put(17,6){\circle{0.3}}
 \put(18,6){\circle{0.3}}
 \put (19,6){\circle{0.3}}
 \put (20,6) {\circle*{0.3}}
\put (21,6) {\circle{0.3}}
\put (5.5,5.5) {\line(1,0){8}}
 \put (13.5,5.5) {\line(0,1){0.5}}
 \put (5.5,5.5) {\line(0,1){0.5}}
\put (13.6,5.5) {\line(1,0){8}}
 \put (13.6,5.5) {\line(0,1){0.5}}
 \put (21.6,5.5) {\line(0,1){0.5}}
\put (6,6) {\line(3,2){3}} \put (7,6) {\line(1,1){2}}
\put(9,6){\line(0,1){2}}
 \put (11,6) {\line(1,6) {1}}
 \put (13,6) {\line(-1,6) {1}}
 \put (20,6) {\line (-1,3) {2}}
 \put (11,6) {\line(3,2) {3}}
 \put (13,6) {\line(1,2) {1}}
 \put (20,6) {\line (-3,1) {6}}
\put (10.5,5.5) {\line(0,1){0.5}}
\put(8, 5.1) {$\frak{X}_-\setminus\frak{X}_0$}
\put(11.6,5.1){$\frak{X}_0$} \put(17.6,5.1){$\frak{X}_+$}
\put(14,8.3) {$Z_+$}
 \put(9.3, 8.3) {$Z_-$} \put(12.3, 12)
{$Z_+^{I}$}\put(18.5,12){$Z_+^{II}$}
\end{picture}
\\

\begin{center}
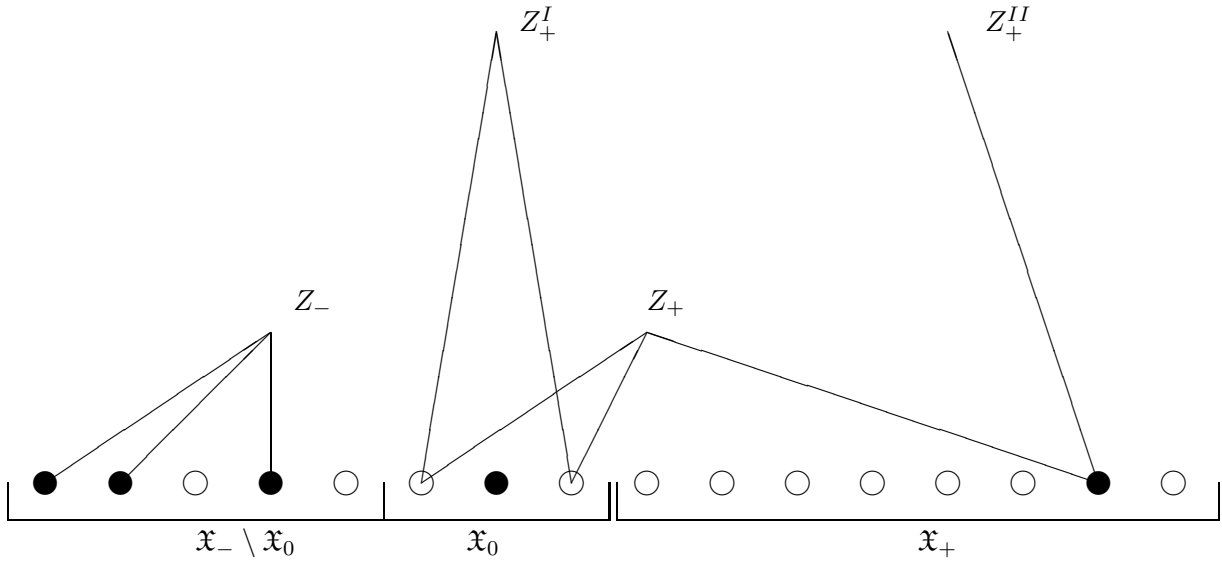
 Fig 3. Balanced particle-hole configurations. $Z_-$
consists of three negative particles, $\arrowvert
Z_-\arrowvert=3$. $Z_+$ consists of one positive particle,
$\arrowvert Z_+^{II}\arrowvert=1$, and two holes in $\frak{X}_0$,
$\arrowvert Z_+^I\arrowvert=2$
\end{center}
\newpage
\setlength{\evensidemargin}{-1.9cm}
\setlength{\oddsidemargin}{-1.9cm}
\begin{center}
\begin{table}
\begin{tabular}{|c|c|c|}
\hline
&  &  \\
Splitting of $\frak{X}$ & $\frak{X}=
  (\frak{X}_-\arrowvert \frak{X}_+)=\left(\frak{X}_-
  \setminus\frak{X}_0\sqcup \frak{X}_0\;\arrowvert\; \frak{X}_+\right)
  $
   &  $\frak{X}=(\hat{\frak{X}}_-\arrowvert \hat{\frak{X}}_+)=\left(\frak{X}_-
  \setminus\frak{X}_0\arrowvert \frak{X}_0\sqcup \frak{X}_+\right)$\\
  & & \\
& $\arrowvert\frak{X}_0\arrowvert=S$ & \\
& & \\
 \hline
  &  &  \\
Point Configurations & $X=(X_-\arrowvert X_+)=
\left(X_-^{I}\sqcup X_-^{II}\arrowvert X_+\right )$  & $Z=(Z_-\arrowvert Z_+)=\left(Z_-\arrowvert Z_+^{I}\sqcup Z_+^{II}\right)$  \\
& & \\
 & $X_-^I=X_-\cap (\frak{X}_-\setminus\frak{X}_0)$,\;
$X_-^{II}=X_-\cap \frak{X}_0$ & $Z_+^I=Z_+\cap \frak{X}_0$,\;
$Z_+^{II}=Z_+\cap\frak{X}_+$\\
 &  &  \\
\hline
   &  &  \\
   $L$-ensembles & $L=\left(\begin{array}{cc}
     0 & A \\
     -A^{T} & 0 \\
   \end{array}\right)$ & $\hat{L}=\left(\begin{array}{cc}
     0 & \hat{A} \\
     -\hat{A}^{T} & 0 \\
   \end{array}\right)$ \\
   & & \\
   & $A(x,y)=\frac{h(x)h(y)}{x-y}$ &
   $\hat{A}(x,y)=\frac{\hat{h}(x)\hat{h}(y)}{x-y}$\\
   & &  \\
  \hline
  & &  \\
   Relations  & $X_-^{I}=Z_-$ & $Z_- =X_-^{I}$ \\
   between configurations & $X_-^{II}=\frak{X}_0\setminus Z_+^I$ & $Z_+^{I} =\frak{X}_0\setminus X_-^{II}$ \\
      & $X_+=Z_{+}^{II}$ & $Z_+^{II} =X_+$ \\
   & & \\
  \hline
\end{tabular}
\caption{$L$-ensemble and $\hat{L}$-ensemble}
\end{table}
\end{center}
\newpage
\begin{center}
\begin{small}
\begin{table}
\begin{tabular}{|c|c|c|c|c|}
\hline
& & & & \\
& $L$-ensemble & $\triangle_N$-ensemble & $\hat{L}$-ensemble & $\hat{\triangle}_{N+S}$-ensemble\\
& & & & \\
\hline
& & & & \\
 Splitting of & $\frak{X}= (\frak{X}_-\arrowvert \frak{X}_+)$ &
$\frak{X}= (\frak{X}_-\arrowvert \frak{X}_+)$ & $\frak{X}=
(\hat{\frak{X}}_-\arrowvert \hat{\frak{X}}_+)$ & $\frak{X}=
(\hat{\frak{X}}_-\arrowvert
\hat{\frak{X}}_+)$\\
& & & & \\
$\frak{X}$ & $=\left(\frak{X}_-
  \setminus\frak{X}_0\sqcup \frak{X}_0\;\arrowvert\; \frak{X}_+\right)$
  & & $=\left(\frak{X}_-
  \setminus\frak{X}_0\;\arrowvert\;\frak{X}_0 \sqcup \frak{X}_+\right)$ &
   \\
& & & & \\
& $\arrowvert\frak{X}_0\arrowvert=S$; $\arrowvert\frak{X}_+\arrowvert=N$ & & $\arrowvert\hat{\frak{X}}_+\arrowvert=N+S $ & \\
& & & & \\
\hline
& & & & \\
Point & $X=(X_-\arrowvert X_+)$ & $X^{\triangle}=(X_-\arrowvert\frak{X}_+\setminus X_+)$ & $Z=(Z_-\arrowvert Z_+)$ & $Z^{\triangle}=(Z_-\arrowvert\hat{\frak{X}}_+\setminus Z_+)$ \\
& & & & \\
configurations & $X_-=X\cap \frak{X}_-$ & & $Z_-=Z\cap \hat{\frak{X}}_-$ & \\
& $X_+=X\cap \frak{X}_+$ & & $Z_+=Z\cap \hat{\frak{X}}_+$ & \\
& & & & \\
\hline
& & & & \\
Number of points & arbitrary $\leq N$ & $N$ & arbitrary $\leq N +S $ & $N+S$\\
& & & & \\
\hline
& & & & \\
Weight & $h$ & $f$ & $\hat{h}$ & $\hat{f}=f$ \\
& & & & \\
\hline
\end{tabular}
\caption{$L$-ensemble and polynomial ensemble}
\end{table}
\end{small}
\end{center}

\end{document}